\documentclass[prc,aps,tightenlines,eqsecnum,floatfix,showpacs]{revtex4}
\usepackage{dcolumn}
\usepackage{bm}
\usepackage{color}
\usepackage{amssymb}
\usepackage{amsmath}
\usepackage{graphicx}
\usepackage{pstricks}
\usepackage{float}
\begin{document}

\title{Electromagnetic structure of $A$=2 and 3 nuclei in chiral effective field theory}
\author{M.\ Piarulli$^{\rm a}$, L.\ Girlanda$^{\rm b}$, L.E.\ Marcucci$^{\rm c,d}$, S.\ Pastore$^{\rm e}$,
R.\ Schiavilla$^{\,{\rm a,f}}$, and M.\ Viviani$^{\,{\rm d}}$}

\affiliation{
$^{\rm a}$\mbox{Department of Physics, Old Dominion University, Norfolk, VA 23529, USA}\\
$^{\rm b}$\mbox{Department of Mathematics and Physics, University of Salento, and INFN-Lecce, I-73100 Lecce, Italy} \\
$^{\rm c}$\mbox{Department of Physics, University of Pisa, I-56127 Pisa, Italy}\\
$^{\,{\rm d}}$\mbox{INFN-Pisa, I-56127 Pisa, Italy}\\
$^{\,{\rm e}}$\mbox{Physics Division, Argonne National Laboratory, Argonne, IL 60439, USA} \\
$^{\rm f}$\mbox{Jefferson Lab, Newport News, VA 23606, USA}\\
}
\date{\today}

\begin{abstract}
The objectives of the present work are twofold.  The first is to address and resolve
some of the differences present in independent, chiral-effective-field-theory ($\chi$EFT)
derivations up to one loop, recently appeared in the literature, of the nuclear charge and
current operators.  The second objective is to provide a complete set of $\chi$EFT predictions
for the structure functions and tensor polarization of the deuteron, for the charge and magnetic
form factors of $^3$He and $^3$H, and for the charge and magnetic radii of these few-nucleon
systems.  The calculations use wave functions derived from high-order chiral two- and three-nucleon
potentials and Monte Carlo methods to evaluate the relevant matrix elements.  Predictions based
on conventional potentials in combination with $\chi$EFT charge and current operators are also
presented.  There is excellent agreement between theory and experiment for all these observables
for momentum transfers up to $q \lesssim 2.0$--2.5 fm$^{-1}$; for a subset of them, this agreement
extends to momentum transfers as high as $q \simeq 5$--6 fm$^{-1}$.  A complete analysis of the
results is provided.
\end{abstract}


\maketitle

\section{Introduction}
\label{sec:intro}
Nuclear electromagnetic charge and current operators in chiral effective
field theory ($\chi$EFT) up to one loop were derived originally by Park
{\it et al.}~\cite{Park96} in covariant perturbation theory.  Recently,
two independent derivations, based on time-ordered perturbation theory
(TOPT), have appeared in the literature, one by some of the present
authors~\cite{Pastore09,Pastore11} and the other by K\"olling
{\it et al.}~\cite{Koelling09,Koelling11}.  The expressions in
Refs.~\cite{Pastore09,Pastore11} and~\cite{Koelling09} for the two-pion-exchange
charge and current operators are in agreement with each other.
Differences between the expressions reported in
Refs.~\cite{Pastore09,Pastore11} and those in Ref.~\cite{Koelling11}
are found in some of the loop corrections to the one-pion-exchange (OPE) and
short-range currents as well as the minimal currents originating
from four-nucleon contact interactions involving two gradients of the
nucleon fields.  The differences in the loop corrections have their
origin in the different implementations of TOPT adopted in
Refs.~\cite{Pastore09,Pastore11} and Ref.~\cite{Koelling11}, and
relate to the treatment of reducible diagrams.  One of the objectives
of the present work is to resolve some of these differences.  This
is addressed in Sec.~\ref{sec:c-cnt} and Appendices~\ref{app:a1}
and~\ref{app:a2}.

The other objective is to provide predictions for the charge and magnetic
radii and form factors of the deuteron and trinucleons ($^3$He and $^3$H),
by utilizing two- and three-nucleon potentials derived
either in $\chi$EFT or in the conventional framework,  in combination with
the charge and current operators obtained here.  The methods
used to carry out the calculations are discussed in Sec.~\ref{sec:cal},
and a detailed analysis of the results is presented in Sec.~\ref{sec:res}.
This last section is organized into three subsections: the first illustrates
the different strategies adopted for the determination of the low-energy
constants (LEC's) that characterize the current operator up to one loop (no unknown
LEC's enter the one-loop charge operator); the second and third report results,
respectively, for the $A(q)$ and $B(q)$ structure functions and
tensor polarization $T_{20}(q)$ of the deuteron, and for the charge
and magnetic form factors of $^3$He and $^3$H, as well as results for the
charge and magnetic radii of these few-nucleon systems.  The conclusions are
summarized in Sec.~\ref{sec:con}, while details on the evaluation of the
loop integrals entering the charge operator are relegated in Appendix~\ref{app:loops}.

There have been earlier $\chi$EFT studies of the deuteron electromagnetic
structure in Refs.\cite{Walzl01,Phillips03,Phillips07} and, most recently,
in Ref.~\cite{Koelling12}---this latter work has focused on the $B(q)$ structure
function.  To the best of our knowledge, however, the one-loop $\chi$EFT predictions
reported here for the $^3$He and $^3$H elastic form factors are new. 
\section{Nuclear charge and current operators up to one loop}
\label{sec:c-cnt}
The two-nucleon current (${\bf j}$) and charge ($\rho$) operators have been
derived in $\chi$EFT up to one loop (to order $e\, Q$)
in Refs.~\cite{Pastore09} and~\cite{Pastore11}, respectively.  In the following,
we denote the momentum due to the external electromagnetic field with ${\bf q}$, and
define
\begin{eqnarray}
{\bf k}_i&=&{\bf p}_i^\prime-{\bf p}_i \ ,\qquad {\bf K}_i=\left({\bf p}_i^\prime+{\bf p}_i\right)/2 \ ,\\
{\bf k}&=&\left({\bf k}_1-{\bf k}_2\right)/2 \ , \qquad {\bf K}= {\bf K}_1+{\bf K}_2 \ ,
\label{eq:ppp}
\end{eqnarray}
where ${\bf p}_i$ (${\bf p}_i^\prime$) is the initial (final) momentum of nucleon $i$.  We
further define
\begin{equation}
{\bf j}=\sum_{n=-2}^{+1} {\bf j}^{(n)} \ ,
\qquad \rho=\sum_{n=-3}^{+1} \rho^{(n)} \ ,
\end{equation}
where the superscript $n$ in ${\bf j}^{(n)}$ and $\rho^{(n)}$ specifies
the order $e\, Q^n$ in the power counting.
The lowest-order (LO) contributions ${\bf j}^{(-2)}$ and $\rho^{(-3)}$ consist of the
single-nucleon current and charge operators,
respectively:
\begin{eqnarray}
\label{eq:jlo}
{\bf j}^{(-2)}&=&\frac{e}{2\, m_N}
\left[ \,2\, e_{N,1}(q^2) \, {\bf K}_1 
+i\,\mu_{N,1}(q^2)\, {\bm \sigma}_1\times {\bf q }\,\right] \nonumber\\
&&\times \delta({\bf p}^\prime_2-{\bf p}_2)+
 1 \rightleftharpoons 2\ ,
 \end{eqnarray}
 and
 \begin{equation}
 \label{eq:rlo}
\rho^{(-3)} = e\, e_{N,1}(q^2) \, \delta({\bf p}^\prime_2-{\bf p}_2) + 1\rightleftharpoons 2 \ ,
\end{equation}
where $m_N$ is the nucleon mass,
${\bf q}={\bf k}_i$ with $i=1$ or 2 (the $\delta$-functions enforcing
overall momentum conservation ${\bf q}={\bf k}_1$ have been dropped for simplicity here and
in the following),
\begin{eqnarray}
e_{N,i}(q^2) &=& \frac{G_E^S(q^2)+G_E^V(q^2)\, \tau_{i,z}}{2}\ , \nonumber \\ 
 \mu_{N,i}(q^2) &=&\frac{G_M^S(q^2)+G_M^V(q^2)\, \tau_{i,z}}{2} \ ,
\label{eq:ekm}
\end{eqnarray}
and $G^{S/V}_E$ and $G^{S/V}_M$ denote the isoscalar/isovector
combinations of the proton and neutron electric ($E$) and magnetic ($M$)
form factors, normalized as $G^S_E(0)=G^V_E(0)=1$, $G^S_M(0)=0.880
\, \mu_N$, and $G^V_M(0)=4.706\, \mu_N$ in units of the nuclear magneton
$\mu_N$.  The counting $e\, Q^{-2}$ ($e\, Q^{-3}$) of the leading-order current
(charge) operator results from the product of a factor $e\, Q$ ($e\, Q^0$) due to
the coupling of the external electromagnetic field to the individual nucleons, and
the factor $Q^{-3}$ from the momentum $\delta$-function entering this type of
disconnected contributions.  Of course, this counting ignores the fact that the
nucleon form factors themselves also have a power series expansion in $Q$.
Here, they are taken from fits to elastic electron scattering data off the proton
and deuteron~\cite{Hyde04}---specifically, the H\"ohler parametrization~\cite{Hohler76}---rather
than derived consistently in chiral perturbation theory ($\chi$PT)~\cite{Kubis01}.
The calculations of the $A=2$ and 3 nuclei elastic form factors that follow are carried out
in the Breit frame, in which the electron-energy transfer vanishes.  Hence,
the hadronic electromagnetic form factors are evaluated at four-momentum transfer
$q^\mu q_\mu=-q^2$.

At order $n=\! -1$ (NLO) there is a one-pion exchange (OPE) contribution to the current
operator which reads
\begin{eqnarray}
 {\bf j}^{(-1)}&=& -i\, e\frac{g^2_A}{F^2_\pi}\,G^V_E(q^2)\,
 ({\bm \tau}_1 \times {\bm \tau}_2)_z 
 \left( {\bm \sigma}_1 -{\bf k}_1\,\frac{{\bm \sigma}_1\cdot {\bf k}_1} {\omega_{k_1}^2}  \right)\nonumber\\
&&\times \frac{{\bm \sigma}_2\cdot {\bf k}_2}{\omega^2_{k_2}} + 1 \rightleftharpoons 2 \ ,
 \label{eq:nlo1} 
 \end{eqnarray}
where we have defined $\omega_k^2=k^2+m_\pi^2$, $m_\pi$ being the 
pion mass.
However, there are no $n=\!-2$ contributions to the charge operator.  The presence
of the isovector electric form factor $G_E^V$ in ${\bf j}^{(-1)}$ follows from the continuity
equation
\begin{equation}
\label{eq:conty}
{\bf q}\cdot {\bf j}^{(-1)}=\left[\, v^{(0)}_\pi \, ,\, \rho^{(-3)} \,\right ] \ ,
\end{equation}
where $\left[\dots\, ,\, \dots\right]$ denotes the commutator, 
$\rho^{(-3)}$ is the charge operator given in Eq.~(\ref{eq:rlo}), and
$v^{(0)}_\pi$ is the static OPE potential
\begin{equation}
\label{eq:ope}
v^{(0)}_{\pi}({\bf k})=-\frac{g_A^2}{F_{\pi}^2}\,{\bm \tau}_1\cdot{\bm \tau}_2\,
\frac{{\bm \sigma}_1\cdot{\bf k}\,{\bm\sigma}_2\cdot{\bf k}}{\omega_k^2} \ .
\end{equation}
The l.h.s.~of Eq.~(\ref{eq:conty}) is of order $Q^0$, the same as
the r.h.s.~since the commutator
brings in an additional factor $Q^3$ due to the implicit momentum
integrations.  It should be emphasized that the continuity equation 
requires that the same form factor be used to describe the electromagnetic
structure of the hadrons in the longitudinal part of the current operator and
in the charge operator.  However, it places no restrictions on the electromagnetic
form factors which may be used in the transverse parts of the current.  Ignoring
this ambiguity, the choice made here ($G_E^V$) satisfies the ``minimal'' requirement
of current conservation~\cite{Carlson98}.

\begin{figure}
\includegraphics[width=3in]{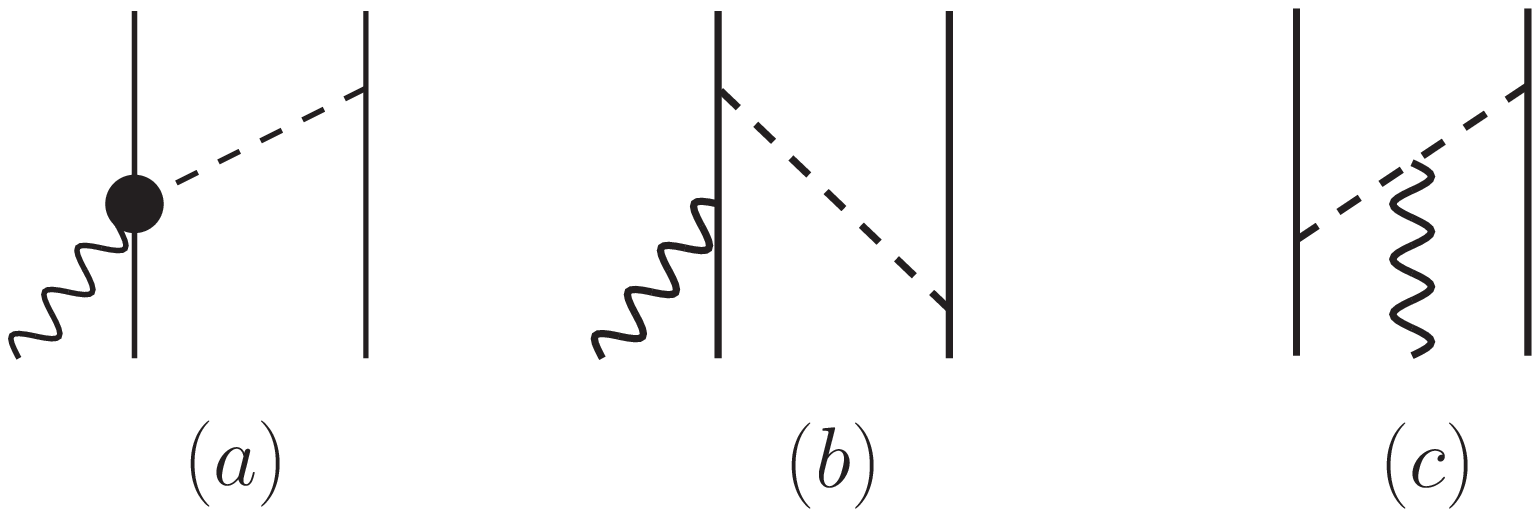}\\
\caption{Diagrams illustrating the two-body charge
operators at order $n=0$ or $e\, Q^0$.  Nucleons, pions
and photons are denoted by solid, dashed, and wavy lines,
respectively.  The solid circle in panel (a) is associated
with a $\gamma\pi N$ vertex of order $e\, Q$.
Only one among the possible time orderings is shown.}
\label{fig:f1}
\end{figure}

Relativistic corrections to the leading order one-body current and charge
operators enter, respectively, at $n=0$ and $n=-1$ (both denoted as N2LO), and are given by
\begin{eqnarray}
\label{eq:j1rc}
 \!\!\!\!{\bf j}^{(0)}&=&-\frac{e}{8 \, m_N^3}
 e_{N,1}(q^2) \, \Big[ 
2\, \left( K_1^2 +q^2/4 \right)  \big( 2\, {\bf K}_1 \nonumber\\
\!\!\!\!&&\!\!\!\!+i\, {\bm \sigma}_1\times {\bf q } \big) 
+ {\bf K}_1\cdot {\bf q}\, \left({\bf q} +2\, i\, {\bm \sigma}_1\times {\bf K }_1 \right)\Big]\nonumber\\ 
\!\!\!\!&&\!\!\!\!- \frac{i\,e}{8 \, m_N^3}
\left[\, \mu_{N,1}(q^2)-e_{N,1}(q^2)\, \right]
 \Big[ {\bf K}_1\cdot {\bf q}\nonumber\\
\!\!\!\!&&\!\!\!\!\times
 \big( 4\, {\bm \sigma}_1\times {\bf K}_1-i\, {\bf q}\big) 
 - \left(  2\, i\, {\bf K}_1 -{\bm \sigma}_1\times {\bf q} \right)\, q^2/2 \nonumber \\
\!\!\!\! &&\!\!\!\!  +2\, \left({\bf K}_1\times {\bf q}\right)
 \, {\bm \sigma}_1\cdot {\bf K}_1 \Big] \, \delta({\bf p}^\prime_2-{\bf p}_2) + 1 \rightleftharpoons 2  \ ,\\
\!\!\!\!\rho^{(-1)}\!\!&=& -\frac{e}{8 \, m_N^2} \, \left[\, 2\, \mu_{N,1}(q^2)-e_{N,1}(q^2)\right] 
\big(q^2 \nonumber\\
\!\!\!\!&&\!\!\!\!+2\, i\, {\bf q}\cdot {\bm \sigma}_1\times {\bf K }_1\big) \, \delta({\bf p}^\prime_2-{\bf p}_2) 
+ 1 \rightleftharpoons 2  \ ,
\label{eq:r1rc}
\end{eqnarray}
while the $n=0$ (N3LO) OPE two-body charge operators,
illustrated in Fig.~\ref{fig:f1}, read
\begin{eqnarray}
\label{eq:pich}
\rho^{(0)}_{\rm a}&=&\frac{e}{2\, m_N} \frac{g_A^2}{F_\pi^2} \left[ G_E^S(q^2)\,{\bm \tau}_1 \cdot {\bm \tau_2}
+ G_E^V(q^2) \,\tau_{2z}\right]\nonumber\\
&&\times \frac{{\bm \sigma}_1 
\cdot {\bf q} \,\, {\bm \sigma}_2 \cdot {\bf k}_2}{\omega^2_{k_2}} + 1 \rightleftharpoons 2 \ , \\
\label{Diagram_dnu2}
 \rho_{\rm b}^{(0)}(\nu)&=&-\frac{e}{4\,m_N}\frac{g_A^2}{F_{\pi}^2}
\frac{{\bm \sigma}_1\cdot{\bf k}_2\,\,
 {\bm \sigma}_2\cdot{\bf k}_2}{\omega_{k_2}^4}
 \Big[(1-\nu) \nonumber\\
&&\times \left[ G_E^S(q^2)\, {\bm \tau}_1\cdot {\bm \tau}_2
 +G_E^V(q^2)\, \tau_{2,z}\right]\,{\bf q}\cdot {\bf k}_2\nonumber\\
&&+2\,i\,G_E^V(q^2)\, ({\bm \tau}_1\times {\bm \tau}_2)_z\,{\bf k}_2\cdot
\big[\,(1-\nu)\,{\bf K}_1 \nonumber\\
&&+(1+\nu)\, {\bf K}_2\, \big]\Big]+1\rightleftharpoons2\ , \\
\rho_{\rm c}^{(0)}&=&i \frac{e}{m_N}\, \frac{g_A^2}{F_\pi^2}\, G_\pi(q^2)\,
\left({\bm \tau}_1\times{\bm \tau}_2\right)_z\,{\bf k}_1\cdot{\bf K}_1\nonumber\\
&&\times\frac{{\bm \sigma}_1\cdot{\bf k}_1\,{\bm \sigma}_2
\cdot{\bf k}_2}{\omega_{k_1}^2\,\omega_{k_2}^2}
 + 1 \rightleftharpoons 2  \ .
\label{eq:rho_gpipi}
\end{eqnarray}
The operator of panel (a) is due to a $\gamma \pi N$ vertex
of order $e\, Q$ originating from the interaction Hamiltonian
\begin{equation}
\frac{e\, g_A}{2\, m_N F_\pi}\int {\rm d}{\bf x}\, 
N^\dagger\, {\bm \sigma}\cdot\left( {\bm \nabla} A^0\right)\, 
 \left( {\bm \tau}\cdot {\bm \pi}+\pi_z \right) N  \ , \nonumber
\end{equation}
derived first by Phillips~\cite{Phillips03}.  In the context of meson-exchange
phenomenology, an operator of precisely this form
results from considering the low-energy limit of the relativistic Born diagrams
associated with virtual pion photo-production amplitudes, see the review paper~\cite{Riska89}
and references therein.  From this perspective, it appears reasonable to include the
nucleon form factors $G_E^S$ and $G_E^V$ in Eq.~(\ref{eq:pich}).

The operator of panel (b) depends on the off-energy-shell extrapolation,
specified by the parameter $\nu$, adopted for the non-static corrections
of order $Q^2$ to the OPE potential~\cite{Friar77},
\begin{equation}
\label{eq:ope2}
v^{(2)}_{\pi}({\bf k},{\bf K};\nu)=\left(1-2\, \nu\right)\,\frac{v_\pi^{(0)}({\bf k})}{\omega_k^2}\,
\frac{\left( {\bf k}\cdot {\bf K}\right)^2}{4\, m_N^2} \ .
\end{equation}
As shown in Ref.~\cite{Friar77} (and within the present approach in Ref.~\cite{Pastore11}), different
off-shell prescriptions for $v^{(2)}(\nu)$
and $\rho^{(0)}(\nu)$ are unitarily equivalent:
\begin{eqnarray}
\!\!\!\!\rho^{(-3)}\!+\!\rho^{(0)}_{\rm b}(\nu)&=&{\rm e}^{-i\, U(\nu)} \left[\rho^{(-3)}+
\rho^{(0)}_{\rm b}(0) \right] \, {\rm e}^{+i\, U(\nu)} \nonumber \\
\!\!\!\!&\simeq& \rho^{(-3)}\!+\!\rho^{(0)}_{\rm b}(0)\!+\!\left[  \rho^{(-3)} , i\, U^{(0)}(\nu) \right] ,
\end{eqnarray}
where the hermitian operator $U(\nu)$ admits the expansion
\begin{equation}
U(\nu) = U^{(0)}(\nu)+U^{(1)}(\nu) + \dots \ ,
\end{equation}
and $U^{(0)}(\nu)$ and $U^{(1)}(\nu)$ (see below) have been constructed, respectively,
in Refs.~\cite{Friar77} and~\cite{Pastore11} (in this last paper, Eqs.~(28) and (55),
which give equivalent momentum-space expressions for $U^{(1)}(\nu)$, contain
a typographical error: the imaginary unit on the l.h.s. should be removed).
Phenomenological potentials, such as the Argonne $v_{18}$ (AV18)~\cite{Wiringa95}, and
$\chi$EFT potentials, such as those recently derived by Entem and Machleidt~\cite{Entem03},
make the choice $\nu=1/2$ in Eq.~(\ref{eq:ope2}), i.e., ignore non static
corrections to the OPE potential.

The operator of panel (c), containing the $\gamma \pi\pi$ vertex, is
obtained by expanding the energy denominators as~\cite{Pastore11}
\begin{equation}
\frac{1}{E_i-E_I-\omega_\pi}= -\frac{1}{\omega_\pi}
\left[ 1 + \frac{E_i-E_I}{\omega_\pi}  + \dots\right] \ ,
\label{eq:deno}
\end{equation}
where $E_I$ denotes $NN$ (or $NN\gamma$) intermediate energies and
$\omega_\pi$ the pion energy (or energies, as the case may be), and 
by noting that the leading (static) corrections vanish, when summed
over the possible six time orderings.  However, the terms proportional
to the ratio $(E_i-E_I)/\omega_\pi$, which is of order $Q$, lead
to the non-static operator given in Eq.~(\ref{eq:rho_gpipi}).  It is
multiplied by the pion form factor $G_\pi(q^2)$, which we parametrize
in vector-meson dominance and consistently with experimental data at 
low momentum transfers as
\begin{equation}
G_\pi(q^2)=\frac{1}{1+q^2/m_\rho^2} \ ,
\end{equation}
where $m_\rho$ is the $\rho$-meson mass.
\subsection{Current operators at order $n=1$ ($e\, Q$)}
\label{sec:jsloop}

\begin{figure}
\includegraphics[width=3in]{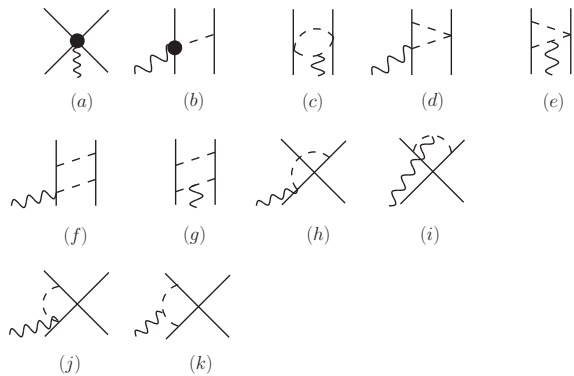}\\
\caption{Diagrams illustrating the two-body current
operators at order $n=1$ or $e\, Q$.  Nucleons, pions
and photons are denoted by solid, dashed, and wavy lines,
respectively.  The solid circle in panel (b) is associated
with a $\gamma\pi N$ vertex of order $e\, Q^2$.
Only one among the possible time orderings is shown.}
\label{fig:f2}
\end{figure}

The currents at order $e\, Q$ (N3LO) are illustrated diagrammatically in Fig.~\ref{fig:f2},
and consist of: (i) terms generated by minimal substitution in the four-nucleon
contact interactions involving two gradients of the nucleon fields as well as by non-minimal
couplings to the electromagnetic field; (ii) OPE terms induced by
$\gamma \pi N$ interactions beyond leading order; and (iii) one-loop
two-pion-exchange (TPE) terms.  We discuss them below.

The contact minimal and non minimal currents, denoted by the
subscripts \lq\lq min'' and \lq\lq nm'' respectively, are written as
\begin{eqnarray}
\label{eq:jmin}
{\bf j}^{(1)}_{\rm a, min}&=&\frac{i\, e}{16}\, G_E^V(q^2)\, \left({\bm \tau}_1\times{\bm \tau}_2\right)_z\,
\Big[ (C_2+3\, C_4+C_7)\, {\bf k}_1 \nonumber\\
&&+( C_2-C_4-C_7)\, {\bf k}_1 \,\,{\bm \sigma}_1\cdot{\bm \sigma}_2 \nonumber\\
&& +C_7\, {\bm \sigma}_1\cdot ({\bf k}_1-{\bf k}_2)\,\, {\bm \sigma}_2 \Big]
-\frac{i\, e}{4}\, e_{N,1}(q^2)\,C_5\, \nonumber\\
 &&\times ({\bm \sigma}_1+{\bm \sigma}_2)
\times {\bf k}_1 + 1 \rightleftharpoons 2 \ , \\
\label{eq:nmcounter}
{\bf j}^{(1)}_{\rm a, nm}&=& - i\,e \Big[  G_E^S(q^2)\,C_{15}^\prime\, {\bm \sigma}_1 
+ G_E^V(q^2)\, C_{16}^\prime\nonumber\\
 &&\times (\tau_{1,z} - \tau_{2,z})\,{\bm \sigma}_1  \Big]\times {\bf q}  
+ 1 \rightleftharpoons 2 \ .
\end{eqnarray}
The expression above for ${\bf j}^{(1)}_{\rm a,min}$ is
the Fierz-transformed version of the current given in Eq.~(3.11) of Ref.~\cite{Pastore09},
see App.~\ref{app:a1} for a derivation.  We note that the first three terms in Eq.~(\ref{eq:jmin})
agree with the first line of Eq.~(5.3) of K\"olling {\it et al.}~\cite{Koelling11}, while
the term proportional to $C_5$ differs by the isoscalar
piece, which, however, can be absorbed in a redefinition of $C_{15}^\prime$.
The low-energy constants (LEC's) $C_1,\dots, C_7$,
which also enter the two-nucleon contact potential, have been constrained by fitting $np$
and $pp$ elastic scattering data and the deuteron binding energy.  We take their values
from the Machleidt and Entem 2011 review paper~\cite{Entem03}.  The LEC's $C_{15}^\prime$
and $C_{16}^\prime$ (and $d_8^\prime$, $d_9^\prime$, and $d_{21}^\prime$ below) are determined
by fitting measured photo-nuclear observables of the $A=2$ and $3$ systems, as
discussed in Sec.~\ref{sec:res}.
Finally, we observe that there is no {\it a priori} justification for the use
of $G_E^S/G_E^V$ (or $G_M^S/G_M^V$) in the non-minimal
contact current, and these form factors are included in order to provide a reasonable fall-off
with increasing $q^2$ for the strength of this current. 

The isovector (IV) OPE current at N3LO is given by
\begin{eqnarray}
\label{eq:cdlt}
{\bf j}^{(1)}_{\rm b,IV}&=&i\, e\, \frac{g_A}{F_\pi^2} \frac{G_{\gamma N\Delta}(q^2)}{\mu_{\gamma N\Delta}}\,
\frac{{\bm \sigma}_2 \cdot {\bf k}_2}{\omega_{k_2}^2}\,
\bigg[  d_8^\prime \tau_{2,z} \,{\bf k}_2 \nonumber\\
&&-d_{21}^\prime ({\bm \tau}_1\times{\bm \tau}_2)_z\, {\bm \sigma}_1\times {\bf k}_2  \bigg] \times {\bf q} + 1\rightleftharpoons 2 \ ,
\end{eqnarray}
and depends on the two (unknown) LEC's $d_8^\prime$ and $d_{21}^\prime$.
They can be related~\cite{Pastore09} to the $N$-$\Delta$ transition axial
coupling constant and magnetic moment (denoted as $\mu_{\gamma N\Delta}$) in a resonance
saturation picture, which justifies the use of the $\gamma N\Delta$
electromagnetic form factor for this term.  It is parametrized as
\begin{equation}
G_{\gamma N \Delta}(q^2)= \frac{\mu_{\gamma N \Delta} }
{( 1+q^2/\Lambda_{\Delta,1}^2 )^2
\sqrt{1+q^2/\Lambda_{\Delta,2}^2} } \ ,
\end{equation}
where $\mu_{\gamma N \Delta}$ is taken as 3 $\mu_N$ from
an analysis of $\gamma N$ data
in the $\Delta$-resonance region~\cite{Carlson86}.  This
analysis also gives $\Lambda_{\Delta,1}$=0.84 GeV and
$\Lambda_{\Delta,2}$=1.2 GeV.  
The isoscalar (IS) piece of the OPE current depends on the LEC $d^\prime_9$ mentioned earlier,
\begin{equation}
{\bf j}^{(1)}_{\rm b,IS}= i\, e\, \frac{g_A}{F_\pi^2} \, d^\prime_9\,G_{\gamma\pi\rho}(q^2) \, 
 {\bm \tau}_1\cdot{\bm \tau}_2\,\frac{{\bm \sigma}_2 \cdot {\bf k}_2}{\omega_{k_2}^2}\,
  {\bf k}_2 \times {\bf q}
+ 1\rightleftharpoons 2 \ ,
\label{eq:cp}
\end{equation}
and, again in a resonance saturation picture, reduces to the well known
$\gamma\pi\rho$ current~\cite{Pastore09}.  Accordingly, we have accounted for
the $q^2$ fall-off of the electromagnetic vertex by including a $\gamma \pi \rho$
form factor, which in vector-meson dominance is parametrized as
\begin{equation}
G_{\gamma\pi\rho}(q^2)=\frac{1}{1+q^2/m_\omega^2} \ ,
\end{equation}
$m_\omega$ is the $\omega$-meson mass.
We can now clarify the differences in these tree-level currents as reported here and
in Ref.~\cite{Koelling11}.  We first note that the relations between the primed $d_i^{\, \prime}$
and $d_i$ in Ref.~\cite{Pastore09} should have read: 
$d_8^{\, \prime}= -8\, d_8$, $d_9^{\, \prime}= -8\, d_9$, $d_{21}^{\, \prime}=2\, d_{21}-d_{22} $.
The term proportional to $d_{22}$ originates from the Lagrangian
\begin{equation}
{\rm term} \propto d_{22} = \frac{2\, e}{F_\pi}\, d_{22} \, N^\dagger S^\mu
\left[ \partial^\nu({\bm \tau}\times{\bm \pi})_z\, F_{\mu\nu} \right] N \ .
\end{equation}
K\"olling {\it et al.}~\cite{Koelling11} integrate it by parts to obtain
\begin{eqnarray}
{\rm term} \propto d_{22} &=&-\frac{2\, e}{F_\pi}\, d_{22} \Bigg[
 (\partial^\nu N)^\dagger S^\mu
\, ({\bm \tau}\times{\bm \pi})_z\, F_{\mu\nu}  N \nonumber\\
&&+ N^\dagger S^\mu ({\bm \tau}\times{\bm \pi})_z\, F_{\mu\nu} (\partial^\nu N) \Bigg] \ ,
\end{eqnarray}
while the authors of Ref.~\cite{Pastore09} use the equations of motion
for the electromagnetic field tensor at leading order, $\partial^\nu F_{\mu\nu}=0$,
to express it as
\begin{equation}
 {\rm term} \propto d_{22} =\frac{2\, e}{F_\pi}\, d_{22} \, N^\dagger \, S^\mu\,
F_{\mu\nu}\,({\bm \tau}\times\partial^\nu{\bm \pi})_z\,  N .
\end{equation}
These two different treatments lead to the $d_{22}$ current as given in Ref.~\cite{Pastore09}
and~\cite{Koelling11}.  They differ by a term proportional to $({\bm \sigma}_1\times{\bf q})\times{\bf q}$,
which does not contribute to the magnetic moment ($M1$) operator ${\bm \mu}=-(i/2)\, {\bm \nabla}_q \times {\bf j}\big|_{q=0}$~\cite{Pastore09}.  Similarly, the term proportional to $f_5(q)$ in Eq.~(4.28)
of Ref.~\cite{Koelling11} does not give any contribution to ${\bm \mu}$, since $f_5(q) \propto q^2$ for small
$q$.   The term proportional to $f_6(q)$ is included in Eq.~(\ref{eq:nlo1}) provided $g^2_A \longrightarrow
g^2_A\left( 1-d_{18}\, m_\pi^2/g_A\right)$.  The value adopted here for $g_A$ is obtained from
two-nucleon scattering data (Sec.~\ref{sec:res}).
Therefore, for processes induced by $M1$ transitions, such as the $n\, d$ and $n\, ^3$He radiative
captures at thermal neutron energies studied in Ref.~\cite{Girlanda10a} or the magnetic scattering
under consideration in this work, the differences above are irrelevant.

The one-loop TPE currents, diagrams (c)--(k) of Fig.~\ref{fig:f2}, are written as
\begin{eqnarray}
\label{eq:jloop}
{\bf j}^{(1)}_{\rm loop}\!\! &=&\!\! -i\, e\, G_E^V(q^2)\, ({\bm \tau}_1\times {\bm \tau}_2)_z \,{\bm \nabla}_{\! k}\,
F_1(k) +i\,e\, G_E^V(q^2)\,\tau_{2,z}\nonumber\\
&&\!\!\times\!\! \left[F_0(k) \,{\bm \sigma}_1 - F_2(k)\, \frac{{\bf k}\,{\bm \sigma}_1\cdot{\bf k}}{k^2}\right]\times {\bf q} 
+ 1 \rightleftharpoons 2 \ ,
\end{eqnarray}
where the functions $F_i(k)$ are
\begin{eqnarray}
F_0(k)&=&\frac{g_A^2}{8\,\pi^2F_{\pi}^4}\,\Bigg[
1 -2\, g_A^2+\frac{  8\,g_A^2\, m_\pi^2 }{k^2+4\, m_\pi^2}
+G(k)\bigg[ 2-2\, g_A^2 \nonumber\\
&&-\frac{  4\,(1+g_A^2)\, m_\pi^2 }{k^2+4\, m_\pi^2} 
+\frac{16\, g_A^2 \,m_\pi^4 }{(k^2+4\, m_\pi^2)^2} \bigg] \Bigg]\ ,
\label{eq:f0k} \\
F_1(k)&=&\frac{1}{96 \, \pi^2\,F_{\pi}^4}\,
G(k) \bigg[4 m_{\pi}^2(1+4 g_A^2-5 g_A^4)\nonumber\\
&&+k^2(1+10 g_A^2 - 23 g_A^4)
-\frac{48\, g_A^4 m^4_\pi}
{4\,  m^2_\pi+k^2}\bigg] \ ,
\label{eq:f1k}\\
F_2(k)&=&\frac{g_A^2}{8\,\pi^2F_{\pi}^4}\,\Bigg[
2-6\, g_A^2+ \frac{  8\,g_A^2\, m_\pi^2 }{k^2+4\, m_\pi^2}
+G(k) \bigg[4\, g_A^2\nonumber\\
&&-\frac{  4\,(1+3\, g_A^2)\, m_\pi^2 }{k^2+4\, m_\pi^2} 
+\frac{16\, g_A^2 \,m_\pi^4 }{(k^2+4\, m_\pi^2)^2} \bigg] \Bigg]\ , 
\label{eq:f2k}
\end{eqnarray}
and the loop function $G(k)$ is defined as
\begin{equation}
G(k)=\frac{\sqrt{4\,m_{\pi}^2+k^2}}{k}\ln 
\frac{\sqrt{4\,m_{\pi}^2+k^2}+k}{\sqrt{4\,m_{\pi}^2+k^2}-k} \ .
\label{eq:loopf}
\end{equation}
The expression above results from expanding ${\bf j}^{(1)}_{\rm loop}({\bf q},{\bf k})$
in a power series in ${\bf q}$ 
as ${\bf j}^{(1)}_{\rm loop}({\bf q},{\bf k})=
{\bf j}^{(1)}_{\rm loop}(0,{\bf k})-i\, {\bf q}\times {\bm \mu}^{(0)}({\bf k}) +\dots$,
where ${\bm \mu}^{(0)}$ is the magnetic dipole operator, and ${\bf j}^{(1)}_{\rm loop}(0,{\bf k})$,
which corresponds to the first term in Eq.~(\ref{eq:jloop}), satisfies
current conservation with the TPE potential $v^{(2)}_{2\pi}({\bf k})$
(of order $Q^2$), since
\begin{eqnarray}
\!\!\!\!\!\!\!\!\!\left[ v^{(2)}_{2\pi}({\bf k})\, ,\, \rho^{(-3)} \right]
&=&e\, \left[ v^{(2)}_{2\pi}({\bf k}-{\bf q}/2)\, ,\,  e_{N,1} \right] \!+\!
 1\!\rightleftharpoons \! 2 \nonumber \\
\!\!\!\!\!\!\!\!\!\!\!&\simeq&-i\, e\, ({\bm \tau}_1\times{\bm \tau}_2)_z\, {\bf q}\cdot {\bm \nabla}_k F_1(k)\!+\! 1\!\rightleftharpoons \! 2
\end{eqnarray}
to leading order in ${\bf q}$.  In fact, the current ${\bf j}_{\rm loop}^{(1)}(0,{\bf k})$
is proportional to the electric dipole operator, and does not contribute to elastic
electromagnetic transitions, such as those of interest here. 

Finally, we note that a more careful analysis, detailed in Appendix~\ref{app:a2}, of
the loop short-range currents corresponding to diagrams (h)-(k) in Fig.~\ref{fig:f2}
shows that they vanish, in contrast to that which was reported in Ref.~\cite{Pastore09}
and in agreement with the result of Ref.~\cite{Koelling11}.
\subsection{Charge operators at order $n=1$ ($e\, Q$)}
\label{sec:cloop}

\begin{figure}
\includegraphics[width=4in]{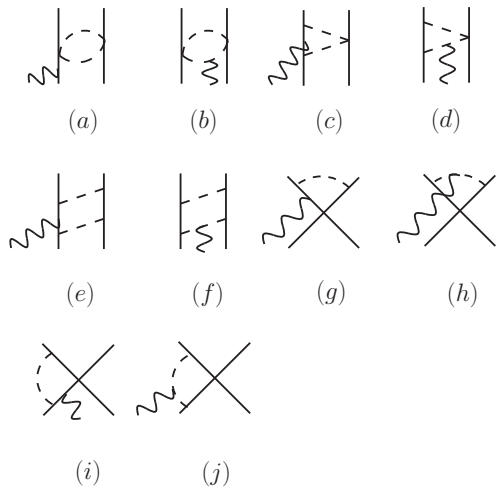}\\
\caption{Diagrams illustrating two-body charge operators
entering at order $n=1$ or $e\, Q$.  Nucleons, pions, and photons
are denoted by the solid, dashed, and wavy lines, respectively.
Only one among the possible time orderings is shown.}
\label{fig:f3}
\end{figure}

The two-body charge operators at one loop (N4LO) are illustrated in Fig.~\ref{fig:f3},
and have been derived in Ref.~\cite{Pastore11}.  The contributions
from diagrams of type (a)-(b) and (g)-(h) vanish, and after carrying out the loop
integrations (discussed in App.~\ref{app:loops}), those from diagrams of type (c)-(f) and (i)-(j) read:
 \begin{eqnarray}
 \label{eq:c_c}
\!\!\!  \rho_{\rm c}^{(1)}\!&=&\!-e\,\frac{1}{2\,\pi}\,\frac{g_A^2}{F_{\pi}^4}\,G_E^V(q^2)\, \tau_{2,z}
 \int_{0}^{1/2}\!{\rm d}x\, \bigg[4\, L(x,k_2)\nonumber\\
&&-\frac{m_{\pi}^2}{L(x,k_2)}\bigg]+1\rightleftharpoons 2\ , \\
\label{eq:c_d}
\!\!\! \rho_{\rm d}^{(1)}\!&=&\!e\,\frac{1}{2\,\pi}\,\frac{g_A^2}{F_{\pi}^4}\, G_\pi(q^2)\, \tau_{2,z}
 \int_{0}^{1/2}\!{\rm d}x\, \bigg[ 4\, L(x,k_1)\nonumber\\
&&-\frac{m_{\pi}^2}{L(x,k_1)}\bigg]+1\rightleftharpoons 2\ ,
\end{eqnarray}
\vspace{-1cm}
\begin{widetext}
\begin{eqnarray}
 \label{eq:c_e}
\!\!\!  \rho_{\rm e}^{(1)}(\nu)\! &=&\!-e\,\frac{1}{16\,\pi}\,\frac{g_A^2}{F_{\pi}^4}\,G_E^V(q^2)\,
\int_{0}^{1/2}\!{\rm d}x\, \Bigg[\left[4\,\tau_{2,z}+\nu\, \left({\bm \tau}_1\times{\bm \tau}_2\right)_z\right]
 \bigg[-24\, L(x,k_2)
+\frac{k_2^2+8\, m_\pi^2}{L(x,k_2)} +\frac{m_\pi^4}{L^3(x,k_2)} \bigg]\nonumber\\
&&+\left[4\,\tau_{1,z}-\nu\, \left({\bm \tau}_1\times{\bm \tau}_2\right)_z\right] 
\frac{\left({\bm\sigma}_2\times {\bf k}_2 \right)\cdot 
\left({\bm\sigma}_1\times{\bf k}_2\right)}{L(x,k_2)}\Bigg] +1\rightleftharpoons 2\ ,\\
\label{eq:c_f}
\!\!\!\rho^{(1)}_{\rm f}\!&=&\!-e\frac{1}{8\,\pi} \frac{g_A^4}{F_\pi^4}\, G_\pi(q^2)
\int_0^1{\rm d}x\, x \int_{-1/2}^{1/2}{\rm d} y \Bigg[-2\,\tau_{1,z}\,
\bigg[-15\,\lambda(x,y)
+\frac{1}{\lambda(x,y)} 
\Big[3\,{\bf A} \cdot({\bf B}+{\bf C})
+({\bf A}+{\bf B})\cdot({\bf A}+{\bf C})\nonumber\\
&&+({\bm \sigma}_1\times{\bf A})\cdot({\bm \sigma}_2\times{\bf A})
-({\bm\sigma}_1 \times{\bf A})\cdot({\bm\sigma}_2\times{\bf C})\nonumber-({\bm\sigma}_1\times{\bf B})
\cdot({\bm\sigma}_2\times{\bf A})
+({\bm\sigma}_1\times{\bf B})\cdot({\bm \sigma}_2\times{\bf C})\Big] \nonumber\\
&&+\frac{1}{\lambda^3(x,y)}\Big[({\bf A}\cdot{\bf B})({\bf A} \cdot {\bf C} )
+{\bm\sigma}_1\cdot({\bf A}\times{\bf B})
\,{\bm\sigma}_2\cdot({\bf A}\times{\bf C})\Big]\bigg]
+\frac{1}{\lambda(x,y)}\,({\bm\tau}_1\times{\bm\tau}_2)_z\bigg[
-3\,{\bm\sigma}_2\cdot({\bf A}\times {\bf C}) \nonumber\\
&&-{\bf B} \cdot({\bm\sigma}_2\times{\bf A})
+({\bf A}+{\bf B})
\cdot({\bm\sigma}_2\times{\bf C})
-\frac{1}{\lambda^2(x,y)} {\bf A}\cdot{\bf B}\,{\bm\sigma}_2\cdot({\bf A}\times{\bf C})
\bigg]\Bigg]+ 1\rightleftharpoons 2 \ ,
\end{eqnarray}
\end{widetext}
\begin{eqnarray}
  \label{eq:c_i}
\!\!\rho_{\rm i}^{(1)}&=&e\,\frac{1}{\pi}\frac{g_A^2}{F_{\pi}^2}\, C_T\, G_E^V(q^2)\, 
  \tau_{1,z}\,{\bm \sigma_1}\cdot\ {\bm \sigma_2}\, m_{\pi}+1\rightleftharpoons 2\ , \\
  \label{eq:c_j}
\!\!\rho_{\rm j}^{(1)}\!\!&=&\!\!-e\,\frac{1}{\pi}\frac{g_A^2}{F_{\pi}^2}\,C_T\, G_\pi(q^2)\, \tau_{1,z}
\int_{0}^{1/2} {\rm d}x \bigg[ \, \frac{ 3\,L^2(x,q)-m_{\pi}^2}{L(x,q)}
{\bm \sigma}_1\cdot {\bm \sigma}_2 \nonumber \\
&& -\frac{1/4-x^2}{L(x,q)}\, {\bm \sigma}_1\cdot{\bf q}\,\,{\bm \sigma}_2\cdot{\bf q} \bigg]
+1\rightleftharpoons 2  \ ,
\end{eqnarray}
where we have defined
\begin{equation}
L^2(x,p)= (1/4-x^2)\,{\bf p}^2+m_\pi^2 \ ,
\end{equation}
\begin{equation}
\label{eq:ldefxy}
 \lambda^2(x,y)=
 x\, {\bf q}^2/4 -\big[ x\, y\, {\bf q}-\left(1-x\right)
 {\bf k}\, \big ]^2+\left(1-x\right){\bf k}^2 +m_{\pi}^2 \ ,
\end{equation}
\begin{eqnarray}
\label{eq:ABCdef}
\bf A&=&- x\left(y\, {\bf q}+{\bf k}\right)\,,\\
\bf B&=&\left( 1-2\,x\,y\right){\bf q}/2+\left( 1-x\right)\,{\bf k} \ ,\\
\bf C&=&-\left( 1+2\,x\,y\right){\bf q}/2+\left( 1-x\right)\,{\bf k} \ .
\end{eqnarray}
It is easily verified that the charge operators (c)+(d), (e)+(f), and
(i)+(j) vanish at \hbox{${\bf q}=0$}.  Finally, we note that the form
of the operator (e) depends on the off-the-energy-shell
prescription adopted for the non-static corrections to the TPE
potential.  As in the OPE case, however, these different forms for
the TPE non-static potential and accompanying charge operator
are unitarily equivalent~\cite{Pastore11}, in particular
$\rho^{(1)}_{\rm e}(\nu)= \rho^{(1)}_{\rm e}(0)+
\left[ \, \rho^{(-3)}\, , \, i\, U^{(1)}(\nu) \,\right]$.  In closing,
we note that the re-analysis, outlined in Appendix~\ref{app:a2},
of the loop corrections to the short-range charge operators
illustrated in panels (g)-(j)  has led to expressions which are
different from those reported originally in Ref.~\cite{Pastore09}.
They also differ from those in Ref.~\cite{Koelling11}.
\section{Calculation}
\label{sec:cal}
The deuteron charge ($G_C$), magnetic ($G_M$), and
quadrupole ($G_Q$) form factors are obtained from~\cite{Schiavilla02}
\begin{eqnarray}
G_C(q)&=&\frac{1}{3} \sum_{M=\pm 1,0} \langle d;M\mid \rho(q\,\hat{\bf z})
\mid d;M\rangle \ ,   \\
G_M(q)&=&  \frac{1}{\sqrt{2\, \eta}} \,{\rm Im}\left[\,  \langle d;1\mid j_y(q\,\hat{\bf z})
\mid  d;0\rangle \, \right]\ , \\
G_Q(q)&=& \frac{1}{2\, \eta}
\big[ \langle d;0\mid \rho(q\,\hat{\bf z}) \mid d;0\rangle \nonumber\\
       &&-\langle d;1\mid \rho(q\,\hat{\bf z}) \mid d;1\rangle \big] \ ,
\end{eqnarray}
where $\mid d;M\rangle$ is the deuteron state with spin projection
$J_z=M$, $\rho$ and $j_y$ denote, respectively, the charge
operator and $y$
component of the current operator, the momentum transfer ${\bf q}$ is
taken along the $z$-axis (the spin quantization axis), and $\eta=(q/2\,m_d)^2$
($m_d$ is the deuteron mass).  They are normalized as
\begin{equation}
\label{eq:norm0}
 G_C(0)=1\ ,\,  G_M(0)=(m_d/m_N)\, \mu_d\ ,\,  G_Q(0)=m_d^2\, Q_d \ ,
\end{equation}
where $\mu_d$ and $Q_d$ are the deuteron magnetic moment
(in units of $\mu_N$) and quadrupole moment, respectively.
Expressions relating the form factors to the measured structure functions
$A$ and $B$, and tensor polarization $T_{20}$ are given in
Ref.~\cite{Schiavilla02}.  The calculations are carried out
in momentum space~\cite{Schiavilla02} with techniques
similar to those described in some detail below for the trinucleons.

The charge and magnetic form factors of the trinucleons are derived from 
\begin{eqnarray}
F_C(q) &=& \frac{1}{Z} \, \langle +\! \mid \rho(q\, \hat{\bf z}) \mid\! + \rangle \ , \\
F_M(q) &=& -\frac{2\, m_N}{q} \, 
{\rm Im} \left [ \,\langle -\!\mid j_y(q\, \hat{\bf z}) \mid\!  + \rangle \, \right] \ ,
\end{eqnarray}
with the normalizations
\begin{equation}
F_C(0)=1 \ , \qquad F_M(0)= \mu \ ,
\end{equation}
where $\mu$ is the magnetic moment (in units of $\mu_N$).  Here $\mid \pm\rangle$
represent either the $^3$He state or $^3$H state in spin projections $J_z=\pm 1/2$.
In momentum space, the one-body electromagnetic operators in Sec.~\ref{sec:c-cnt} have the generic form
\begin{equation}
O_{\rm 1b}({\bf q})=\sum_{ {\rm cyclic}\, l,m,n} \overline{\delta}({\bf k}_l-{\bf q})\,\, 
\overline{\delta}({\bf k}_m)\,\, \overline{\delta}({\bf k}_n)\,\, O_{\rm 1b}({\bf k}_l,{\bf K}_l)  \ ,
\end{equation}
and their matrix elements can be written as
\begin{eqnarray}
\langle O_{\rm 1b}({\bf q}) \rangle &=&\sum_{ {\rm cyclic}\, l,m,n}
\int_{{\bf p}_l,{\bf p}_m, {\bf p}_n} 
\psi_{M^\prime}^\dagger({\bf p}_l+{\bf q}/2,{\bf p}_m,{\bf p}_n)\nonumber\\
&&\times O_{\rm 1b}({\bf q},{\bf p}_l)\,  \psi_M({\bf p}_l-{\bf q}/2,{\bf p}_m,{\bf p}_n)  \ ,
\label{eq:m.e.}
\end{eqnarray}
where we have defined
\begin{equation}
\int_{{\bf p}_i} = \int\frac{{\rm d} {\bf p}_i}{(2\,\pi)^3} \,\,\,\, {\rm and}\,\,\,\, 
\overline{\delta}(\dots) = (2\, \pi)^3\, \delta(\dots) \  .
\end{equation}
For an assigned configuration $({\bf p}_l,{\bf p}_m,{\bf p}_n)$, the wave functions
are expanded on a basis of $8\times 3$ spin-isospin states for the three nucleons as
\begin{equation}
\psi({\bf p}_l,{\bf p}_m,{\bf p}_n)=\sum_{a=1}^{24} \psi_a({\bf p}_l,{\bf p}_m,{\bf p}_n)\, \mid\!a\rangle\ ,
\end{equation}
where the components $\psi_a$ are complex functions and the basis states
 (for $^3$H, for example)
$\mid a\rangle = \mid (p\uparrow)_1,(n\uparrow)_2, (n\uparrow)_3\rangle$,
$\mid (n\uparrow)_1,(p\uparrow)_2, (n\uparrow)_3\rangle$, and so on.  The spin-isospin
algebra for the overlaps
\begin{equation}
\psi^\dagger \, O\, \psi =\sum_{a,b=1}^{24} \psi_a^*\, O_{ab}\, \psi_b \ ,
\end{equation}
is carried out with the techniques developed in Ref.~\cite{Schiavilla89}.   Monte
Carlo (MC) methods are used to evaluate the integrations in Eq.~(\ref{eq:m.e.}) by sampling
momenta from a (normalized) probability density $\mid \psi_M({\bf p}_l,{\bf p}_m,{\bf p}_n) \mid^2$ according
to the Metropolis algorithm.

The two-body operators in Sec.~\ref{sec:c-cnt} have the momentum-space representation
\begin{eqnarray}
O_{\rm 2b}({\bf q})&=&\sum_{ {\rm cyclic}\, l,m,n}  
\overline{\delta}({\bf K}_{lm}-{\bf q})\,\,  \overline{\delta}({\bf k}_n)\nonumber\\
&&\times O_{\rm 2b}({\bf K}_{lm}/2+{\bf k}_{lm},
{\bf K}_{lm}/2-{\bf k}_{lm}) \ ,
\end{eqnarray}
where the momenta ${\bf K}_{lm}={\bf k}_l+{\bf k}_m$ and ${\bf k}_{lm}=({\bf k}_l-{\bf k}_m)/2$.
These operators have power law behavior at large momenta, and need to be regularized.
This is accomplished by introducing a momentum cutoff function of the form
\begin{equation}
C_\Lambda(k_{lm}) ={\rm e}^{-(k_{lm}/\Lambda)^4} \ ,
\label{eq:ctf}
\end{equation}
with the parameter $\Lambda$ in the range (500--600) MeV (see discussion in Sec.~\ref{sec:res}).
The matrix elements are expressed as
\begin{widetext}
\begin{eqnarray}
\langle O_{\rm 2b}({\bf q}) \rangle&=&\sum_{ {\rm cyclic}\, l,m,n} \int_{{\bf k}_{lm}}
\int_{{\bf p}_l,{\bf p}_m, {\bf p}_n} 
\psi_{M^\prime}^\dagger({\bf p}_l+{\bf q}/4+{\bf k}_{lm}/2,{\bf p}_m+{\bf q}/4-{\bf k}_{lm}/2,{\bf p}_n)\,
C_\Lambda(k_{lm})\nonumber\\
&& \times O_{\rm 2b}({\bf q},{\bf k}_{lm}) \, 
\psi_M({\bf p}_l-{\bf q}/4-{\bf k}_{lm}/2,{\bf p}_m-{\bf q}/4+{\bf k}_{lm}/2,{\bf p}_n)\ .
\label{eq:me2}
\end{eqnarray}
\end{widetext}
The spin-isospin algebra is handled as above, while the multidimensional integrations are
efficiently done by a combination of MC and standard quadratures techniques.  We write
\begin{equation}
\langle O_{\rm 2b}({\bf q}) \rangle
\!\!= \!\!\int \!\!{\rm d}\hat{\bf k}\!\!
\int_{{\bf p}_l,{\bf p}_m, {\bf p}_n}\!\!\!\!\!\!\!\!\!F(\hat{\bf k},{\bf p}_l,{\bf p}_m, {\bf p}_n)
\simeq \frac{1}{N_c} \sum_{c=1}^{N_c} \frac{F(c)}{W(c)} \ ,
\end{equation}
where $c$ denotes configurations $(\hat{\bf k},{\bf p}_l,{\bf p}_m,{\bf p}_n)$
(total number $N_c$)
sampled with the Metropolis algorithm from the probability density
$W(c)=\mid \psi_M({\bf p}_l,{\bf p}_m,{\bf p}_n) \mid^2\!\!/(4\, \pi)$,
i.e., uniformly over the $\hat{\bf k}$ directions.  For each such
configuration $c$, the function $F$ is obtained by Gaussian
integration over the magnitude $k_{lm}$ (as well as the parameters
$x$ and $y$ for the case of the charge operators at one loop)
\begin{eqnarray}
F(c)\!\!&=&\!\!\!\!\!\sum_{ {\rm cyclic}\, l,m,n} \!\!\frac{1}{(2\, \pi)^3} 
\!\!\int_0^\infty {\rm d}k_{lm} \, k_{lm}^2
\sum_{a,b=1}^{24}  \psi_a^*(\dots k_{lm}\hat{\bf k} \dots) \nonumber\\
&&\times O_{{\rm 2b},ab}({\bf q},k_{lm}\hat{\bf k})\, 
\psi_b(\dots k_{lm}\hat{\bf k} \dots) \ .
\end{eqnarray}
Convergence in these Gaussian integrations requires of the order of 20--30 points,
in the case of $k_{lm}$ distributed over a non-uniform grid up to $2\, \Lambda$
or so, while $N_c$ of the order of $100,000$ is sufficient to reduce the statistical
errors in the MC integrations, which are of the order of a few \% at the highest
$q$ values (and considerably smaller at lower $q$).   These MC errors are further
reduced by taking appropriate linear combinations of
the matrix elements of the electromagnetic operators
using different $\hat{\bf q}$ directions and different spin projections for
the initial and final states.
The trinucleons wave functions are obtained with the hyperspherical 
harmonics (HH) expansion discussed in 
Refs.~\cite{Kievsky97,Viviani06,Kievsky08}.  This method can be applied
in either coordinate- or momentum-space.  Below, we briefly 
review its momentum-space implementation.
\subsection{The hyperspherical harmonics method in momentum-space}
\label{subsec:hh}

The trinucleon wave functions with total angular momentum $J J_z$
are written as
\begin{equation}
|\psi^{JJ_z}\rangle=\sum_\mu c_\mu \,|\psi^{JJ_z}_\mu\rangle \ ,
\label{eq:psi}
\end{equation}
where $|\psi^{JJ_z}_\mu\rangle$ is a suitable complete set of 
states, and $\mu$ is an index denoting the set of quantum numbers 
necessary to specify the basis elements (see below). 
By applying the  Rayleigh-Ritz variational principle, 
the problem of determining $c_\mu$ and the ground-state energy $E_0$ of the system
is reduced to a generalized eigenvalue problem.

In momentum space we define the  Jacobi momenta as
\begin{equation}
{\bf k}_{2p}=\left({\bf p}_j-{\bf p}_i\right)/\sqrt{2} \ ,\,\,
{\bf k}_{1p}=\sqrt{2/3}\big[{\bf p}_k-\left({\bf p}_i+{\bf p}_j\right)/2 \big] \ , 
  \label{eq:jacm3}
\end{equation}
where ${\bf p}_i$ denotes the momentum of nucleon $i$ and
$p$ specifies a given permutation of the three nucleons, with
$p=1$ corresponding to the ordering 1,2,3.  We introduce 
a hyper-momentum $K$ and a set of 
angular and hyper-angular variables as 
\begin{equation}
  K=\left(k_{1p}^2+k_{2p}^2\right)^{1/2} \ ,   \qquad
    \Omega^{(K)}_p=\left[ \hat{{\bf k}}_{2p}, \hat{{\bf k}}_{1p}\,;\phi_{p}\right] \ , 
  \label{eq:hyperq}
\end{equation}
where $\tan{\phi_{p}}=k_{1p}/k_{2p}\,$.  In terms of these variables,
the basis functions $|\psi^{JJ_z}_\mu\rangle$ are defined as
\begin{equation}
  |\,\psi^{JJ_z)}_\mu\,\rangle=
   g_{ G \,l }(K) {\cal Y}_{ \{G\} }\!\left(\Omega^{(K)}\right)\ , 
  \label{eq:qexp}
\end{equation}
where ${\cal Y}_{ \{G\} }\!\left(\Omega^{(K)}\right)$
are written as~\cite{Marcucci09}
\begin{eqnarray}
  {\cal Y}_{ \{G\} }(\Omega^{(K)})&=& \sum_{p=1}^{3}\left[ 
  Y^{LL_z}_{ [G] }\!\left(\Omega^{(K)}_p\right) 
  \otimes \left[S_2\otimes \frac{1}{2}\right]_{S S_z} \right]_{J J_z}\nonumber\\
 &&\times \left[T_2\otimes\frac{1}{2}\right]_{T T_z}   \ , \label{eq:hha3}
\end{eqnarray}
and the sum is over the three even permutations.  
The spins (isospins) of nucleons $i$ and $j$ are coupled to 
$S_2$ ($T_2$), which is then coupled to the spin (isospin) 
of the third nucleon to give a state with total spin $S$ (isospin $TT_z$).
The total orbital angular momentum $L$ and total 
spin $S$ are coupled to the total angular momentum  $JJ_z$.  
The functions $Y^{LL_z}_{[G]}(\Omega^{(K)}_p)$ with definite values of 
$LL_z$ are the hyperspherical-harmonics functions, and 
are written as~\cite{Kievsky97}
\begin{eqnarray}
  Y^{LL_z}_{ [G] }\!\left(\Omega^{(K)}_p\right)\!\!& =&\!\!
  \biggl[Y_{\ell_2}(\hat{\bf k}_{2p}) \otimes
   Y_{\ell_1}(\hat{\bf k}_{1p}) \biggr]_{LL_z}
  N_{[G] }\,(\cos\phi_p)^{\ell_2}\nonumber \\
&&\times (\sin\phi_p)^{\ell_1}\,
  P_{n}^{\ell_1+\frac{1}{2},\ell_2+\frac{1}{2}}(\cos 2\phi_p) \ ,
\label{eq:hh3}
\end{eqnarray}
where $Y_{\ell_1}(\hat{\bf k}_{1p})$ and $Y_{\ell_2}(\hat{\bf k}_{2p})$ are 
spherical harmonics, $N_{[G]}$ is a normalization factor, and 
$P_{n}^{\ell_1+\frac{1}{2},\ell_2+\frac{1}{2}}(\cos 2\phi_p)$ denotes the Jacobi 
polynomial of degree $n$.
The grand angular quantum number $G$ is 
defined as $G=2\, n+\ell_1+\ell_2$.
The subscripts $\{G\}$ and $[G]$ in Eqs.~(\ref{eq:qexp})--(\ref{eq:hh3})
stand, respectively, for  
$\{G\}\equiv\{\ell_1,\ell_2,L,S_2,T_2,S,T;n\}$ and 
$[G]\equiv[\ell_1,\ell_2;n]$, and $\mu$ in Eq.~(\ref{eq:psi})
stands for $\mu\equiv\{G\}\, l$.
Finally, the functions $g_{G\, l}(K)$ in Eq.~(\ref{eq:qexp}) are defined as
 \begin{equation}
   g_{G\, l}(K)=\frac{(-i)^G}{K^2} \,\int_0^\infty d\rho\, \rho^{3}\,
   J_{G+2}(K\rho)\, f_{l}(\rho) \ ,
\label{eq:vg}
\end{equation}
where $J_{G+2}(K\rho)$ are Bessel functions and the 
functions $f_l(\rho)$ are related to Laguerre polynomials
$L^{(5)}_l(\gamma\rho)$ via
\begin{equation}
 f_l(\rho)=\gamma^{3} \sqrt{l!/(l+5)!}\,\,\, 
 L^{(5)}_l(\gamma\rho)\,\,{\rm e}^{-\gamma\rho/2} \ .
 \label{eq:fllag}
\end{equation}
The non-linear parameter $\gamma$ is 
variationally optimized.
With this form of $f_l(\rho)$, 
the corresponding functions $g_{G\, l}(K)$ can easily be calculated, 
and are explicitly given in Ref.~\cite{Viviani06}.
The form adopted for $g_{G \, l}(K)$ is such that the momentum-space basis is simply
the Fourier transform of the coordinate-space one~\cite{Kievsky08}.

\section{Results}
\label{sec:res}
This section consists of three subsections.  In the first one,
we discuss various strategies for the determination of the unknown
LEC's $d_8^\prime$, $d_9^\prime$, $d_{21}^\prime$,
$C_{15}^\prime$, and $C_{16}^\prime$ entering the current operator at N3LO.
In contrast, the charge operator up to N4LO only depends
on the nucleon axial coupling constant $g_A$, pion decay amplitude $F_\pi$,
and nucleon mass and magnetic moments. The values adopted in the present
work for $g_A$ and $F_\pi$ are, respectively, 1.28 and 184.6 MeV, which
give a $\pi N$ coupling constant ($g_{\pi NN}$) of 13.6, as obtained in
analyses of $NN$ elastic scattering data at energies below
the pion production threshold~\cite{deSwart98}.  The two-body operators
are regularized via the cutoff function in Eq.~(\ref{eq:ctf}), and
$\Lambda$ values of 500 MeV and 600 MeV are considered.

In the second and third subsections we present results, respectively,
for the deuteron $A(q)$ and $B(q)$ structure functions and tensor
polarization $T_{20}(q)$, and for the charge and magnetic form factors
of $^3$H and $^3$He, along with results for the static properties
of these few-nucleon systems including the deuteron quadrupole moment,
the deuteron and trinucleons charge and magnetic radii and
magnetic moments.  The $A=2$ calculations use either the Argonne
$v_{18}$ (AV18)~\cite{Wiringa95}  or chiral potentials at order $Q^4$
with cutoff set at 500 MeV (N3LO) or 600 MeV (N3LO$^*$)~\cite{Entem03}.
Of course, the $A=3$ calculations also include
three-nucleon potentials---the Urbana-IX model~\cite{Pudliner95} in combination with
the AV18, and the chiral N2LO potential~\cite{Navratil07} in combination
with either the N3LO or N3LO$^*$.  The LEC's $c_D$ and $c_E$ (in standard notation)
in the chiral three-nucleon potential have been constrained by
reproducing the $^3$H/$^3$He binding energies and the tritium
Gamow-Teller matrix element~\cite{Marcucci12} in each case.  With the
AV18/UIX Hamiltonian, the $^3$H and $^3$He binding energies are found to be
8.487 MeV and 7.747 MeV, respectively.

The calculations are carried out in configuration space in the first
subsection, and in momentum space---with the methods outlined in
Sec.~\ref{sec:cal}---in the following two subsections.  We have checked
that the $r$- and $p$-space versions of the computer codes produce
identical results up to to tiny differences due to numerics and
to numerically non-equivalent implementations of the momentum
cutoff function in Eq.~(\ref{eq:ctf}) in these $r$- and $p$-space
calculations.  The hadronic electromagnetic form factors entering
the one- and two-body charge and current operators are those
specified in Sec.~\ref{sec:c-cnt}.  The matrix
elements of these operators are evaluated in the Breit frame with
Monte Carlo methods.  The number of sampled configurations
is of the order of 10$^6$ for the deuteron and 10$^5$ for the
$A=3$ systems.  The statistical errors, which are not shown
in the results that follow, are typically $\lesssim 1$\% over
the whole momentum-transfer range, and in fact much less
than 1\% for $q \lesssim 2$ fm$^{-1}$.

\subsection{Determination of the LEC's}
\label{sec:lecs}

As already remarked, the LEC's $C_i$, $i=1,\dots,7$, in
the minimal contact current, corresponding to $\Lambda$
cutoffs of 500 and 600 MeV, are taken from fits to
$NN$ scattering data~\cite{Entem03}.  In reference to
the LEC's entering the OPE and non-minimal contact
currents at N3LO, it is convenient to introduce the adimensional
set $d_i^{S,V}$ (in units of the cutoff $\Lambda$) as
\begin{eqnarray}
\!\!\!\!\!\!\!\!&&C_{15}^\prime=d_1^S/\Lambda^4  \ , \qquad d_9^\prime=d_2^S/\Lambda^2 , \nonumber \\
\!\!\!\!\!\!\!\!&&C_{16}^\prime=d_1^V/\Lambda^4 \ , \qquad d_8^\prime=d_2^V/\Lambda^2\ ,\qquad
d_{21}^\prime=d_3^V/\Lambda^2 \ ,\nonumber\\
&&
\end{eqnarray}
where the superscript $S$ or $V$ on the $d^{S,V}_i$ characterizes the isospin of the
associated operator, i.e.,~whether it is isoscalar or isovector.  The isoscalar
$d_i^{S}$, listed in Table~\ref{tb:tds}, have been fixed by reproducing the experimental
deuteron magnetic moment $\mu_d$ and isoscalar combination $\mu_S$ of the
trinucleon magnetic moments.  The LEC $d_1^S$ multiplying the contact
current is rather large, but not unreasonably large, while the LEC $d_2^S$
is quite small.  
\begin{table}
\caption{Adimensional values of the isoscalar LEC's corresponding
to cutoffs $\Lambda=500$ MeV and 600 MeV obtained
for the N3LO/N2LO and N3LO$^*$/N2LO$^*$ Hamiltonians; the values in
parentheses are relative to the AV18/UIX Hamiltonian.}
\begin{tabular}{c|c|c|}
$\Lambda$  & $d_1^S$  & $d_2^S\times 10 $ \\
\hline
500 &  4.072 (2.522) & 2.190 (--1.731)   \\
600  & 11.38 (5.238) & 3.231 (--2.033) \\
\hline
\end{tabular}
\label{tb:tds}
\end{table}
The cumulative contributions to $\mu_d$ and $\mu_S$
are reported in Table~\ref{tb:muds}.  The NLO and N3LO-loop magnetic moment
operators are isovector, and therefore do not contribute to these isoscalar observables.
At N3LO the only non-vanishing contributions are those associated
with the OPE and minimal (min) and non-minimal (nm) contact currents. 
Of course, the last row in Table~\ref{tb:muds}
reproduces the experimental values for $\mu_d$ and $\mu_S$.
\begin{widetext}
\begin{center}
\begin{table*}
\caption{Cumulative contributions to the
deuteron and trinucleons isoscalar magnetic moments in units of $\mu_N$,
corresponding to cutoffs $\Lambda=500$ MeV and 600 MeV obtained
for the N3LO/N2LO and N3LO$^*$/N2LO$^*$ Hamiltonians; the contributions in
parentheses are relative to the AV18/UIX Hamiltonian.  The experimental
values for the deuteron and trinucleons isoscalar magnetic moments
are $0.8574 \, \mu_N$ and $0.4257\, \mu_N$, respectively.}
\begin{tabular}{c||c|c||c|c||}
    & \multicolumn{2}{c||}{$\mu_d$} & \multicolumn{2}{c||}{$\mu_S$}\\
    \hline
 $\Lambda$          & 500  & 600 & 500 & 600  \\
\hline
LO          &  0.8543 (0.8472) &   0.8543 (0.8472) & 0.4222 (0.4104)  & 0.4220 (0.4104)  \\
N2LO        &  0.8471 (0.8400) &   0.8474 (0.8400) &  0.4143 (0.4027) & 0.4155 (0.4027) \\
N3LO(min)     &  0.8725 (0.8739) &   0.8806 (0.8760) &  0.4501 (0.4455) & 0.4611 (0.4483) \\
N3LO(nm) & 0.8548 (0.8593) &  0.8538 (0.8626) & 0.4247 (0.4269) & 0.4235 (0.4313)     \\
N3LO(OPE)      &  0.8574 (0.8574) &    0.8574 (0.8574) & 0.4257 (0.4257) & 0.4257 (0.4257) \\
\hline
\end{tabular}
\label{tb:muds}
\end{table*}
\end{center}
\end{widetext}

The isovector LEC $d_3^V$ is taken as $d_2^V/4$ by assuming
$\Delta$ dominance.  The three different sets of 
remaining LEC's $d_1^V$ and $d_2^V$ reported
in Table~\ref{tb:tdv} have been determined in the following way.
In set I $d_1^V$ and $d_2^V$ have been constrained to reproduce the
experimental values of the $np$ radiative capture cross section $\sigma_{np}$
at thermal neutron energies and the isovector combination $\mu_V$ of the
trinucleons magnetic moments.  This procedure, however, leads to unreasonably
large values for both LEC's, and is clearly unacceptable.  In particular, it makes
the contributions of the associated magnetic dipole operators unnaturally large, and, 
as shown in Table~\ref{tb:mudv}, totally spoils the expected convergence
pattern.  This pathology is especially severe in the case of the AV18/UIX
Hamiltonian model.
\begin{widetext}
\begin{center}
\begin{table}
\caption{Adimensional values of the isovector LEC's corresponding
to cutoffs $\Lambda=500$ MeV and 600 MeV obtained
for the N3LO/N2LO and N3LO$^*$/N2LO$^*$ Hamiltonians; the values in
parentheses are relative to the AV18/UIX Hamiltonian. Note that
$d_3^V=d_2^V/4$ in all cases; see text for further explanations.}
\begin{tabular}{c||c|c||c|c||c|c||}
$\Lambda$  & $d_1^V$(I) & $d_2^V$(I) & $d_1^V$(II) & $d_2^V$(II) & $d_1^V$(III) & $d_2^V$(III)  \\
\hline
500 &  10.36 (45.10)  & 17.42 (35.57) & --13.30 (--9.339) & 3.458  & --7.981 (--5.187)  &  3.458  \\
600 &  41.84 (257.5) &  33.14 (75.00) & --22.31 (--11.57) & 4.980  &   --11.69 (--1.025) & 4.980  \\
\hline
\end{tabular}
\label{tb:tdv}
\end{table}
\end{center}
\end{widetext}
In sets II and III $d_2^V$ is assumed to be saturated by the  $\Delta$ resonance,
i.e.
\begin{equation}
d_2^V = \frac{4\, \mu_{\gamma N\Delta} \, h_A\, \Lambda^2}{9\, m_N \,( m_\Delta-m_N)} \ ,
\end{equation}
where $m_\Delta-m_N=294$ MeV, $h_A/F_\pi=f_{\pi N\Delta}/m_\pi$
with $f^2_{\pi N\Delta}/(4\,\pi)=0.35$ as obtained by equating the first-order
expression of the $\Delta$-decay width to the experimental value, and the transition magnetic
moment $\mu_{\gamma N\Delta}=3\, \mu_N$~\cite{Carlson86}---a similar strategy has been
implemented in a number of calculations, based on the $\chi$EFT magnetic moment
operator derived in Ref.~\cite{Park96}, of the $np$, $nd$, and
$n^3$He radiative captures, and magnetic moments of $A=2$ and 3 nuclei~\cite{Park00}.
On the other hand, the LEC $d_1^V$
multiplying the contact current is fitted to reproduce either $\sigma_{np}$
in set II or $\mu_V$ in set III.  Both alternatives still lead to somewhat
large values for this LEC, but we find the degree of unnaturalness tolerable
in this case.   We observe that there are no three-body
currents at N3LO~\cite{Girlanda10a}, and therefore it is reasonable to fix the strength of this
$M1$ operator by fitting a three-nucleon observable such as $\mu_V$.
\begin{widetext}
\begin{center}
\begin{table}
\caption{Cumulative contributions to the
$np$ radiative capture cross section in mb and trinucleons isovector magnetic moment in units of $\mu_N$,
corresponding to cutoffs $\Lambda=500$ MeV and 600 MeV obtained
for the N3LO/N2LO and N3LO$^*$/N2LO$^*$ Hamiltonians; the contributions in
parentheses are relative to the AV18/UIX Hamiltonian.  See text for further explanations.
The experimental values for the $np$ cross section and trinucleons isovector magnetic
moment are $(332.6\pm 0.7)$ mb and $-2.553\, \mu_N$, respectively.}
\begin{tabular}{c||c|c||c|c||} 
& \multicolumn{2}{c||}{$\sigma_{np}$} & \multicolumn{2}{c||}{$\mu_V$} \\
\hline
  $\Lambda$        & 500  & 600 & 500 & 600  \\
\hline
LO                      & 305.8 (304.6) &  304.6 (304.6) & --2.193 (--2.159) & --2.182 (--2.159)\\
NLO                   & 320.6 (319.3) &  318.9 (320.9) & --2.408 (--2.382) & --2.392 (--2.413)\\
N2LO                 & 319.2 (317.7) &  317.6 (319.2) & --2.384 (--2.359) & --2.370 (--2.390)\\
N3LO(loop)        & 321.3 (320.9) &  320.5 (322.4) & --2.430 (--2.418) & --2.432 (--2.448) \\
N3LO(min) & 321.3 (320.9) &  320.5 (322.4) & --2.413 (--2.406) & --2.415 (--2.437)\\
\hline
N3LO(nm, $d_1^V$-I) & 315.2 (287.4)  & 305.7 (242.7)  & --2.297 (--1.782) & --2.142 (--0.9029)   \\
N3LO(OPE, $d_2^V$-I) & 332.6 (332.6)  & 332.6 (332.6) & --2.553  (--2.553) & --2.553 (--2.553) \\
\hline
N3LO(nm, $d_1^V$-II)     & 329.1 (328.1) &  328.5 (326.2) & --2.562 (--2.535) & --2.561 (--2.506) \\
N3LO(OPE, $d_2^V$-II) & 332.6 (332.6) &  332.6 (332.6) & --2.612 (--2.610) & --2.622 (--2.616) \\
\hline
N3LO(nm, $d_1^V$-III)     & 326.0  (324.9) &  324.7 (322.7) & --2.502 (--2.478) & --2.491 (--2.443) \\
N3LO(OPE, $d_2^V$-III)    & 329.4 (329.4) &  328.8 (329.1) & --2.553 (--2.553) & --2.553 (--2.553) \\
\hline
\end{tabular}
\label{tb:mudv}
\end{table}
\end{center}
\end{widetext}
Cumulative contributions to $\sigma_{np}$ and $\mu_V$ are
listed in Table~\ref{tb:mudv}.  At N3LO, we have
identified separately those due only to loop currents
labeled as N3LO(loop), and those from loop+minimal contact
currents labeled as N3LO(min).  The experimental values
for $\sigma_{np}$ and $\mu_V$ are
reproduced with set I, row labeled N3LO(OPE, $d_2^V$-I),
while only $\sigma_{np}$ or $\mu_V$ are reproduced with
set II or III, rows labeled N3LO(OPE, $d_2^V$-II) or
N3LO(OPE, $d_2^V$-III).   Indeed, the  N3LO(OPE, $d_2^V$-II or III)
results provide predictions for $\mu_V$ or $\sigma_{np}$,
respectively.  These predictions are within 3\% for $\mu_V$ and
1\% for $\sigma_{np}$ of the experimental values, and exhibit a weak cutoff and Hamiltonian-model 
dependence.

In Ref.~\cite{Girlanda10a} the $d_i^{S,V}$ were determined using the same procedure adopted
here for set I.  However, the values reported in that work are drastically different
from those obtained in the present one.  These differences are due to several
factors: i) in Ref.~\cite{Girlanda10a} the $M1$ operator derived from
Eq.~(\ref{eq:jloop}) included an isovector loop correction proportional to the LEC's
$C_S$ and $C_T$, which turns out to vanish in a more
careful analysis of the relevant diagrams (the loop short-range
currents discussed in Appendix~\ref{app:a2});
ii) in Ref.~\cite{Girlanda10a} the values for the LECs $C_1,\dots,C_7$ were taken from a
chiral potential obtained at $Q^2$ (NLO)~\cite{Pastore09} rather than
at $Q^4$ (N3LO)~\cite{Entem03} as in the present case; iii) in Ref.~\cite{Girlanda10a} the
minimal contact current is the Fierz-transformed version of that given in Eq.~(\ref{eq:jmin})
(see discussion in Appendix~\ref{app:a1}).  However, this Fierz equivalence is spoiled by the regularization
procedure, i.e.~by the inclusion of the same cutoff function $C_\Lambda(k)$ for
both.  Hence the contribution of this current in the present work is different from that obtained
in Ref.~\cite{Girlanda10a}.

\subsection{Static properties and form factors of the deuteron}
\label{sec:deut}
%
\begin{center}
\begin{table*}
\caption{Cumulative contributions to the 
deuteron root-mean-square charge radius and quadrupole moment
corresponding to cutoffs $\Lambda=500$ and 600 MeV obtained
with the N3LO and N3LO$^*$ Hamiltonians; results in parentheses
are relative to the AV18 Hamiltonian.  The experimental values
for $r_d$ and $Q_d$ are 1.9734(44) fm~\cite{Mohr05} and 0.2859(3) fm$^2$~\cite{Bishop79}, respectively.}
\begin{tabular}{c||c|c||c|c||}
& \multicolumn{2}{c||}{$r_d$ (fm)}& \multicolumn{2}{c||}{$Q_d$ (fm$^2$)} \\
\hline
  $\Lambda$         & 500     &     600 &      500         &    600           \\
\hline
LO                         &   1.976 (1.969) & 1.968 (1.969)       &  0.2750 (0.2697) &  0.2711 (0.2697) \\
N2LO                     &   1.976 (1.969) & 1.968 (1.969)     &  0.2731 (0.2680) &  0.2692 (0.2680) \\
N3LO(OPE)           &    1.976 (1.969) & 1.968 (1.969)   &  0.2863 (0.2818) &  0.2831 (0.2814) \\
N3LO($\nu=1/2$)   &  1.976 (1.969) & 1.968 (1.969)  &  0.2851 (0.2806) &  0.2820 (0.2802) \\
\hline
\end{tabular}
\label{tb:qmd}
\end{table*}
\end{center}

The deuteron root-mean-square charge radius and quadrupole moment, obtained with the
chiral and AV18 potentials and cutoff parameters $\Lambda=500$ MeV
and 600 MeV, are listed in Table~\ref{tb:qmd}.  We denote
the leading order ($n=-3$ in the notation of Sec.~\ref{sec:c-cnt})
term of Eq.~(\ref{eq:rlo}) with LO, the $n=-1$ relativistic
correction of Eq.~(\ref{eq:r1rc}) with N2LO, and the
$n=0$ terms of Eqs.~(\ref{eq:pich}) and~(\ref{Diagram_dnu2})--(\ref{eq:rho_gpipi})
with N3LO(OPE) and N3LO($\nu$), respectively.  The remaining
charge operators at N4LO ($n=1$),
being isovector, do not contribute to these observables (and
corresponding form factors).  The N3LO/N3LO$^*$ and AV18 potentials
neglect retardation corrections in their OPE component, which
corresponds to setting $\nu=1/2$ in Eq.~(\ref{eq:ope2}).  Note that
the isoscalar piece of the N3LO($\nu$) charge operator scales as $1-\nu$,
and contributes less than 0.5\% of the LO result for $\nu=1/2$.
The N2LO and N3LO corrections to $r_d$, which is well reproduced by
theory, are negligible.  The chiral potential predictions for
$Q_d$ are within 1\% of the experimental value, while the AV18
ones underestimate it by about 2\%.  Variation of the cutoff
in the (500--600) MeV range leads to about 1\% (negligible) changes in the
N3LO/N3LO$^*$ (AV18) results.  The LO and N2LO charge operators do not include
the cutoff function and the AV18 results are independent of $\Lambda$.  This is
not the case for the results corresponding to the N3LO and N3LO$^*$ potentials
because of their intrinsic $\Lambda$ dependence.

\begin{widetext}
\begin{center}
\begin{figure}
\includegraphics[width=7in]{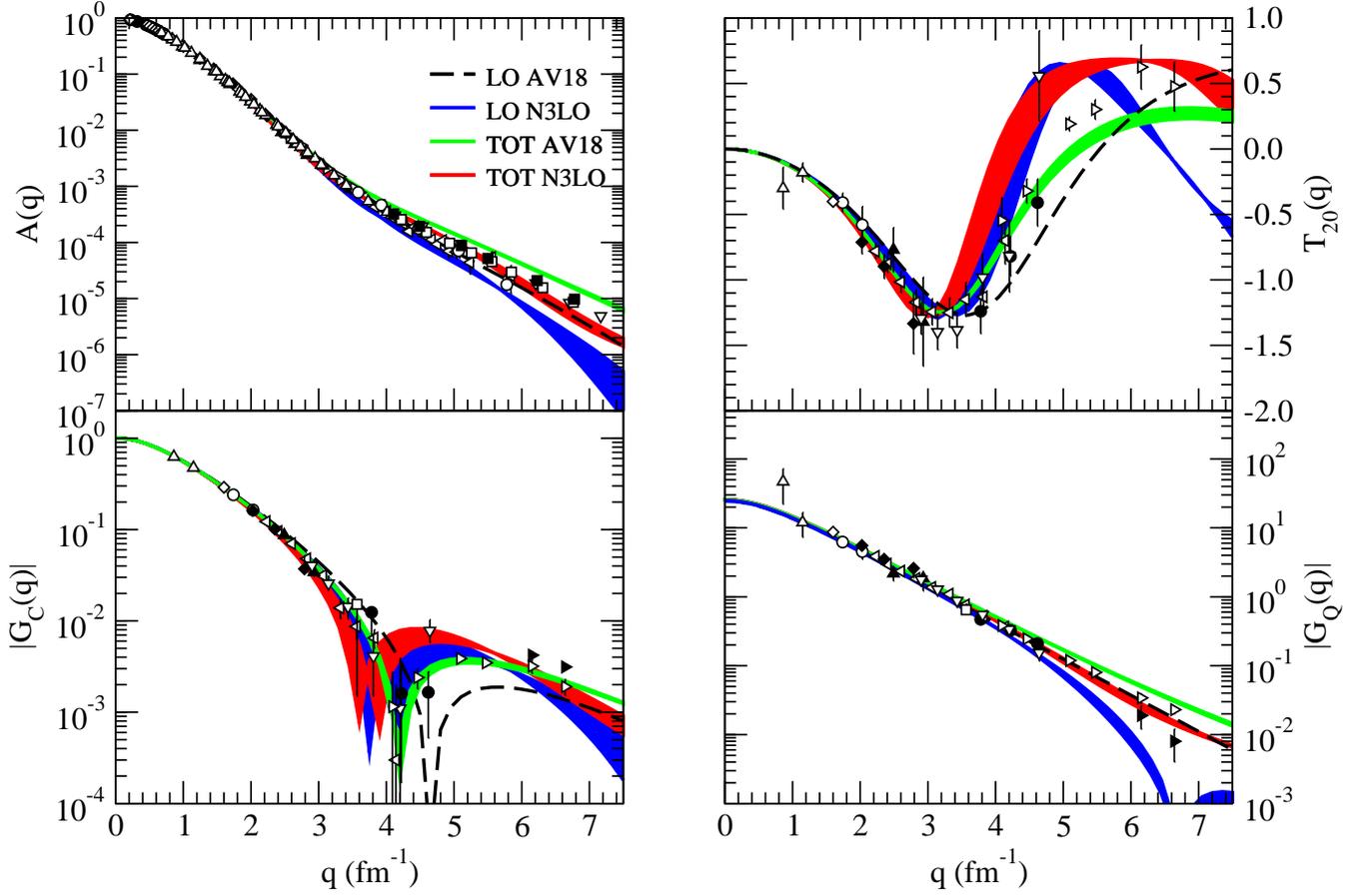}\\
\caption{(Color online). The deuteron $A(q)$ structure function and tensor polarization
$T_{20}(q)$ (top panels), and charge and quadrupole form factors $G_C(q)$
and $G_Q(q)$ (bottom panels), obtained at leading order (LO) and with inclusion of charge
operators up to N3LO (TOT), is compared with experimental data from
Refs.~\cite{STANFORD65,ORSAY66,CEA69,DESY71,MONTEREY73,SLAC75,
MAINZ81,BONN85,SACLAY90,JLABHALLA99,JLABHALLC99,BATES84,BATES91,
BATES2011,VEPP85,VEPP86,VEPP90,VEPP2003,
BONN91,NIKHEF96,NIKHEF99,JLAB2000}.
Predictions corresponding to $\nu=1/2$ and cutoffs $\Lambda$ in the range
500--600 MeV are displayed by the bands.}
\label{fig:f4}
\end{figure}
\end{center}
\end{widetext}
\vspace{1cm}

The deuteron $A(q)$ structure function and tensor polarization
$T_{20}(q)$, obtained at LO and by including corrections up to N3LO
in the charge operator, are compared to data in Fig.~\ref{fig:f4}, top panels.
In this figure (as well as in those that follow)
the momentum-transfer range goes up to $q=7.5$ fm$^{-1}$, much beyond
the $\simeq 3$--4 $m_\pi$ upper limit, where one would naively expect this comparison
to be meaningful, given that the present theory retains up to TPE mechanisms.
On the other hand, we note that the next (non-vanishing) isoscalar contributions
only enter at N5LO ($n=2$)~\cite{Phillips07}, and are therefore suppressed by
two powers of $Q$ relative to those at N3LO.

The $A(q)$ structure function is well reproduced by theory up to
$q \simeq 3$ fm$^{-1}$.  At higher momentum transfers, the N3LO results
based on the AV18 tend to overestimate the data---a feature also seen
in the conventional approach of Ref.~\cite{Schiavilla02}---while
those based on the chiral potentials still provide a
good fit to the data.  The cutoff dependence is weak at low $q$,
but becomes more pronounced as $q$ increases.

Similar considerations hold for the $T_{20}(q)$ observable, although in
this case the N3LO results derived from the chiral potentials
overpredict the data for $q \gtrsim 3$ fm$^{-1}$, while those from the
AV18 fit reasonably well the data up to $q \simeq 4.5$ fm$^{-1}$.  In contrast,
the conventional approach~\cite{Schiavilla02} (also based on the AV18,
of course) reproduces very well the measured $T_{20}$ over the whole
$q$-range.  The OPE charge operator in that work has the same
structure as the present N3LO(OPE) one, but includes a much harder
cutoff than adopted here.  Furthermore, the calculation of Ref.~\cite{Schiavilla02}
also retains short-range (isoscalar)
mechanisms associated with $\rho$-meson exchange and $\gamma\pi\rho$ transition,
which in $\chi$EFT are presumably subsumed in contact operators at N5LO~\cite{Phillips07}.
We note that in both $A(q)$ and $T_{20}(q)$ a small magnetic contribution,
discussed separately below, is accounted for (the electron scattering
angle in $T_{20}$ is set at 70$^\circ$).

The charge and quadrupole form factors extracted from the unpolarized and tensor polarized
deuteron data are compared to results obtained in LO and by including corrections up to N3LO
in Fig.~\ref{fig:f4}, bottom panels.  The $G_C(q)$ and $G_Q(q)$
form factors calculated with deuteron wave functions from the chiral
potentials are in qualitative agreement with
predictions obtained by Phillips~\cite{Phillips07} at the same chiral order (although
the $Q^4$ potentials used in that study are from Ref.~\cite{Epelbaum05} rather
than from Ref.~\cite{Entem03} as in the present work).  The spread in the N3LO results
due to cutoff variations observed here is similar to that reported
in Ref.~\cite{Phillips07} for both $G_C(q)$ and $G_Q(q)$.  However, the central values
for these observables in the momentum-transfer region $q \gtrsim 2.5$ fm$^{-1}$ reported in
that work appear to underestimate the data appreciably.  This is not the case here, particularly
for $G_Q(q)$, for which the N3LO predictions provide an excellent fit to the measured values
(up to $q \simeq 6$ fm$^{-1}$).  These differences likely arise from differences
in the deuteron wave functions obtained in Ref.~\cite{Entem03} and Ref.~\cite{Epelbaum05}
(see Fig.~16 in the 2011 review paper~\cite{Entem03} for a comparison).  Indeed, for these
same reasons, the AV18 results are in better agreement with data for $G_C(q)$ 
in the diffraction region than the N3LO/N3LO$^*$, while the reverse is true for $G_Q(q)$
at $q \gtrsim 3$ fm$^{-1}$.  The AV18 deuteron wave function, particularly its D-wave component,
is markedly different from the N3LO/N3LO$^*$ (see again Fig.~16 in the Machleidt and Entem review~\cite{Entem03}).
\begin{table}
\caption{Individual contributions to the
monopole form factor $G_C(q)$ corresponding to cutoff $\Lambda=500$ MeV
for the N3LO Hamiltonian; $\nu=1/2$ and $(-x)$ stands for $10^{-x}$.}
\begin{tabular}{c||l|l|l|l|}
q\,(fm$^{-1}$)	&    LO	   &     N2LO 	&   N3LO(OPE) 	& N3LO($\nu$)  \\
\hline
0.0		&\,\,\,1.00		&\,\,\,0.00		&\,\,\,0.00	       &\,\,\,0.00	\\	
0.3		&\,\,\,0.945		&--0.340(--3)		&--0.211(--3)	       &--0.600(--6)	\\
0.6		&\,\,\,0.792 		&--0.113(--2)		&--0.799(--3)	       &--0.500(--6)	\\
0.9		&\,\,\,0.614		&--0.196(--2)		&--0.165(--2)		&\,\,\,0.530(--5)\\
1.2		&\,\,\,0.452		&--0.255(--2)		&--0.260(--2)		&\,\,\,0.218(--4)\\
1.5		&\,\,\,0.321		&--0.280(--2)		&--0.349(--2)		&\,\,\,0.515(--4)\\
1.8		&\,\,\,0.220		&--0.273(--2)		&--0.422(--2)		&\,\,\,0.932(--4)\\
2.1		&\,\,\,0.146		&--0.242(--2)		&--0.470(--2)		&\,\,\,0.143(--3)\\
2.4		&\,\,\,0.920(--1)	&--0.197(--2)		&--0.493(--2)		&\,\,\,0.195(--3)\\
2.7		&\,\,\,0.547(--1)	&--0.145(--2)		&--0.493(--2)		&\,\,\,0.245(--3)\\
3.0		&\,\,\,0.295(--1)	&--0.923(--3)		&--0.474(--2)		&\,\,\,0.287(--3)\\
3.3		&\,\,\,0.131(--1)	&--0.448(--3)		&--0.441(--2)		&\,\,\,0.319(--3)\\
3.6		&\,\,\,0.295(--2)	&--0.518(--4)		&--0.400(--2)		&\,\,\,0.339(--3)\\
3.9		&--0.278(--2)		&\,\,\,0.250(--3)	&--0.356(--2)		&\,\,\,0.348(--3)\\
4.2		&--0.556(--2)		&\,\,\,0.453(--3)	&--0.311(--2)		&\,\,\,0.346(--3)\\
4.5		&--0.645(--2)		&\,\,\,0.566(--3)	&--0.269(--2)		&\,\,\,0.335(--3)\\
4.8		&--0.621(--2)		&\,\,\,0.602(--3)	&--0.230(--2)		&\,\,\,0.318(--3)\\
5.1		&--0.539(--2)		&\,\,\,0.579(--3)	&--0.196(--2)		&\,\,\,0.297(--3)\\
\hline
\end{tabular}
\label{tb:gc}
\end{table}
\begin{table}
\caption{Same as in Table~\ref{tb:gc}, but for the quadrupole form factor $G_Q(q)$, normalized
at $q=0$ as in Eq.~(\ref{eq:norm0}).}
\begin{tabular}{c||l|l|l|l|}
q\,(fm$^{-1}$)	&    LO	   &     N2LO 	&   N3LO(OPE) 	& N3LO($\nu$)  \\
\hline
0.0		&24.8		&--0.172			&1.20			&--0.108		\\	
0.3		&23.5		&--0.176			&1.19			&--0.986(--1)		\\
0.6		&19.7 		&--0.182			&1.13			&--0.968(--1)		\\
0.9		&15.3		&--0.184			&1.04			&--0.936(--1)		\\
1.2		&11.4		&--0.179			&0.930			&--0.890(--1)		\\
1.5		&8.29		&--0.167			&0.810			&--0.829(--1)		\\
1.8		&5.92		&--0.150			&0.690			&--0.757(--1)		\\
2.1		&4.18		&--0.131			&0.576			&--0.677(--1)		\\
2.4		&2.92		&--0.111			&0.473			&--0.596(--1)		\\
2.7		&2.03		&--0.917(--1)		&0.383			&--0.515(--1)		\\
3.0		&1.39		&--0.740(--1)		&0.307			&--0.440(--1)		\\
3.3		&0.947		&--0.584(--1)		&0.244			&--0.371(--1)		\\
3.6		&0.635		&--0.449(--1)		&0.193			&--0.310(--1)		\\
3.9		&0.418		&--0.335(--1)		&0.152			&--0.257(--1)		\\
4.2		&0.268		&--0.242(--1)		&0.119			&--0.212(--1)		\\
4.5		&0.167       	&--0.168(--1)		&0.933(--1)		&--0.174(--1)		\\
4.8		&0.997(--1)	&--0.111(--1)		&0.731(--1)		&--0.142(--1)		\\
5.1		&0.568(--1)	&--0.686(--2)		&0.574(--1)		&--0.116(--1)		\\
\hline
\end{tabular}
\label{tb:gq}
\end{table}

The individual contributions corresponding to
$\Lambda=500$ MeV and the N3LO potential are listed in Tables~\ref{tb:gc} and~\ref{tb:gq}
for $q$ values in the range (0.0--5.1) fm$^{-1}$.  The N2LO (N3LO) charge operators are
proportional to $1/m_N^2$ ($1/m_N$), and therefore vanish in the static limit.  The N3LO(OPE)
correction is the leading one for $q \gtrsim 1.5$ fm$^{-1}$, and is responsible for shifting
the zero in the LO $G_C(q)$ to lower $q$.  However, this correction
interferes constructively with the LO contribution in the case of $G_Q(q)$.  The $\nu$-dependent
retardation correction N3LO($\nu$) is found to be negligible, which allows one to conclude
that violations of the unitary equivalence between the OPE potential and associated
charge operator is of little numerical import (for $\nu=0$--1).

\begin{figure}
\includegraphics[width=3.5in]{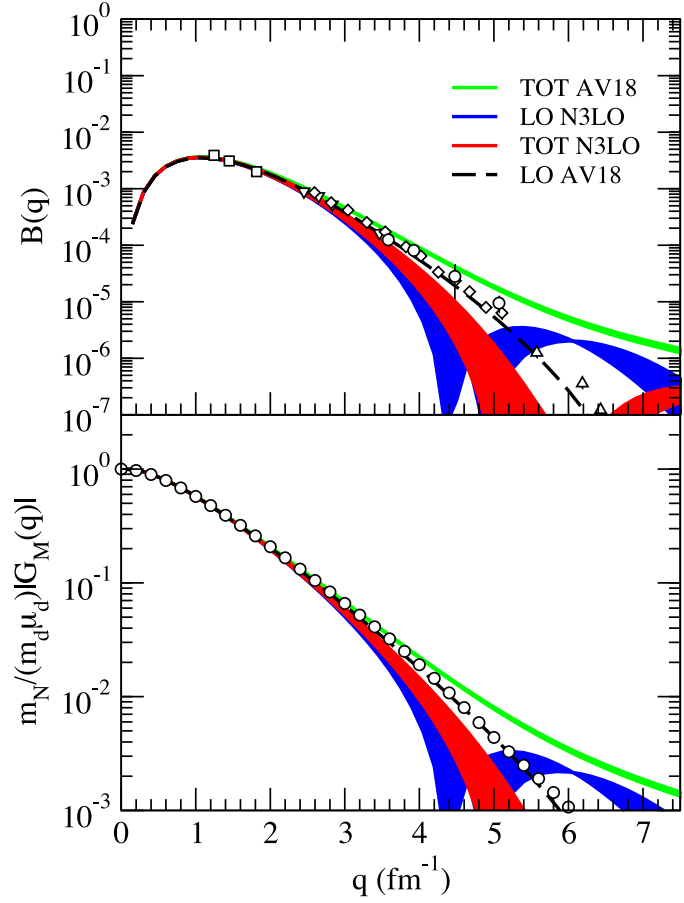}\\
\caption{(Color online). The deuteron $B(q)$ structure function (top panel)
and magnetic form factor $G_M(q)$ (bottom panel), obtained at leading order (LO) and
with inclusion of current operators up to N3LO (TOT), is compared with the experimental data
from Refs.~\cite{STANFORD65,MAINZ81,SACLAY85,BONN85,SLAC90,SICK2001}.
Predictions corresponding to cutoffs $\Lambda$ in the range 500--600 MeV are displayed by the bands.}
\label{fig:f5}
\end{figure}

The deuteron magnetic moment is one of the two observables utilized to fix the
LEC's entering the isoscalar current operators at N3LO, denoted as N3LO(nm) and
N3LO(OPE) in Sec.~\ref{sec:lecs} and Table~\ref{tb:muds}.  The structure
function $B(q)$ and magnetic form factor $G_M(q)$, obtained with the AV18 and
chiral potentials, and currents at LO and by including corrections up to N3LO,
are compared to data in Fig.~\ref{fig:f5}.  There is generally
good agreement between theory and experiment for $q$ values up to $\simeq 2$ fm$^{-1}$.
At higher $q$'s, the results corresponding to the chiral (AV18) potential
under-predict (over-predict) the data significantly when the current includes up to
N3LO corrections.  In particular, the diffraction seen in the data at $q \simeq 6.5$ fm$^{-1}$
is absent in the AV18 calculations, and is shifted to lower $q$ values in the N3LO/N3LO$^*$ ones.
There are large differences between the N3LO/N3LO$^*$ and AV18 results with the LO current,
which simply reflect differences in the S- and D-wave components of the deuteron wave functions
corresponding to these potentials.  The cutoff dependence is large for the chiral potentials, while it
remains quite modest for the AV18 over the whole momentum transfer range.  This is consistent
with the rather different sensitivity of the LEC's $d_1^S$ and $d_2^S$ to variations of $\Lambda$ in the (500-600)
MeV range obtained with either the chiral potential or AV18, see Table~\ref{tb:tds}.
There is a mismatch in the chiral counting between the potentials of Ref.~\cite{Entem03} at order $Q^4$
and the present current at order $e\, Q$.  This becomes obvious  when considering current conservation,
which for these potentials would require accounting for terms up to order $e\, Q^3$ in the
current, well beyond available derivations~\cite{Pastore09,Koelling09,Koelling11} at this time.

\vspace{1.5cm}

\begin{widetext}
\begin{center}
\begin{table}
\caption{Individual contributions to the
magnetic form factor $G_M(q)$,  normalized
at $q=0$ as in Eq.~(\ref{eq:norm0}), corresponding to cutoff $\Lambda=500$ MeV
for the N3LO Hamiltonian; $(-x)$ stands for $10^{-x}$.}
\begin{tabular}{c||l|l|l|l|l|}
q\,(fm$^{-1}$)	&    LO	   &     N2LO 	&   N3LO(min) 	& N3LO(nm) 	& N3LO(OPE)  \\
\hline
0.0		&\,\,\,1.71		&--0.144(--1)		&0.508(--1)	&--0.353(--1)		&0.483(--2)		\\
0.3		&\,\,\,1.62		&--0.150(--1)		&0.508(--1)	&--0.353(--1)		&0.480(--2)		\\
0.6		&\,\,\,1.36		&--0.159(--1)		&0.495(--1)	&--0.343(--1)		&0.473(--2)		\\
0.9		&\,\,\,1.07		&--0.166(--1)		&0.474(--1)	&--0.329(--1)		&0.461(--2)		\\
1.2		&\,\,\,0.807		&--0.166(--1)     	&0.447(--1)	&--0.310(--1)		&0.444(--2)		\\
1.5		&\,\,\,0.590		&--0.159(--1)		&0.416(--1)	&--0.289(--1)		&0.423(--2)		\\
1.8		&\,\,\,0.422		&--0.144(--1)		&0.382(--1)	&--0.265(--1)		&0.399(--2)		\\
2.1		&\,\,\,0.296		&--0.125(--1)		&0.347(--1)	&--0.241(--1)		&0.372(--2)		\\
2.4		&\,\,\,0.202		&--0.104(--1)		&0.312(--1)	&--0.217(--1)		&0.344(--2)		\\
2.7		&\,\,\,0.134		&--0.818(--2)		&0.279(--1)	&--0.194(--1)		&0.315(--2)		\\
3.0		&\,\,\,0.860(--1)	&--0.608(--2)		&0.248(--1)	&--0.172(--1)		&0.286(--2)		\\
3.3		&\,\,\,0.522(--1)	&--0.418(--2)		&0.219(--1)	&--0.152(--1)		&0.258(--2)		\\
3.6		&\,\,\,0.291(--1)	&--0.254(--2)		&0.193(--1)	&--0.134(--1)		&0.231(--2)		\\
3.9		&\,\,\,0.138(--1)	&--0.120(--2)		&0.169(--1)	&--0.117(--1)		&0.206(--2)		\\
4.2		&\,\,\,0.415(--2)	&--0.159(--3)		&0.148(--1)	&--0.103(--1)		&0.183(--2)		\\
4.5		&--0.152(--2)		&\,\,\,0.581(--3)	&0.130(--1)	&--0.899(--2)		&0.162(--2)		\\
4.8		&--0.444(--2)		&\,\,\,0.105(--2)	&0.113(--1)	&--0.785(--2)		&0.143(--2)		\\
5.1		&--0.554(--2)		&\,\,\,0.128(--2)	&0.988(--2)	&--0.686(--2)		&0.126(--2)		\\
\hline
\end{tabular}
\label{tb:gm}
\end{table}
\end{center}
\end{widetext}

The AV18 results obtained here for $B(q)$ are similar to those reported in the conventional framework of
Ref.~\cite{Schiavilla02} (see curve labeled IA+$\rho\pi\gamma$-NR in Fig.~5).  In that work, the current
included the standard impulse-approximation (IA) term---the LO current in $\chi$EFT---and the two-body
term from $\rho\pi\gamma$ transitions.  The size, and in fact sign, of the $\rho\pi\gamma$ contribution
were found to depend on whether the current was derived by retaining the fully relativistic (R) structure
of the associated Feynman amplitude, or only the leading-order term in its non-relativistic (NR)
expansion---in this latter case, it is essentially the N3LO(OPE) current of Eq.~(\ref{eq:cp}).  Indeed, the
$\rho\pi\gamma$ contribution had the same (opposite) sign as the IA when
it was evaluated with the NR (R) current, and the IA+$\rho\pi\gamma$(NR) results
overestimated the data by an amount similar to that shown in Fig.~\ref{fig:f5}. 

Recently, a calculation of the deuteron magnetic structure, based on the same $\chi$EFT utilized
here, has appeared in the literature~\cite{Koelling12}.  It uses chiral potentials at order $Q^2$ derived in
Ref.~\cite{Epelbaum05}, and a different strategy from that adopted here for constraining the two LEC's in the
isoscalar N3LO current.  One of them is still fixed by reproducing $\mu_d\, $; the other, however,
is determined by a fit to $B(q)$ data up to $q\simeq 2$ fm$^{-1}$.  Predictions
for this observable in $q=(2$--4) fm$^{-1}$ seem to overestimate the data at the
highest $q$ values ($q \gtrsim 3.5$ fm$^{-1}$), but display much less
cutoff dependence than obtained here.  This is clearly due to the different way in which
the LEC's are constrained in the two calculations. 

Finally, in Table~\ref{tb:gm} we list the individual contributions to $G_M(q)$ obtained
with the N3LO potential and cutoff $\Lambda=500$ MeV.  The notation is as follows:
LO is the leading-order ($e\, Q^{-2}$) current of Eq.~(\ref{eq:jlo}); N2LO is the
relativistic correction of order $n=0$ ($e\, Q^0$) in Eq.~(\ref{eq:j1rc}); N3LO(min), N3LO(nm), and
N3LO(OPE) are the corrections of order $n=1$ ($e\, Q$) in Eqs.~(\ref{eq:jmin}), (\ref{eq:nmcounter}),
and~(\ref{eq:cp}), respectively.  The N3LO(min) and N3LO(nm) contributions from the
minimal and non-minimal contact currents cancel to a large extent, and their combined effect
is comparable to the N3LO(OPE) contribution.  This interplay among different
corrections, however, depends strongly on $\Lambda$ and the Hamiltonian model
considered.

\subsection{Static properties and form factors of the trinucleons}
\begin{widetext}
\begin{center}
\begin{figure*}
\includegraphics[width=7in]{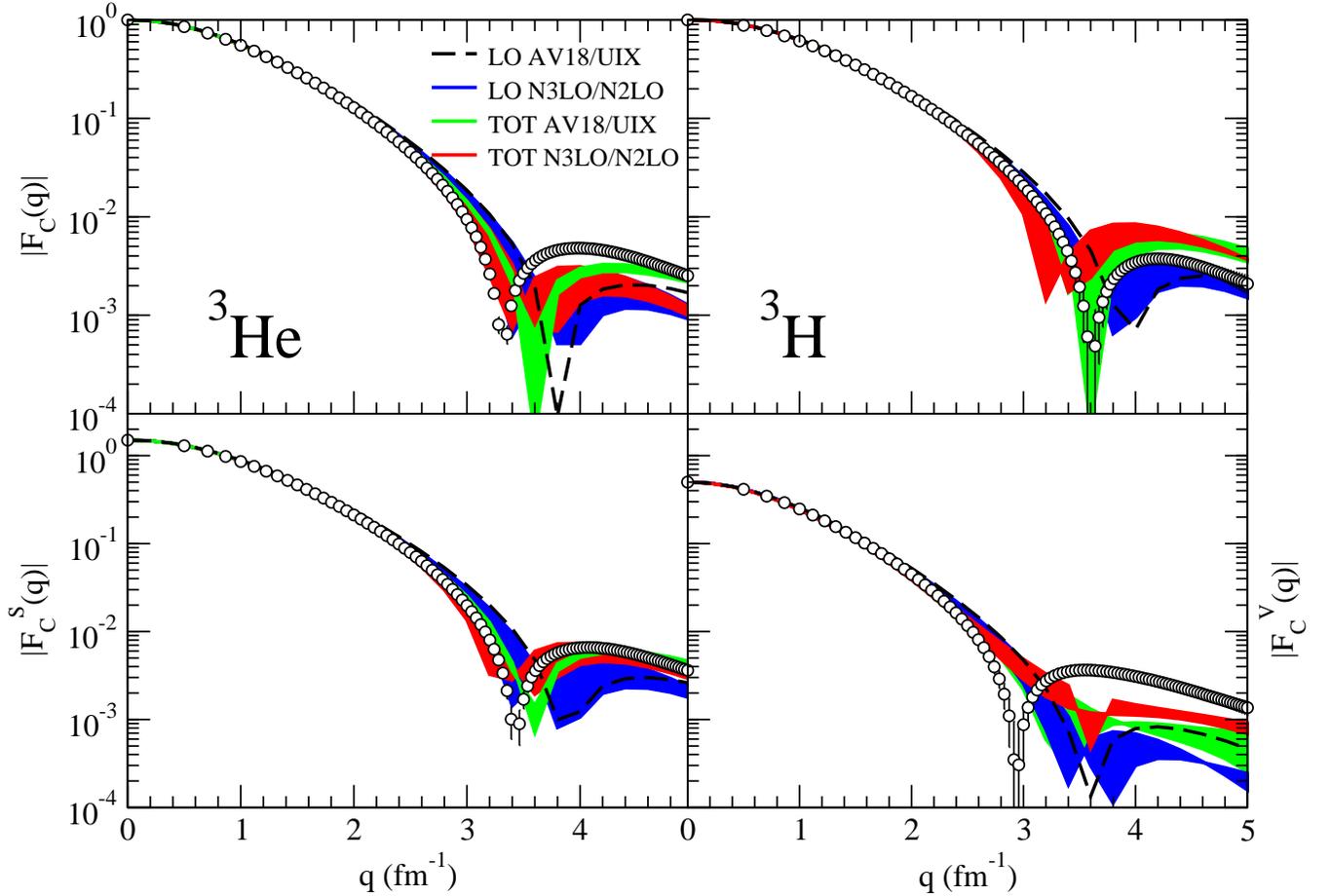}\\
\caption{(Color online). The $^3$He and $^3$H charge form factors (top panels),
and their isoscalar and isovector combinations (bottom panels),
obtained at leading order (LO)  and with inclusion of charge operators up to N4LO (TOT), is compared with
experimental data~\cite{AMROUN}.  Predictions corresponding to $\nu=1/2$ and cutoffs
$\Lambda$ in the range (500--600) MeV are displayed by the bands.}
\label{fig:f15}
\end{figure*}
\end{center}
\end{widetext}

\begin{widetext}
\begin{center}
\begin{table}
\caption{Cumulative contributions in fm to the $^3$He and $^3$H root-mean-square
charge radii corresponding to $\nu=1/2$ and cutoffs $\Lambda=500$ MeV and 600 MeV, obtained
with the N3LO/N2LO and N3LO$^*$/N2LO$^*$ Hamiltonians; results in parentheses
are relative to the AV18/UIX Hamiltonian.  The experimental values
for the $^3$He and $^3$H charge radii are~\cite{SICK2001}
$(1.959\pm0.030)$ fm and $(1.755\pm 0.086)$\,fm, respectively.}
\begin{tabular}{c||c|c||c|c||}
& \multicolumn{2}{c||}{$^3$He}& \multicolumn{2}{c||}{$^3$H} \\
\hline
  $\Lambda$         & 500     &     600 &      500         &    600           \\
\hline
LO               &  1.966 (1.950)  & 1.958 (1.950) &  1.762 (1.743) &  1.750 (1.743) \\
N2LO             &  1.966 (1.950)  & 1.958 (1.950) &  1.762 (1.743) &  1.750 (1.743) \\
N3LO             &  1.966 (1.950)  & 1.958 (1.950) &  1.762 (1.743) &  1.750 (1.743) \\
N4LO             &  1.966 (1.950)  & 1.958 (1.950) &  1.762 (1.743) &  1.750 (1.743) \\
\hline
\end{tabular}
\label{tb:crh3a}
\end{table}
\end{center}
\end{widetext}
The notation for the various components of the charge operator
is the same as given at the beginning of Sec.~\ref{sec:deut},
except that now the one-loop (isovector) corrections at N4LO
contribute too, since the $^3$He and $^3$H nuclei have predominantly
total isospin $T=1/2$.
As a matter of fact, the hyperspherical harmonics wave functions
utilized to represent their ground states also include
small $T=3/2$ admixtures due to isospin-symmetry breaking terms 
induced by the electromagnetic and strong interactions.

There are no unknown LEC's entering the charge operator up to N4LO, and
the predicted root-mean-square charge radii of $^3$He and $^3$H, obtained
with the N3LO/N2LO and AV18/UIX combinations of two- and three-nucleon potentials
and cutoffs in the (500--600) MeV range, are listed in Table~\ref{tb:crh3a}.
Corrections at N2LO, N3LO, and N4LO are negligible---the corresponding operators
vanish at $q=0$.  The spread between the N3LO/N2LO ($\Lambda=500$ MeV) and
N3LO$^*$/N2LO$^*$ ($\Lambda=600$ MeV) results at LO is about 0.5\%, which
is much smaller, particularly for $^3$H, than the experimental error. 
The predicted radii for both Hamiltonian models are within 0.5\% of the
current experimental central values.

\begin{figure}
\includegraphics[width=3.35in]{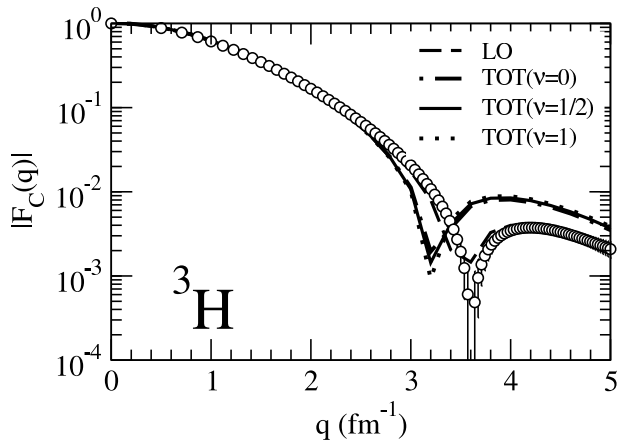}\\
\caption{The $^3$H charge form factor obtained with the N3LO/N2LO Hamiltonian,
cutoff $\Lambda=500$ MeV, and charge operators up to N4LO corresponding
to $\nu=0$, 1/2, and 1.}
\label{fig:f10}
\end{figure}

The calculated charge form factors of $^3$He and $^3$H, and their
isoscalar and isovector combinations $F_C^S(q)$ and $F_C^V(q)$,
normalized, respectively, to 3/2 and 1/2 at $q=0$, are compared
to data in Fig.~\ref{fig:f15}.  The agreement between
theory and experiment is excellent for $q \lesssim 2.5$ fm$^{-1}$.
At larger values of the momentum transfer, there is a significant
sensitivity to cutoff variations in the results obtained
with the chiral potentials. This cutoff dependence is large at
LO and is reduced, at least in $^3$He, when corrections up to N4LO are included.
These corrections have opposite sign than the LO, and tend to shift
the zeros in the form factors to lower momentum transfers, bringing
theory closer to experiment in the diffraction region. 

As already remarked, the chiral (and conventional) two-nucleon potentials
utilized in the present study ignore retardation corrections in their OPE and
TPE components, which corresponds to the choice $\nu=1/2$ in the non-static
pieces of the corresponding potentials and accompanying charge operators
in Eqs.~(\ref{Diagram_dnu2}) and~(\ref{eq:c_e})~\cite{Pastore11}.  Figure~\ref{fig:f10}
is meant to illustrate how inconsistencies between the potential and charge
operator impact predictions for the $^3$H form factor, by presenting results
obtained with the N3LO/N2LO Hamiltonian ($\nu=1/2$), cutoff $\Lambda=500$ MeV,
and N3LO and N4LO corrections with $\nu=0$ and 1 in the charge operator.  Their
effect is negligible.

\begin{figure}
\includegraphics[width=3.35in]{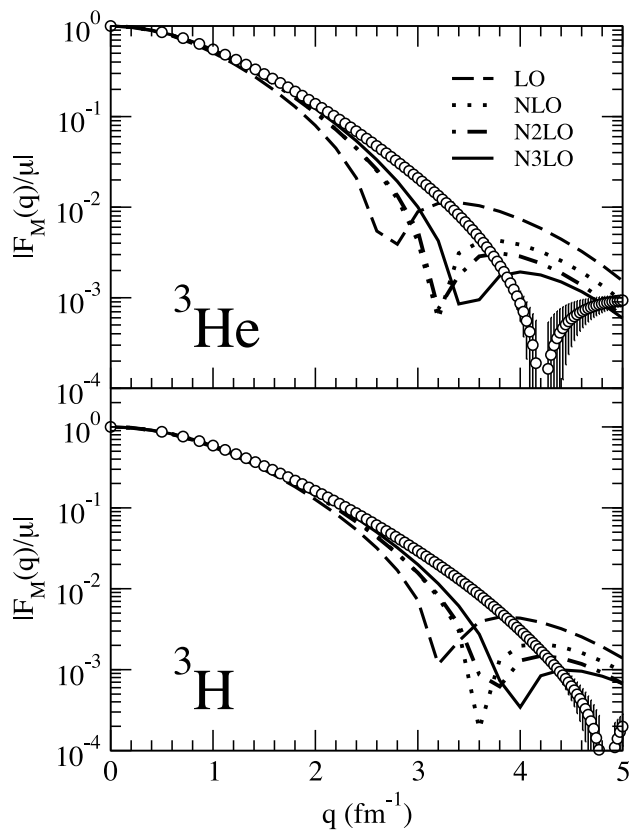}\\
\caption{Cumulative contributions to the $^3$He and $^3$H charge form factors,
obtained with the N3LO/N2LO Hamiltonian, cutoff $\Lambda=500$ MeV, and $\nu=1/2$,
from the components of the charge operator order by order.}
\label{fig:f16}
\end{figure}
\vspace{1cm}

In Fig.~\ref{fig:f16}, we show cumulatively the LO,
N2LO, N3LO, and N4LO contributions to the charge form factors of $^3$He
and $^3$H.  The N2LO are smallest, while the N3LO and N4LO turn out to
be comparable.  This is illustrated explicitly in Tables~\ref{tb:fche3}
and~\ref{tb:fch3}, where we list the individual contributions of the various
terms entering at each order.  In these tables, we denote with N3LO(OPE)
and N3LO($\nu$) the operators in Eqs.~(\ref{eq:pich}) and (\ref{Diagram_dnu2})--(\ref{eq:rho_gpipi}),
respectively; with N4LO(cd), N4LO(ef; $\nu$), N4LO(ij) those in Eqs.~(\ref{eq:c_c})--(\ref{eq:c_d}),
Eqs.~(\ref{eq:c_e})--(\ref{eq:c_f}), and Eqs.~(\ref{eq:c_i})--(\ref{eq:c_j}).
Among the corrections at N3LO the OPE term---column N3LO(OPE)---illustrated by
panel (a) in Fig.~\ref{fig:f1} is dominant, while among those at N4LO the TPE
terms---columns N4LO(ef; $\nu$) and N4LO(ij)---illustrated by panels (e)-(f)
and (i)-(j) in Fig.~\ref{fig:f3} are dominant.  The N3LO(OPE)
and N4LO(ef; $\nu$) and N4LO(ij) contributions are of similar magnitude,
indeed there is no hint of suppression in going from N3LO to N4LO, as one
would have naively expected on the basis of power counting.

The $^3$He contributions in Table~\ref{tb:fche3} have been divided by
the number of protons $Z=2$ in order to have the form factor normalized
to one at $q=0$.  The N4LO charge operators are isovector and, if $^3$He
and $^3$H were pure $T=1/2$ states, then $2 \times $N4LO($^3$He)=--N4LO($^3$H).
That this equality is not exactly satisfied reflects the fact that the
present $^3$He and $^3$H wave functions are not simply the charge mirror
of each other---that is, $\left(\prod_i \tau_{i,x}\right)\,|^3$He$\rangle
\neq |^3$H$\rangle$, where $\tau_{i,x}$ is the $x$-component of nucleon
$i$ isospin operator.
\begin{widetext}
\begin{center}
\begin{table}
\caption{Individual contributions to the $^3$He charge form factor, obtained
with the N3LO/N2LO Hamiltonian, cutoff $\Lambda=500$ MeV, and $\nu=1/2$; ($-x$) stands for $10^{-x}$.}
\begin{tabular}{c||l|l|l|l|l|l|l|}
q (fm$^{-1})$  & LO & N2LO & N3LO(OPE)& N3LO($\nu$) & N4LO(cd) & N4LO(ef; $\nu$) &N4LO(ij)\\
\hline
0.0	&\,\,1.00		&\,\,0.00		&\,\,0.00		&\,\,0.00	      &\,\,0.00		&\,\,0.00	 	&0.00\\ 
0.2	&\,\,0.983	&--0.561(--3)	&--0.152(--3)&\,\,0.100(--4)&--0.477(--5)	&\,\,0.904(--4)	&0.301(--4)\\
0.6	&\,\,0.807 	&--0.406(--2)	&--0.125(--2)&\,\,0.870(--4)&\,\,0.108(--3)	&\,\,0.785(--4)	&0.346(--3)\\
1.0	&\,\,0.562	&--0.760(--2)	&--0.289(--2)&\,\,0.205(--3)&\,\,0.296(--3)	&\,\,0.126(--4)	&0.835(--3)\\
1.4	&\,\,0.342	&--0.876(--2)	&--0.434(--2)&\,\,0.324(--3)&\,\,0.503(--3)	&--0.134(--3)	&0.132(--2)\\
1.8	&\,\,0.185	&--0.758(--2)	&--0.513(--2)&\,\,0.409(--3)&\,\,0.680(--3)	&--0.347(--3)	&0.167(--2)\\
2.2	&\,\,0.892(--1)	&--0.525(--2)	&--0.516(--2)&\,\,0.447(--3)&\,\,0.796(--3)	&--0.585(--3)	&0.184(--2)\\
2.6	&\,\,0.366(--1)	&--0.288(--2)	&--0.463(--2)&\,\,0.433(--3)&\,\,0.841(--3)	&--0.796(--3)	&0.184(--2)\\
3.0	&\,\,0.111(--1)&--0.107(--2)	&--0.380(--2)&\,\,0.385(--3)&\,\,0.821(--3)	&--0.938(--3)	&0.170(--2)\\
3.4	&\,\,0.623(--3)	&--0.969(--5)	&--0.292(--2)&\,\,0.315(--3)&\,\,0.750(--3)	&--0.989(--3)	&0.148(--2)\\
3.8	&--0.258(--2)	&\,\,0.488(--3)	&--0.213(--2)&\,\,0.241(--3)&\,\,0.644(--3)	&--0.949(--3)	&0.122(--2)\\
4.2	&--0.276(--2)	&\,\,0.577(--3)	&--0.148(--2)&\,\,0.174(--3)&\,\,0.522(--3)	&--0.831(--3)	&0.957(--3)\\
4.6	&--0.200(--2)	&\,\,0.476(--3)	&--0.994(--3)&\,\,0.117(--3)&\,\,0.395(--3)	&--0.660(--3)	&0.702(--3)\\
5.0	&--0.119(--2)	&\,\,0.318(--3)	&--0.642(--3)&\,\,0.730(--4)&\,\,0.274(--3)	&--0.469(--3)	&0.476(--3)\\
\hline
\end{tabular}
\label{tb:fche3}
\end{table}
\begin{table}
\caption{Same as in Table~\ref{tb:fche3}, but for $^3$H.}
\begin{tabular}{c||l|l|l|l|l|l|l|}
q (fm$^{-1})$  & LO & N2LO & N3LO(OPE)& N3LO($\nu$) & N4LO(cd) & N4LO(ef; $\nu$) &N4LO(ij)\\
\hline
0.0	&\,\,1.00	&\,\,0.00		&\,\,0.00		&\,\,0.00			&\,\,0.00	&\,\,0.00	&0.00\\ 
0.2	&\,\,0.991	 	&\,\,0.647(--3)	&--0.185(--3)	&\,\,0.110(--4)		&\,\,0.103(--4)	&--0.190(--3)	&--0.633(--4)\\
0.6   &\,\,0.844	 	&\,\,0.462(--2)	&--0.152(--2)	&\,\,0.880(--4)		&--0.229(--3)	&--0.136(--3)	&--0.729(--3)\\
1.0	&\,\,0.621	 	&\,\,0.844(--2)	&--0.356(--2)  &\,\,0.245(--3)		&--0.628(--3)	&\,\,0.459(--4) &--0.176(--2)\\
1.4	&\,\,0.402	 	&\,\,0.940(--2)	&--0.543(--2)	&\,\,0.462(--3)		&--0.107(--2)	&\,\,0.396(--3)	 &--0.278(--2)\\
1.8	&\,\,0.231	 	&\,\,0.784(--2)	&--0.653(--2)	&\,\,0.686(--3)	 	&--0.144(--2)	&\,\,0.874(--3)	&--0.352(--2)\\
2.2	&\,\,0.118	 	&\,\,0.520(--2)	&--0.671(--2)	&\,\,0.858(--3)		&--0.169(--2)	&\,\,0.138(--2)	&--0.389(--2)\\
2.6	&\,\,0.517(--1)	        &\,\,0.270(--2)	&--0.615(--2)	&\,\,0.944(--3)		&--0.179(--2)	&\,\,0.182(--2)	&--0.389(--2)\\
3.0	&\,\,0.174(--1)	        &\,\,0.903(--3)	&--0.519(--2)	&\,\,0.940(--3)		&--0.174(--2)	&\,\,0.210(--2)	&--0.361(--2)\\
3.4	&\,\,0.204(--2)	        &  --0.972(--4)	&--0.411(--2)	&\,\,0.867(--3)		&--0.159(--2)	&\,\,0.218(--2) &--0.314(--2)\\
3.8	&--0.328(--2)		&  --0.483(--3)	&--0.310(--2)	&\,\,0.615(--3)		&--0.136(--2)	&\,\,0.206(--2)	&--0.259(--2)\\
4.2	&--0.403(--2)		&  --0.508(--3)	&--0.225(--2)	&\,\,0.543(--3)		&--0.110(--2)	&\,\,0.178(--2)	&--0.201(--2)\\
4.6	&--0.313(--2)		&  --0.383(--3)	&--0.159(--2)	&\,\,0.483(--3)		&--0.825(--3)	&\,\,0.139(--2)	&--0.147(--2)\\
5.0	&--0.196(--2)		&  --0.232(--3)	&--0.108(--2)	&\,\,0.364(--3)		&--0.565(--3)	&\,\,0.942(--3)	&--0.985(--3)\\
\hline
\end{tabular}
\label{tb:fch3}
\end{table}
\end{center}
\end{widetext}
\begin{widetext}
\begin{center}
\begin{table}
\caption{Cumulative contributions in fm to the $^3$He and $^3$H root-mean-square
magnetic radii corresponding to cutoffs $\Lambda=500$ MeV and 600 MeV, obtained
with the N3LO/N2LO and N3LO$^*$/N2LO$^*$ Hamiltonians; results in parentheses
are relative to the AV18/UIX Hamiltonian.  Predictions corresponding
to sets I, II, and II of isovector LEC's $d_1^V$ and $d_2^V$ in Table~\ref{tb:tdv}
are listed.  The experimental values
for the $^3$He and $^3$H magnetic radii are~\cite{SICK2001}
$(1.965\pm0.153)$ fm and $(1.840\pm 0.181)$\,fm, respectively.}
\begin{tabular}{c||c|c||c|c||}
& \multicolumn{2}{c||}{$^3$He}& \multicolumn{2}{c||}{$^3$H} \\
\hline
  $\Lambda$    & 500     &     600 &      500         &    600           \\
\hline
LO           &  2.098 (2.092) &  2.090 (2.092) &  1.924 (1.918) &  1.914 (1.918) \\
NLO          &  1.990 (1.981) &  1.983 (1.974) &  1.854 (1.847) &  1.845 (1.841) \\
N2LO         &  1.998 (1.992) &  1.989 (1.984) &  1.865 (1.859) &  1.855 (1.854) \\
\hline
N3LO(I)      &  1.924 (1.931) &  1.910 (1.972) &  1.808 (1.800) &  1.796 (1.819) \\
\hline
N3LO(II)     &  1.901 (1.890) &  1.883 (1.896) &  1.789 (1.774) &  1.773 (1.778) \\
\hline
N3LO(III)    &  1.927 (1.915) &  1.913 (1.924) &  1.808 (1.792) &  1.794 (1.797) \\
\hline
\end{tabular}
\label{tb:mrh3}
\end{table}
\end{center}
\end{widetext}

Moving on to the magnetic structure of the trinucleons, we
note that the isoscalar combination $\mu_S$ of $^3$He and $^3$H magnetic
moments is used to fix one of the two (isoscalar) LEC's entering the
current at N3LO.  Both the isovector combination $\mu_V$ and the $np$ radiative
capture cross section $\sigma_{np}$ are used to fix the isovector LEC's in set I
of the N3LO currents, while in sets II and III one of these LEC's is fixed by $\Delta$ dominance,
and the other is determined by reproducing $\sigma_{np}$ ($\mu_V$) in set II (III),
see Tables~\ref{tb:muds} and~\ref{tb:mudv}.  By construction, then, the
$^3$He and $^3$H magnetic moments are exactly reproduced in sets I and III,
while in set II they are calculated to be, respectively, --2.186 (--2.196) $\mu_N$ and
3.038 (3.048) $\mu_N$ with the N3LO/N2LO (N3LO$^*$/N2LO$^*$) Hamiltonian
and $\Lambda=500$ (600) MeV, and similar results with the AV18/UIX Hamiltonian.
These should be compared to the experimental values of --2.127 $\mu_N$ and
2.979 $\mu_N$.

\vspace{1cm}

\begin{widetext}
\begin{center}
\begin{figure}
\includegraphics[width=7in]{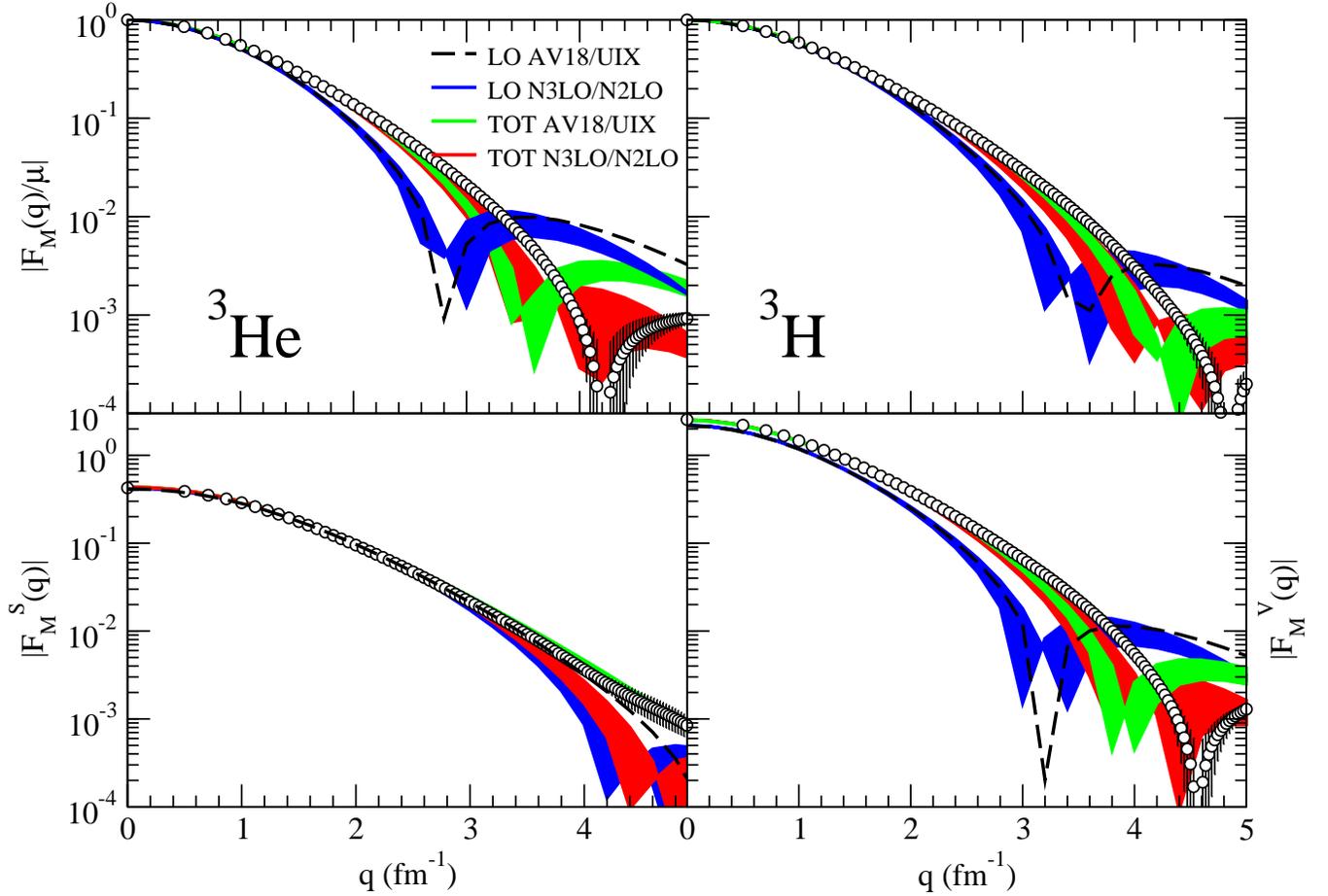}\\
\caption{(Color online). The $^3$He and $^3$H magnetic form factors (top panels),
and their isoscalar and isovector combinations (bottom panels),
obtained at leading order (LO) 
and with inclusion of current operators up to N3LO (TOT) corresponding
to the LEC's $d_1^S$ and $d_2^S$ in Table~\ref{tb:tds} and to set III of
isovector LEC's $d_1^V$ and $d_2^V$ in Table~\ref{tb:tdv}, are compared with
experimental data~\cite{AMROUN}.  Predictions relative to cutoffs
$\Lambda$ in the range (500--600) MeV are displayed by the bands.}
\label{fig:f17}
\end{figure}
\end{center}
\end{widetext}

The $^3$He and $^3$H magnetic radii corresponding to sets I-III are given
in Table~\ref{tb:mrh3}.  The predicted values are consistent with experiment,
although the measurements have rather large errors (10\% for $^3$H).
Their spread as $\Lambda$ varies in the (500--600) MeV range is at the
1\% level or less.  A recent quantum Monte Carlo study~\cite{Pastore12}, using wave functions derived
from conventional two- and three nucleon potentials (the AV18 and Illinois 7 model~\cite{Pieper08}) and set III 
of $\chi$EFT currents, has led to predictions for magnetic moments and transitions
in nuclei with $A \leq 9$ in excellent agreement with the measured values.
Therefore in the following, unless stated otherwise, we adopt set III of isovector
LEC's.  We disregard set I for the reasons already explained in Sec.~\ref{sec:lecs}.

The magnetic form factors of $^3$He and $^3$H and their isoscalar
and isovector combinations $F_M^S(q)$ and $F_M^V(q)$, normalized
respectively as $\mu_S$ and $\mu_V$ at $q=0$, at LO and with
inclusion of corrections up to N3LO in the current, are displayed
in Fig.~\ref{fig:f17}.  As is well known from
studies based on the conventional meson-exchange framework (see
the review~\cite{Carlson98} and references therein), two-body
currents are crucial for ``filling in'' the zeros obtained in
the LO calculation due to the interference between the S- and
D-state components in the ground states of these nuclei.
For $q \lesssim 2$ fm$^{-1}$ there is excellent agreement between
the present $\chi$EFT predictions and experiment.  However, as the
momentum transfer increases, even after making allowance for the
significant cutoff dependence, theory tends to underestimate the data,
in particular it predicts the zeros in both form factors occurring
at significantly lower values of $q$ than observed.  Thus, the
first diffraction region remains problematic for the present theory, confirming
earlier conclusions derived from studies in the conventional
framework~\cite{Marcucci98,Marcucci05}.

\begin{figure}
\includegraphics[width=3.35in]{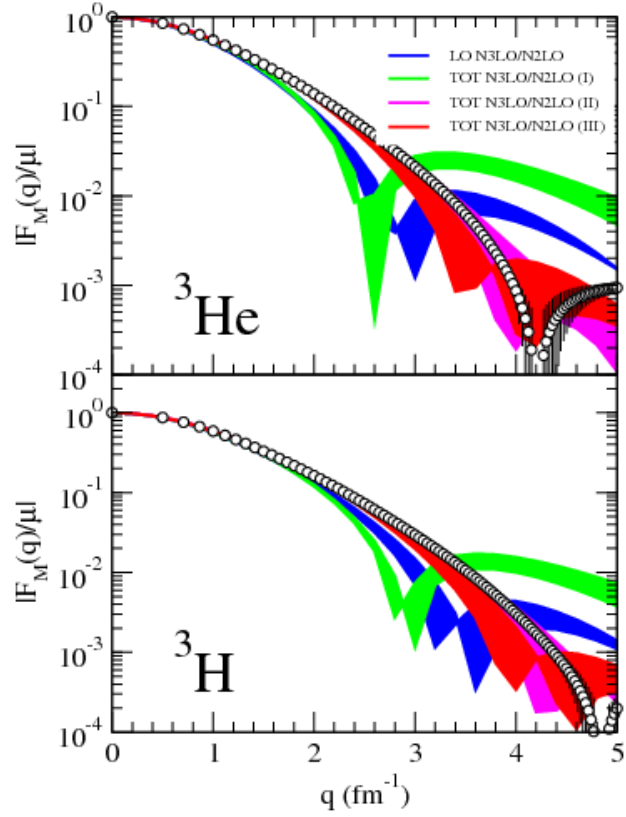}\\
\caption{(Color online). The $^3$He and $^3$H magnetic form factors,
obtained at leading order (LO) 
and with inclusion of current operators up to N3LO (TOT) corresponding
to sets I, II, and III of isovector LEC's $d_1^V$ and $d_2^V$ in Table~\ref{tb:tdv},
is compared with experimental data~\cite{AMROUN}.
Predictions, relative to the N3LO/N2LO Hamiltonian and corresponding to
cutoffs $\Lambda=500$--600 MeV, are displayed by the bands.}
\label{fig:f18}
\end{figure}
Figure~\ref{fig:f18} illustrates the sensitivity of the
N3LO predictions on the different ways in which the isovector LEC's are
constrained in sets I, II, III.  The set I results are strongly at variance with
data.  Set II leads to two-body current contributions larger than in set III,
and consequently, in contrast to set III, the corresponding form factors
reproduce the data in the diffraction region.  However, the cutoff
variation of the results is considerably larger than for set III, as reflected
in the change of the LEC $d_1^V$ for $\Lambda=500$--600 MeV
in Table~\ref{tb:tdv}.  Furthermore, set II overestimates $\mu_V$ by
about 3\%.

\begin{figure}
\includegraphics[width=3.35in]{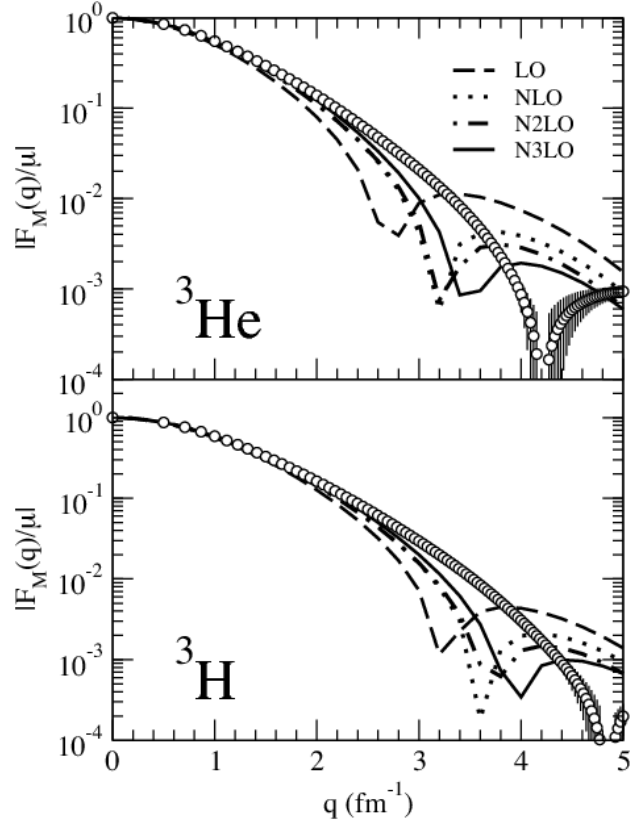}\\
\caption{Cumulative contributions to the $^3$He and $^3$H magnetic form factors,
obtained with the N3LO/N2LO Hamiltonian and cutoff $\Lambda=500$ MeV,
from the components of the current operator order by order.  Set III
is adopted for the isovector LEC's $d_1^V$ and $d_2^V$ in Table~\ref{tb:tdv}.} 
\label{fig:f19}
\end{figure}

\begin{widetext}
\begin{center}
\begin{table}
\caption{Individual contributions to the $^3$He magnetic form factor, obtained
with the N3LO/N2LO Hamiltonian, cutoff $\Lambda=500$ MeV, and set III
of isovector LEC's; ($-x$) stands for $10^{-x}$.}
\begin{tabular}{c||l|l|l|l|l|l|l|}
q\,(fm$^{-1}$)	& LO & NLO & N2LO & N3LO (L) &  N3LO (min) & N3LO(nm) & N3LO(OPE) \\
\hline
0.2	&--1.72	       	&--0.196		&\,\,0.162(--1)		&--0.385(--1)	&0.511(--1)	&--0.109		&--0.488(--1)\\
0.6	&--1.37	       	&--0.184		&\,\,0.160(--1)	&--0.358(--1)	&0.478(--1)		&--0.101		&--0.414(--1)\\
1.0	&--0.898     	&--0.160		&\,\,0.142(--1)		&--0.310(--1)	&0.421(--1)		&--0.877(--1)	&--0.293(--1)\\
1.4	&--0.495      	&--0.128		&\,\,0.105(--1)	&--0.253(--1)	&0.352(--1)		&--0.716(--1)	&--0.164(--1)\\
1.8	&--0.228      	&--0.934(--1)	&\,\,0.595(--2)		&--0.196(--1)	&0.280(--1)		&--0.554(--1)	&--0.584(--2)\\
2.2	&--0.794(--1)	&--0.632(--1)	&\,\,0.189(--2)		&--0.145(--1)	&0.214(--1)		&--0.410(--1)	&\,\,0.106(--2)\\
2.6	&--0.964(--2)	&--0.402(--1)	&--0.912(--3)  		&--0.103(--1)	&0.159(--1)	&--0.292(--1)	&\,\,0.448(--2)\\
3.0	&\,\,0.158(--1)&--0.241(--1)	&--0.234(--2)		&--0.709(--2)	&0.114(--1)		&--0.201(--1)	&\,\,0.542(--2)\\
3.4	&\,\,0.199(--1)&--0.138(--1)	&--0.266(--2)		&--0.473(--2)	&0.796(--2)		&--0.134(--1)	&\,\,0.496(--2)\\
3.8	&\,\,0.159(--1)&--0.756(--2)	&--0.231(--2)  		&--0.306(--2)	&0.543(--2)		&--0.869(--2)	&\,\,0.392(--2)\\
4.2	&\,\,0.103(--1)	&--0.396(--2)	&--0.170(--2)  		&--0.191(--2)	&0.360(--2)		&--0.543(--2)	&\,\,0.281(--2)\\
4.6	&\,\,0.576(--2)	&--0.198(--2)	&--0.108(--2)  		&--0.115(--3)	&0.231(--2)		&--0.326(--2)	&\,\,0.185(--2)\\
5.0	&\,\,0.272(--2)	&--0.929(--3)	&--0.584(--3)   	&--0.658(--3)	&0.143(--2)		&--0.186(--2)	&\,\,0.113(--2)\\
\hline
\end{tabular}
\label{tb:fmhe3}
\end{table}
\end{center}
\end{widetext}
\begin{widetext}
\begin{center}
\begin{table}
\caption{Same as in Table~\ref{tb:fmhe3}, but for $^3$H.}
\begin{tabular}{c||l|l|l|l|l|l|l|}
q\,(fm$^{-1}$)	& LO & NLO & N2LO & N3LO (L) &  N3LO (min) & N3LO(nm) & N3LO(OPE) \\
\hline
0.2	&\,\,2.56		&0.199		&--0.325(--1)	&0.413(--1)	&\,\,0.178(--1)&0.659(--1)	&\,\,0.517(--1)\\
0.6   &\,\,2.11		&0.188		&--0.319(--1)	&0.380(--1)	&\,\,0.163(--1)&0.612(--1)	&\,\,0.439(--1)\\
1.0	&\,\,1.47		&0.164		&--0.290(--1) 	&0.333(--1)	&\,\,0.138(--1)&0.530(--1)	&\,\,0.312(--1)\\
1.4	&\,\,0.894	&0.131		&--0.232(--1)	&0.272(--1)	&\,\,0.107(--1)&0.432(--1)	&\,\,0.176(--1)\\
1.8	&\,\,0.477	&0.964(--1)	&--0.159(--1)	&0.211(--1)	&\,\,0.763(--2)&0.334(--1)	&\,\,0.627(--2)\\
2.2	&\,\,0.221	&0.656(--1)	&--0.894(--2)	&0.156(--1)	&\,\,0.502(--2)&0.247(--1)	&--0.119(--2)\\
2.6	&\,\,0.834(--1)&0.418(--1)	&--0.354(--2)	&0.111(--1)	&\,\,0.299(--2)&0.176(--1)	&--0.450(--2)\\
3.0	&\,\,0.188(--1)&0.252(--1)	&\,\,0.117(--3)&0.764(--2)	&\,\,0.155(--2)&0.122(--1)	&--0.615(--2)\\
3.4	&--0.591(--2)	&0.145(--1)	&\,\,0.156(--2)&0.508(--2)	&\,\,0.607(--3)&0.815(--2)	&--0.576(--2)\\
3.8	&--0.117(--1)	&0.796(--2)	&\,\,0.201(--2)&0.327(--2)	&\,\,0.353(--4)&0.531(--2)	&--0.469(--2)\\
4.2	&--0.101(--1)	&0.419(--2)	&\,\,0.177(--2)&0.203(--2)	&--0.275(--3)	&0.337(--2)	&--0.350(--2)\\
4.6	&--0.666(--2)	&0.210(--2)	&\,\,0.128(--2)&0.121(--2)	&--0.415(--3)	&0.207(--2)	&--0.243(--2)\\
5.0	&--0.365(--2)	&0.993(--3)	&\,\,0.777(--3)&0.681(--3)	&--0.451(--3)	&0.124(--2)	&--0.159(--2)\\
\hline
\end{tabular}
\label{tb:fmh3}
\end{table}
\end{center}
\end{widetext}
Figure~\ref{fig:f19} exhibits cumulatively the LO,
NLO, N2LO, and N3LO contributions to the $^3$He and $^3$H magnetic form
factors, obtained with the N3LO/N2LO Hamiltonian and cutoff $\Lambda=500$ MeV.
Tables~\ref{tb:fmhe3} and~\ref{tb:fmh3} list the individual components of
these contributions at selected values of $q$.  The notation is as
follows: with LO we denote the one-body current in Eq.~(\ref{eq:jlo});
with NLO the OPE currents in Eq.~(\ref{eq:nlo1}); with N2LO the relativistic
correction to the one-body current in Eq.~(\ref{eq:j1rc}); with N3LO(loop)
the one-loop current in Eq.~(\ref{eq:jloop}); with N3LO(min) the ``minimal''
contact current in Eq.~(\ref{eq:jmin}); with N3LO(nm) the ``non-minimal''
contact current in Eq.~(\ref{eq:nmcounter}); and finally with N3LO(OPE) the
OPE currents at N3LO given in Eqs.~(\ref{eq:cdlt}) and~(\ref{eq:cp}).
The NLO and N3LO(loop) are purely isovector, while the remaining operators
have both isoscalar and isovector terms.
As in the case of the charge form factors, the expected suppression of the
N$n$LO corrections as $(q/\Lambda_\chi)^n$, where we have taken $Q \sim q$
as the ``low-momentum'' scale and $\Lambda_\chi=700$--800 MeV as the
chiral-symmetry breaking scale, does not appear to be satisfied (not even at the
smallest $q$ values).

\section{Conclusions}
\label{sec:con}
In the first part of this study (Sec.~\ref{sec:c-cnt} and Appendices~\ref{app:a1} and~\ref{app:a2}),
we have clarified the origin of some of the differences in the N3LO and N4LO
corrections to the current and charge operators, reported in Ref.~\cite{Pastore09,Pastore11} and
in Ref.~\cite{Koelling11}.  In contrast to the authors of Ref.~\cite{Koelling11}, we have not yet
provided a complete derivation of the contributions associated with loop corrections to tree-level
(OPE) current and charge operators (although some were discussed in Ref.~\cite{Pastore09}); in
particular, we have not carried out a full-fledged renormalization of these operators in our
formalism.  However, as pointed out in Sec.~\ref{sec:jsloop}, the renormalized OPE current in
Eq.~(4.28) of Ref.~\cite{Koelling11} leads to the same magnetic moment operator obtained from the
currents in Eqs.~(\ref{eq:nlo1}), (\ref{eq:cdlt}), and~(\ref{eq:cp}) of the present work, with the
understanding, of course, that the LEC's entering these equations are assumed
to have been renormalized.  There remain differences
in the pion-loop corrections to the short range charge operator,
Eq.~(5.5) of Ref.~\cite{Koelling11} and Eqs.~(\ref{eq:c_i})--(\ref{eq:c_j}), the latter
presumably due to the different ways in which non-iterative pieces of reducible
contributions are isolated in the two formalisms.  The authors of Refs.~\cite{Koelling09,Koelling11}
use TOPT in combination with the unitary transformation method~\cite{Okubo54}
to decouple, in the Hilbert space of pions and nucleons, the states consisting of nucleons
only from those including, in addition, pions.  In contrast, we construct a potential
such that, when iterated in the Lippmann-Schwinger equation, leads to a $T$-matrix matching,
order by order in the power counting, the $\chi$EFT amplitude calculated in TOPT~\cite{Pastore11,Pastore08}.

In the second part of this study, we have provided predictions for the
static properties, including charge and magnetic radii and magnetic moments, and elastic
form factors of the deuteron and trinucleons.  The wave functions describing these nuclei
were derived from either $\chi$EFT or conventional two- and three-nucleon potentials.
The matrix elements of the $\chi$EFT charge and current operators were evaluated
in momentum-space with Monte Carlo methods.

The $\chi$EFT calculations (based on the N3LO potential) and
the hybrid ones (based on the AV18) reproduce very well the
observed electromagnetic structure of the deuteron for momentum
transfers $q$ up to 2--3 fm$^{-1}$.  In some cases, as in the $A(q)$
structure function, the agreement between the experimental and
$\chi$EFT calculated values extends up to $q\lesssim 6$ fm$^{-1}$, a much
higher momentum transfer than one would naively expect the present
expansion to be valid for.  On the other hand, the measured $B(q)$ structure function
is significantly under-predicted (over-predicted) for $q \gtrsim 3$ fm$^{-1}$
in the $\chi$EFT (hybrid) calculations.  The $\chi$EFT results, in contrast
to the hybrid ones, have a rather large cutoff dependence.  This cutoff
dependence originates, in the hybrid calculations, solely from that in the
N3LO current, while in the $\chi$EFT calculation it also reflects
the $\Lambda$ dependence intrinsic to the potential (the N3LO for
$\Lambda=500$ MeV or N3LO$^*$ for $\Lambda=600$ MeV).

The calculated $^3$He and $^3$H charge form factors are in excellent
agreement with data up to $q \lesssim 3$ fm$^{-1}$.  However, the
observed positions of the zeros are not generally well reproduced
by theory, and the measured $^3$He ($^3$H) form factor in the region
of the secondary maximum at $q \simeq 4$ fm$^{-1}$ is underestimated
(overestimated) in both $\chi$EFT and hybrid calculations.  A glance
at the $F_C^S(q)$ and $F_C^V(q)$ in Fig.~\ref{fig:f15}
suggests that two-body isovector contributions to the charge operator should
be considerably larger (in magnitude) than presently calculated, in order
to shift the zero in $F_C^V(q)$ to smaller $q$.

The isovector currents at N3LO depend on two LEC's ($d_1^V$ and $d_2^V$),
which have been fixed in one of three different ways: by reproducing
the experimental $np$ radiative capture cross section $\sigma_{np}$ and
isovector magnetic moment $\mu_V$ of the trinucleons simultaneously (set I);
by using $\Delta$ dominance to constrain $d_2^V$ and by determining
$d_1^V$ so as to fit either $\sigma_{np}$ (set II) or $\mu_V$ (set III).
Set I is not seriously considered for the reasons explained in Sec.~\ref{sec:lecs}.
The $^3$He and $^3$H magnetic form factors calculated with N3LO currents
corresponding to set III, while in excellent agreement with data for
$q \lesssim 3$ fm$^{-1}$, under-predict them at higher momentum transfers.
On the other hand, set II N3LO currents in the $\chi$EFT
calculations (based on the N3LO/N2LO and N3LO$^*$/N2LO$^*$ Hamiltonians)
would lead to significantly better agreement with data over the whole
range of momentum transfers (see Fig.~\ref{fig:f18}),
but would overestimate the observed $\mu_V$ by $\simeq 3$\%.

\section*{Acknowledgments}
R.S. would like to thank the T-2 group in the Theoretical Division at LANL,
and especially J.\ Carlson and S.\ Gandolfi, for the support and warm hospitality
extended to him during a sabbatical visit in the Fall 2012, during which
part of this work was completed. 
The work of R.S. is supported by the U.S.~Department of Energy, Office of
Nuclear Physics, under contract DE-AC05-06OR23177.  The calculations were
made possible by grants of computing time from the National Energy Research
Scientific Computing Center.
%
%
%
%
%
\appendix
\section{Minimal contact currents}
\label{app:a1}
In this appendix, we show the equivalence between the minimal contact
current in Eq.~(3.11) of Ref.~\cite{Pastore09} and that given in
Eq.~(\ref{eq:jmin}) in terms of the known low-energy constants (LEC's)
$C_1, \dots,C_7$.  One way to achieve this is to start from the Lagrangian
given in Eq.~(2.13) of Ref.~\cite{Girlanda10} (additional terms with fixed
coefficients proportional to $1/m_N^2$ have been ignored)
\begin{eqnarray}
{\cal L} &=& -\frac{1}{2} C_S\, O_S  - \frac{1}{2} C_T \, O_T
  - \frac{1}{2} C_1 ( O_1 + 2\, O_2) \nonumber\\
&&+ \frac{1}{8} C_2 (2\, O_2
+ O_3 ) - \frac{1}{2} C_3 ( O_9 + 2\, O_{12}) \nonumber\\
&&- \frac{1}{8} C_4 (O_9 +
O_{14})+ \frac{1}{4} C_5 ( O_6 - O_5)- \frac{1}{2} C_6 (O_7 \nonumber\\
&&+ 2 \,
O_{10})
 -\frac{1}{16} C_7 (O_7 + O_8 + 2\, O_{13})\ ,
\end{eqnarray}
where the operators $O_i$ are the standard set in Table I of Ref.~\cite{Girlanda10},
and then to gauge the gradients as ${\bm \nabla}N\rightarrow {\bm \nabla}N-i\,e\, e_N\, {\bf A}\, N $
to obtain
\begin{eqnarray}
{\bf j}^{(1)}_{\rm a, min}&=&\frac{C_2}{4}\left(\tau_{1,z}-\tau_{2,z}\right)\, \left({\bf K}_1-{\bf K}_2\right)\nonumber\\
&&\!\!\!+\frac{C_4}{4}\left(\tau_{1,z}-\tau_{2,z}\right)\,{\bm \sigma}_1\cdot
{\bm \sigma}_2\,  \left({\bf K}_1-{\bf K}_2\right) \nonumber\\
&&\!\!\!-\frac{i\, C_5}{4}\, ({\bm \sigma}_1+{\bm \sigma}_2)
\times (e_1\,{\bf k}_1+e_2\, {\bf k}_2) \nonumber \\
&&\!\!\!+\frac{C_7}{8}\left(\tau_{1,z}-\tau_{2,z}\right)\,
\big[{\bm \sigma}_1\cdot ({\bf K}_1-{\bf K}_2)\, {\bm \sigma}_2\nonumber\\
&&+
 {\bm \sigma}_2\cdot ({\bf K}_1-{\bf K}_2)\, {\bm \sigma}_1\big]  \ .
 \label{eq:eq16}
\end{eqnarray}
Because of the antisymmetry of two-nucleon states, we have
\begin{equation}
{\bf j}^{(1)}_{\rm a,min} = - P^\tau\, P^\sigma\, P^{\rm space}\, {\bf j}^{(1)}_{\rm a,min}
\end{equation}
where $P^{\rm space}$, $P^\sigma$, and $P^\tau$ are, respectively,
the space, spin and isospin exchange operators.  Making use of the
identities:
\begin{eqnarray}
&&P^{\,{\rm space}}({\bf K}_1-{\bf K}_2) =-({\bf k}_1-{\bf k}_2)/2 \ , \\
&&P^{\, \sigma} {\bm \sigma}_1\cdot{\bm \sigma}_2 =
(3-{\bm \sigma}_1\cdot{\bm \sigma}_2)/2 \ ,\\
&&P^{\, \tau} (\tau_{1,z}-\tau_{2,z}) =i\, ({\bm \tau}_1 \times {\bm \tau}_2)_z \ ,
\end{eqnarray}
\begin{widetext}
 \begin{equation}
P^{\,\sigma}P^{\,{\rm space}} \Big[{\bm \sigma}_1\cdot ({\bf K}_1-{\bf K}_2)\, {\bm \sigma}_2+
 {\bm \sigma}_2\cdot ({\bf K}_1-{\bf K}_2)\, {\bm \sigma}_1 \Big] =
-\frac{1}{2}\Big[ ({\bf k}_1-{\bf k}_2)(1-{\bm \sigma}_1\cdot{\bm \sigma}_2)
+{\bm \sigma}_1\cdot ({\bf k}_1-{\bf k}_2)\, {\bm \sigma}_2
+ {\bm \sigma}_2\cdot ({\bf k}_1-{\bf k}_2)\, {\bm \sigma}_1\Big] \ ,
\end{equation}
\end{widetext}
Eq.~(\ref{eq:jmin}) in the text follows.

An alternative way to proceed is to express the original set of LEC's $C_1^\prime,\dots,C_{14}^\prime$
entering Eq.~(3.11) of Ref.~\cite{Pastore09}
in terms of the twelve independent LEC's $C_1,\dots, C_7,C_1^*, \dots, C_5^*$
(in the notation of Ref.~\cite{Pastore09}), and then to set the $C_i^*=0$, that
is, to ignore the currents induced by these terms, since they are suppressed by $1/m_N^2$.
Substituting
\begin{eqnarray}
&& C_1' = \frac{1}{2} C_1 \ ,\,\,\,
C_2' = C_1 -\frac{1}{4} C_2\ , \,\,\,
C_3' = -\frac{1}{8} C_2\ , \nonumber\\
&&C_4' +C_6' = -\frac{1}{4}C_5 \ ,\,\,\,  C_5'+ C_6'= 0\ , \nonumber\\
&&C_7' +\frac{1}{2} C_{11}'=\frac{1}{2} C_6 + \frac{1}{16} C_7 \ , \,\,\,
C_8' -\frac{1}{2} C_{11}'=\frac{1}{16} C_7\ ,\nonumber \\
&&C_9 '=\frac{1}{2} C_3 + \frac{1}{8} C_4 \ ,\,\,\,
C_{10}' + C_{11}'= C_6 \  , \,\,\,
C_{12}' =  C_3 \ ,\nonumber\\
&&C_{13}' = \frac{1}{8} C_7 \ , \,\,\,
C_{14}' = \frac{1}{8} C_4 \ ,
\label{eq:eq34}
\end{eqnarray}
into Eq.~(3.11) of Ref.~\cite{Pastore09}, we find
\begin{eqnarray}
{\rm Eq.}~(3.11)\,\,{\rm of}\,\, {\rm Ref.}~[1]\,&=& {\rm Eq.}~({\rm A}2)-
i\,e\, C_4^\prime\, (e_1+e_2)\nonumber\\
&&\times ({\bm \sigma}_1+{\bm \sigma}_2)\times {\bf q} \ ,
\end{eqnarray}
and the difference can be absorbed into a redefinition of $C_{15}^\prime$, 
since 
$(\tau_1^z + \tau_2^z) ({\bm\sigma}_1 + {\bm \sigma}_2) \times {\bf q}=0$ 
after antisymmetrization.
Notice also that, in view of the identity $O_4+O_5=O_6+O_{15}$ (which was
derived in Ref.~\cite{Girlanda10}, apart from  $O_{15}$, 
of no relevance there), among the
operators of the sub-leading contact Lagrangian, the operator $O_6$ is
redundant, and indeed the dependence on the associated LEC $C_6^\prime$ 
cancels in the observables.
\section{One-loop short-range current and charge operators}
\label{app:a2}

\begin{figure}
\caption{Time-ordered diagrams illustrating one of the classes
of loop corrections to the single-nucleon current.
Nucleons, pions and photons are denoted by solid, dashed,
and wavy lines, respectively.}
\includegraphics[width=3in]{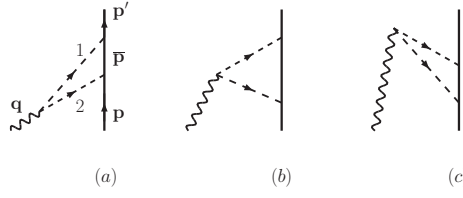}\\
\label{fig:fa1}
\end{figure}

In this appendix, we discuss the contributions associated with panels (h)-(k) in
Fig.~\ref{fig:f2} for the current operator, and (g)-(j) in Fig.~\ref{fig:f3}
for the charge operator.  We begin with the current
operator.  The contributions of diagrams (h) and (j)
in Fig.~\ref{fig:f2} vanish,
while the contribution of diagrams of type (i) was obtained as
(conventions for ${\bf q}$-integrations and $\overline{\delta}$-functions
are the same as in Ref.~\cite{Pastore09})
\begin{eqnarray}
{\bf j}^{(1)}_{\rm i}&=&2\, i\, \frac{e\, g_A^2\, C_T}{F_\pi^2}\, ({\bm \tau}_1\times{\bm \tau}_2)_z
\int_{{\bf q}_1,{\bf q}_2} \overline{\delta}({\bf q}_1+{\bf q}_2-{\bf q})\, \nonumber\\
&&\times\overline{D}(\omega_1,\omega_2)\,
 ({\bf q}_1-{\bf q}_2)\, {\bm \sigma}_1 \cdot {\bf q}_2\, {\bm \sigma}_2 \cdot {\bf q}_1 \ ,
\label{eq:cg}
\end{eqnarray}
where 
\begin{equation}
\overline{D}(\omega_1,\omega_2)=\frac{\omega_1^2+\omega_1\omega_2+\omega_2^2}
{\omega_1^3\omega_2^3(\omega_1+\omega_2)} \ .
\end{equation}
Before analyzing diagram (k), we need to consider the leading and
next-to-leading contributions to the single-nucleon diagrams
shown in Fig.~\ref{fig:fa1}.
For simplicity, we define the vertices
\begin{equation}
\label{eq:vs}
V_1=i \frac{g_A}{F_\pi} \, {\bm \sigma}  \cdot {\bf q}_1\, \tau_a\ ,
V_2=i \frac{g_A}{F_\pi} \, {\bm \sigma}  \cdot {\bf q}_2\, \tau_b\ ,
\end{equation}
\vspace{-0.5cm}
\begin{equation}
V_\gamma =-i\,e\,\epsilon_{abz} \,({\bf q}_1-{\bf q}_2) \ .
\end{equation}
Then the current reads
\begin{eqnarray}
{\bf j}_{\gamma\pi\pi}&=& \frac{V_1 V_2 V_\gamma}{4\, \omega_1\omega_2}
\Bigg[\,\frac{1}{E_i-\omega_1-\overline{E}}\, \frac{1}{E_i-\omega_1-\omega_2-E} \nonumber \\
&&+\frac{1}{E_i-\omega_1-\overline{E}}\, \frac{1}{E_i-\omega_\gamma-\omega_2-\overline{E}} \nonumber\\
&&+\frac{1}{E_i-\omega_\gamma-\omega_1-\omega_2-E^\prime}\, 
\frac{1}{E_i-\omega_\gamma-\omega_2-\overline{E}}\,\, \Bigg] ,\nonumber \\
&&
\end{eqnarray}
where $E_i=E+\omega_\gamma$, and $\overline{E}$ is the energy of the intermediate nucleon of momentum
$\overline{{\bf p}}$.  After expanding the energy denominators as in Eq.~(\ref{eq:deno}) to include
linear terms in the nucleon kinetic energies, we find, up to next-to-leading order included,
\begin{eqnarray}
\!\!\!\!{\bf j}_{\gamma\pi\pi}&=&\frac{V_1 V_2 V_\gamma}{4}\bigg[\frac{2}{\omega_1^2 \omega_2^2}
+D^\prime(\omega_1,\omega_2)\, (E^\prime-\overline{E})\nonumber\\
\!\!\! &&+D(\omega_1,\omega_2) (E-\overline{E})
+D_\gamma(\omega_1,\omega_2)\,\omega_\gamma \bigg] \ ,
\end{eqnarray}
where
\begin{eqnarray}
\label{eq:ds}
D(\omega_1,\omega_2)&=& \frac{2\, \omega_1+\omega_2}{\omega_1^2\omega_2^3(\omega_1+\omega_2)} \ ,\\
D^\prime(\omega_1,\omega_2)&=& \frac{\omega_1+2\, \omega_2}{\omega_1^3\omega_2^2(\omega_1+\omega_2)}\ , \\
D_\gamma(\omega_1,\omega_2)&=&-\frac{\omega_1-\omega_2}{\omega_1^2 \omega_2^2\,(\omega_1+\omega_2)^2} \ ,
\end{eqnarray}
and
\begin{equation}
D \rightleftharpoons D^\prime \,\,\, {\rm with}\,\,
{\bf q}_1 \rightleftharpoons {\bf q}_2\ , \qquad  D+D^\prime = 2\, \overline{D} \ .
\end{equation}

\begin{figure}
\caption{Set of time-ordered diagrams for the contribution
illustrated by the single diagram (k) in Fig.~\ref{fig:f2}.
Notation as in Fig.~\protect{\ref{fig:fa1}}.}
\includegraphics[width=3.35in]{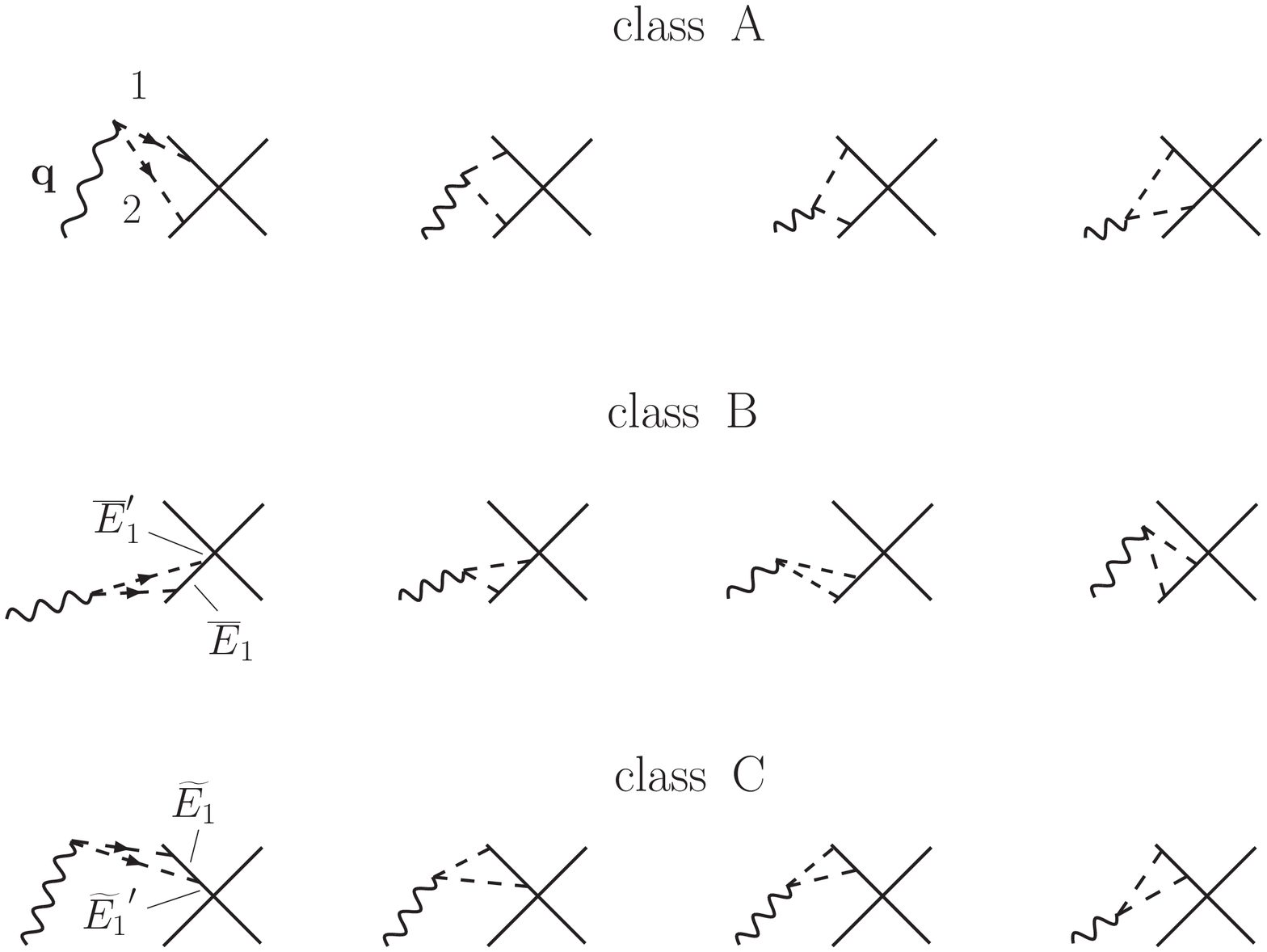}\\
\label{fig:fa2}
\end{figure}

We now proceed to analyze the contributions of diagrams of type (k) in Fig.~\ref{fig:f2}.
To this end, we show in Fig.~\ref{fig:fa2} the complete set of time-ordered
diagrams of the same topology as (k), which we have separated for convenience
into the three classes A, B, and C. Class A consists only of irreducible diagrams,
which at order $n=1$ or $e\, Q$, i.e., in the static limit, lead to 
\begin{eqnarray}
\!\!\!{\rm class \,\, A}\!\!&=&\!\!-\frac{1}{2}\, V_\gamma V_1 V_{\rm CT} V_2 \overline{D}(\omega_1,\omega_2) \nonumber\\
\!\!\!\!&=&\!\!-\frac{1}{4}V_\gamma V_1 V_{\rm CT} V_2
\left[ D(\omega_1,\omega_2) +D^\prime(\omega_1,\omega_2) \right] \ ,\nonumber\\
&&
\end{eqnarray}
where the vertices $V_1$, $V_2$, and $V_\gamma$ are defined as above (with
the spin and isospin matrices now referring to nucleon 1), and
\begin{equation}
\label{eq:vsct}
 V_{\rm CT}= C_S+C_T\, {\bm \sigma}_1\cdot {\bm \sigma}_2 \ .
\end{equation}
On the other hand, to order $e\, Q$ included, class B gives
\begin{widetext}
\begin{eqnarray}
{\rm class \,\, B}&=&\!\!\frac{ V_\gamma V_{\rm CT}V_1 V_2}{4\, \omega_1\omega_2}
\frac{1}{E_i-\overline{E}_1^{\, \prime}-E_2}\Bigg[ \frac{1}{\omega_1(\omega_1+\omega_2)}
\left(1+\frac{E_1+\omega_\gamma-\overline{E}_1}{\omega_1}+\frac{\omega_\gamma}{\omega_1+\omega_2}\right) 
+\frac{1}{\omega_1\omega_2}\bigg(1+\frac{E_1+\omega_\gamma-\overline{E}_1}{\omega_1}\nonumber \\
&&\!\!+\frac{E_1-\overline{E}_1}{\omega_2}\bigg) 
  +\frac{1}{\omega_2(\omega_1+\omega_2)}
\left(1+\frac{E_i-\overline{E}_1^{\, \prime}-E_2}{\omega_1+\omega_2}-\frac{\omega_\gamma}{\omega_1+\omega_2}+
\frac{E_1-\overline{E}_1}{\omega_2}\right)\Bigg]
-\frac{ V_\gamma V_{\rm CT}V_1 V_2}{4\, \omega_1\omega_2} \frac{1}{\omega_2(\omega_1+\omega_2)^2}\nonumber\\
&&
\end{eqnarray}
\end{widetext}
where we have used energy conservation between the initial and final states
$E_1+E_2+\omega_\gamma=E_1^\prime+E_2^\prime$.  We now note that the
irreducible contribution from the last diagram (in class B) is cancelled by
the second term in the next to last line of the above equation, so that we are
left with
\begin{widetext}
\begin{eqnarray}
{\rm class \,\, B}&=&\frac{ V_\gamma V_{\rm CT}V_1 V_2}{4}\frac{1}{E_i-\overline{E}_1^{\, \prime}-E_2}
\Bigg[ \frac{2}{\omega^2_1\omega^2_2}
+\omega_\gamma \, D_\gamma(\omega_1,\omega_2)
+(E_1-\overline{E}_1)\, D(\omega_1,\omega_2) 
+(E_1+\omega_\gamma-\overline{E}_1)\, D^\prime(\omega_1,\omega_2)  \Bigg] \nonumber \\
&=& V_{\rm CT}\, \frac{1}{E_i-\overline{E}_1^{\, \prime}-E_2}\,\, {\bf j}_{\gamma \pi\pi}
+\frac{V_\gamma V_{\rm CT}V_1 V_2}{4}\, D^\prime(\omega_1,\omega_2) \ ,
\end{eqnarray}
\end{widetext}
since $E_1+\omega_\gamma-\overline{E}_1=(E_i-\overline{E}_1^{\,\prime}-E_2) +
(\overline{E}_1^{\,\prime}-\overline{E}_1)$ and 
$V_\gamma$ commutes with each of the remaining vertices.
The first term represents an iteration, while the recoil-corrected class B contribution
is simply given by $V_\gamma V_{\rm CT}V_1 V_2\, D^\prime(\omega_1,\omega_2)/4$.  A similar
analysis for class C leads to the recoil-corrected class C contribution
given by $V_\gamma V_1 V_2 V_{\rm CT}\, D(\omega_1,\omega_2)/4$.  Therefore
combining the contributions from classes A, B, and C, we find
\begin{eqnarray}
{\bf j}^{(1)}_{\rm k}&=&\frac{1}{4} V_\gamma V_1 \,\left[ V_2\, ,\, V_{\rm CT}\right] \, D(\omega_1,\omega_2)\nonumber\\
&&+\frac{1}{4} \left[ V_{\rm CT}\, , \, V_1 \right]\, V_2\, V_\gamma \, D^\prime(\omega_1,\omega_2) \ ,
\end{eqnarray}
or explicitly
\begin{eqnarray}
{\bf j}^{(1)}_{\rm k}&=&2\, i\, \frac{e\, g_A^2\, C_T}{F_\pi^2}\, \tau_{1z}
\int_{{\bf q}_1,{\bf q}_2} \overline{\delta}({\bf q}_1+{\bf q}_2-{\bf q})\, \overline{D}(\omega_1,\omega_2)\nonumber\\
&&\times ({\bf q}_1-{\bf q}_2)\, {\bm \sigma}_2 \cdot {\bf q}_2\times{\bf q}_1
+ 1\rightleftharpoons 2 \ .
\end{eqnarray}
The currents ${\bf j}^{(1)}_{\rm i}$ and ${\bf j}^{(1)}_{\rm k}$
obtained here are in agreement with those in Eq.~(5.2) of~\cite{Koelling11}, but for an overall
factor of 2.  Ultimately, this difference has no impact, since 
\begin{eqnarray}
{\bf j}^{(1)}_{\rm i}+{\bf j}^{1)}_{\rm k}\!\! &\propto&\!\! ({\bm \tau}_1\times{\bm \tau}_2)_z\,
({\bm \sigma}_1\times{\bm \sigma}_2)\times{\bf q}\nonumber\\
\!\!&&\!\!+ \left( 2\, \tau_{1z}\, {\bm \sigma}_2+
2\, \tau_{2z}\, {\bm \sigma}_1\right) \times {\bf q} = 0 \ ,
\end{eqnarray}
which vanishes due to antisymmetry of the two-nucleon states.

In Ref.~\cite{Pastore09} we had not considered the next-to-leading order
contributions to the single-nucleon $\gamma \pi\pi$ vertex when deriving
the one-loop correction to the OPE current shown in Fig.~\ref{fig:fa3}
(see Appendix~E of Ref.~\cite{Pastore09}).  As a consequence we had
failed to isolate the correct non-iterative piece, which had led, in particular,
to a non-hermitian operator.
\begin{figure}
\caption{One-loop correction to the OPE current (only one among
the possible time-orderings is shown).
Notation as in Fig.~\protect{\ref{fig:fa1}}.}
\includegraphics[width=2.5in]{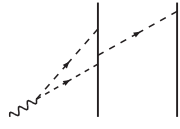}\\
\label{fig:fa3}
\end{figure}
We find that this term is now given by
\begin{widetext}
\begin{eqnarray}
{\bf{j}}^{(1)}_{\rm OPE, loop}&=&i\, e\frac{g_A^4}{F_\pi^4}\, ({\bm \tau}_1\times{\bm \tau}_2)_z
\frac{{\bm \sigma}_2\cdot{\bf k}_2}{\omega^2_{k_2}}\int_{\bf p} {\bf p}\,
\bigg[2\,\frac{\omega_+^2+\omega_-^2+\omega_+\omega_-}{\omega_+^3\omega_-^3(\omega_++\omega_-)}
\Big( {\bm \sigma}_1\cdot{\bf q}\,{\bf k}_2\cdot{\bf p}
-{\bm \sigma}_1\cdot{\bf p}\,{\bf k}_2\cdot{\bf q}\Big)\nonumber\\
&&-\frac{\omega_+-\omega_-}{\omega_+^3\omega_-^3}\,{\bm \sigma}_1\cdot{\bf k}_2\,
(q^2-p^2)\bigg] + 1 \rightleftharpoons 2 \ ,
\end{eqnarray}
\end{widetext}
where $\omega_{\pm}=\sqrt{({\bf q}\pm {\bf p})^2+4\, m_\pi^2}$.

Next, we turn our attention to the charge operator.  In Ref.~\cite{Pastore11}
we showed that the contributions of diagrams (g)-(h) in Fig.~\ref{fig:f3} vanish.
However, in light of the previous considerations, those due to diagrams (i) and (j)
given there need to be revised.  Indeed, an analysis similar to that carried out
above leads to the single-nucleon charge operator (see Fig.~\ref{fig:fa1}) up to
next-to-leading order included
\begin{eqnarray}
\rho_{\gamma\pi\pi}&=&\frac{V_1 V_2 \widetilde{V}_\gamma}{4}
\bigg[\frac{4}{\omega_1\,\omega_2\,(\omega_1+ \omega_2)}
+\widetilde{D}^\prime(\omega_1,\omega_2)\, (E^\prime-\overline{E}) \nonumber\\
&&+\widetilde{D}(\omega_1,\omega_2) (E-\overline{E})
+\widetilde{D}_\gamma(\omega_1,\omega_2)\,\omega_\gamma \bigg] \ ,
\end{eqnarray}
and to a contribution for diagram (j) (see Fig.~\ref{fig:fa2}) which reads
\begin{eqnarray}
\rho^{(1)}_{\rm j}&=&-2\, \frac{e\, g_A^2\, C_T}{F_\pi^2}\, \tau_{1z}
\int_{{\bf q}_1,{\bf q}_2} \overline{\delta}({\bf q}_1+{\bf q}_2-{\bf q})\,
\widetilde{D}(\omega_1,\omega_2)\nonumber\\
&&\times{\bm \sigma}_1 \cdot \left[ 
{\bf q}_1 \times\left( {\bm \sigma}_2\times{\bf q}_2\right) \right]
+ 1\rightleftharpoons 2 \ .
\end{eqnarray}
We have defined $\widetilde{V}_{\gamma}=-i\,e\,\epsilon_{abz}\,$, and
\begin{eqnarray}
\label{eq:dsc}
\widetilde{D}(\omega_1,\omega_2)&=& \frac{3\, \omega_1+\omega_2}{\omega_1^2\, \omega_2^2\,
(\omega_1+\omega_2)} \ ,\\
\widetilde{D}^\prime(\omega_1,\omega_2)&=& \frac{\omega_1+3\, \omega_2}{\omega_1^2\,
\omega_2^2\, (\omega_1+\omega_2)}\ , \\
\widetilde{D}_\gamma(\omega_1,\omega_2)&=&\frac{\omega_1-\omega_2}{\omega_1^2\,
 \omega_2^2\,(\omega_1+\omega_2)^2} \ ,
\end{eqnarray}
with
\begin{equation}
\widetilde{D} \rightleftharpoons \widetilde{D}^\prime \,\,\, {\rm with}\,\,
{\bf q}_1 \rightleftharpoons {\bf q}_2\ ,  \qquad \widetilde{D}+\widetilde{D}^\prime = \frac{4}{\omega_1^2\,
\omega_2^2} \ .
\end{equation}
\begin{figure}
\caption{Set of time-ordered diagrams for the contribution
illustrated by the single diagram (i) in Fig.~\ref{fig:f3}.
Notation as in Fig.~\protect{\ref{fig:fa1}}.}
\includegraphics[width=3.35in]{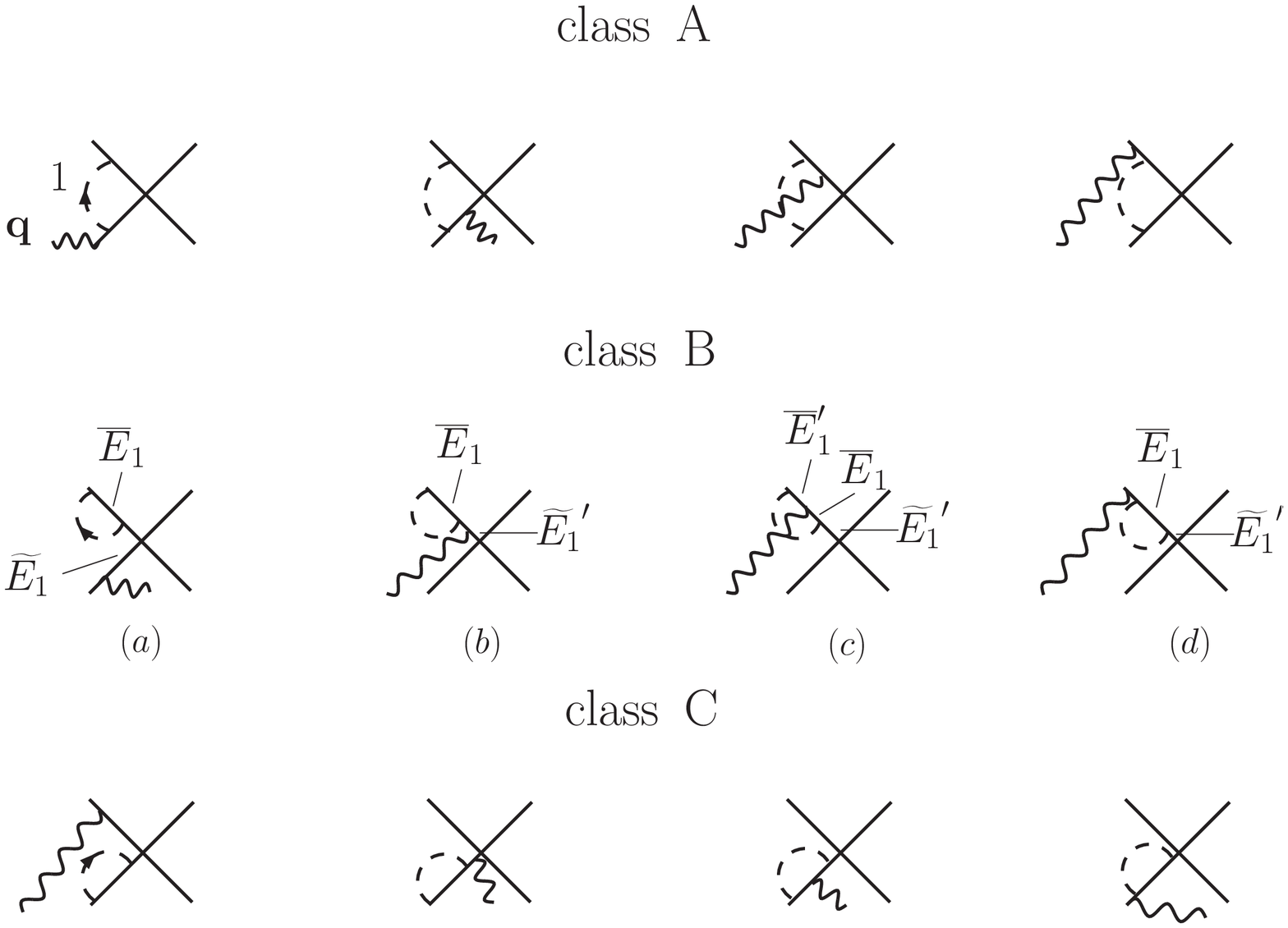}\\
\label{fig:fa4}
\end{figure} 
We now revise the derivation of the charge operator of type (i) illustrated in Fig.~\ref{fig:f3}.
The associated time ordered diagrams are represented in Fig.~\ref{fig:fa4}, and have been
separated into three classes. In Ref.~\cite{Pastore11}, the expression reported in Eq.~(53)
has been obtained by accounting for the recoil corrected class A diagrams only, i.e.
\begin{eqnarray}
\label{eq:clA}
 {\rm class \,\, A} &=& \frac{1}{2\,\omega_1^4}\Big[ V_1 V_{\rm CT} V_1^\prime V_\gamma^\prime
+ V_\gamma^\prime V_1 V_{\rm CT} V_1^\prime  \nonumber\\
&&- V_1 V_{\rm CT} V_\gamma^\prime V_1^\prime
-V_1 V_\gamma^\prime  V_{\rm CT} V_1^\prime\Big] \nonumber\\ 
&=& e\,\frac{2\, g_A^2}{3\, F_\pi^2}\, \tau_{1,z}\left(3 \,C_S - C_T \, {\bm \sigma}_1\cdot{\bm \sigma}_2\right)
\int_{{\bf q}_1} \frac{q_1^2}{\omega_1^4} \ ,\nonumber\\
&&
\end{eqnarray}
where $V_1$ and $V_{\rm CT}$ are given in Eqs.~(\ref{eq:vs}) and~(\ref{eq:vsct}),
respectively, and $V_1^\prime$ and $V_\gamma^\prime$ are defined as
\begin{equation}
 V_1^\prime = - V_1 \ ,\qquad V_\gamma^\prime = \frac{e}{2}\left(1+\tau_{1,z}\right) \ .
\end{equation}

\begin{figure}
\caption{Diagrams for the single-nucleon contributions.
Notation as in Fig.~\protect{\ref{fig:fa1}}.}
\includegraphics[width=2.5in]{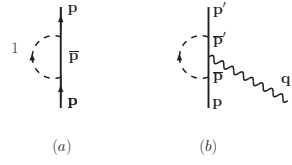}\\
\label{fig:fa5}
\end{figure} 
Classes B and C involve only reducible diagrams.  First, we analyze the single-nucleon
contributions entering the reducible diagrams, represented in Fig.~\ref{fig:fa5}.
We account for leading, next-to-leading, and next-to-next-to-leading order
corrections in the expansion of the energy denominators.  The contributions
$M_N$ and $\rho_\gamma$, associated with panels (a) and (b), respectively, read
\begin{eqnarray}
\!\!\!\!M_N \!\!&=&\!\! - \frac{V_1 \,V_1^\prime}{ 2\, \omega_1^2}\left[1+\frac{E-\overline{E}}{\omega_1}
+\frac{(E-\overline{E})^2}{\omega_1^2}\right]\ , \\
\!\!\!\!\rho_\gamma\!\! &=&\!\! \frac{V_1 \,V_\gamma^\prime \, V_1^\prime}{ 2 \,\omega_1^3}
\left( 1+\frac{E^\prime-\overline{E}^\prime}{\omega_1} + \frac{E-\overline{E}}{\omega_1}\right)\ ,
\end{eqnarray}
where the energies are as indicated in Fig.~\ref{fig:fa5}, and energy conservation
\hbox{($E+\omega_\gamma=E^\prime$)} has been used.  In terms of these,
the contributions of diagram (a) and (b) in class B are given by
\begin{eqnarray}
\!\!\!\!B_{\rm a}\!\! &=&\!\!  M_N \, \frac{1}{E_i-E_1^\prime-E_2^\prime}\, V_{\rm CT}\, 
\frac{1}{E_i-\widetilde{E}_1-E_2}\, V_\gamma^\prime \ ,\\
\!\!\!\!B_{\rm b}\!\!&=&\!\! M_N\, \frac{1}{ E_i-E_1^\prime-E_2^\prime }\, V_\gamma^\prime\,
\frac{1}{ E_i-\widetilde{E}_1^\prime-E^\prime_2 -\omega_\gamma}\, V_{\rm CT} \ ,\nonumber\\
&&
\end{eqnarray}
where we have identified and isolated the nucleon self-energy terms (to be
reabsorbed by mass counter-terms).
Evaluation of panel (c) leads to
\begin{eqnarray}
B_{\rm c} &=&
\frac{V_1 \,V_\gamma^\prime \, V_1^\prime}{ 2\, \omega_1^3}
\left[ 1+\frac{E_1^\prime-\overline{E}_1^\prime }{\omega_1}
 + \frac{\widetilde{E}_1^\prime-\overline{E}_1}{\omega_1}\right]\nonumber\\
&&\times\frac{1}{E_i-\widetilde{E}_1^\prime-E_2^\prime-\omega_\gamma}V_{\rm CT}
+ \frac{V_1 \,V_\gamma^\prime \, V_1^\prime V_{\rm CT} }{ 2\, \omega_1^4}\nonumber \\
&=&\rho_\gamma\, \frac{1}{E_i-\widetilde{E}_1^\prime-E_2^\prime-\omega_\gamma}\, V_{\rm CT}
+\frac{V_1 \,V_\gamma^\prime \, V_1^\prime V_{\rm CT}}{ 2\, \omega_1^4}\ ,\nonumber\\
&&
\end{eqnarray}
where the first term represents an iteration with $\rho_\gamma$ and the
contact interaction, and the second term is the recoil correction contributing
to the two-nucleon charge operator.  Finally, the contribution of panel (d),
in which, in contrast to panels (a) and (b), the self-energy insertion is between
the photon absorption and contact interaction, is expressed as
\begin{widetext}
\begin{eqnarray}
B_{\rm d} &=& - V_\gamma^\prime\,
 \frac{1}{E_i-\widetilde{E}_1^\prime-E_2^\prime-\omega_\gamma}\,
\frac{V_1 \,V_1^\prime}{ 2\, \omega_1^2}\left[1+\frac{\widetilde{E}_1^\prime-\overline{E}_1}{\omega_1}
+\frac{(\widetilde{E}_1^\prime-\overline{E}_1)^2}{\omega_1^2}\right]
\frac{1}{E_i-\widetilde{E}_1^\prime-E_2^\prime-\omega_\gamma} \, V_{\rm CT}\nonumber\\
&-& \frac{V_\gamma^\prime V_1 \,V_1^\prime}{ 2\, \omega_1^3}\left( 1 + 
2\, \frac{\widetilde{E}_1^\prime-\overline{E}_1}{\omega_1}\right)
\frac{1}{E_i-\widetilde{E}_1^\prime-E_2^\prime-\omega_\gamma} V_{\rm CT}
-\frac{V_\gamma^\prime V_1 \,V_1^\prime V_{\rm CT}}{ 2\, \omega_1^4} \nonumber\\
&=& V_\gamma^\prime\, \frac{1}{E_i-\widetilde{E}_1^\prime-E_2^\prime-\omega_\gamma} \,
M_N \,\frac{1}{E_i-\widetilde{E}_1^\prime-E_2^\prime-\omega_\gamma} \, V_{\rm CT}
+\, \widetilde{\rho}_\gamma\,  \frac{1}{E_i-\widetilde{E}_1^\prime-E_2^\prime-\omega_\gamma} \,V_{\rm CT}
-\frac{V_\gamma^\prime V_1 \,V_1^\prime V_{\rm CT}}{ 2\, \omega_1^4}  \ ,
\end{eqnarray}
\end{widetext}
where the last term is a two-nucleon term, and we interpret $\widetilde{\rho}_\gamma$ as a further correction
to the single-nucleon $\gamma N$ vertex.   However, the analysis and
proper interpretation of this type of corrections are beyond the scope of the
present work.

After a similar analysis of the class C diagrams is carried out, we find that the
complete B+C contribution reads
\begin{eqnarray}
 {\rm classes \,\, B + C }\!\! &=&\!\! \frac{1}{2\, \omega_1^4}\Big[
V_1 V_\gamma^\prime  \,V_1^\prime V_{\rm CT}-V_\gamma^\prime V_1   \,V_1^\prime V_{\rm CT}\nonumber\\
\!\!&&\!\!+V_{\rm CT} V_1  V_\gamma^\prime \,V_1^\prime - V_{\rm CT} V_1  V_1^\prime V_\gamma^\prime 
\Big]\nonumber\\
\!\!&=& \!\!-e\,\frac{2 \, g_A^2}{F_\pi^2}\, \tau_{1,z}\left( C_S + C_T \, {\bm \sigma}_1\cdot{\bm \sigma}_2\right)
\int_{{\bf q}_1} \frac{q_1^2}{\omega_1^4} \ ,\nonumber\\
&&
\end{eqnarray}
which combined with Eq.~(\ref{eq:clA}) leads to the type (i) charge operator in Fig.~\ref{fig:f3}
\begin{equation}
 \rho^{(1)}_i=-e\,\frac{8 \, g_A^2 \,C_T}{3\,F_\pi^2}\,\tau_{1,z}\, {\bm \sigma}_1\cdot{\bm \sigma}_2
\int_{{\bf q}_1} \frac{q_1^2}{\omega_1^4} + 1\rightleftharpoons 2\ .
\end{equation}

\section{Loop integrations}
\label{app:loops}

In this appendix, we outline the derivation of the two-body charge operators
at one loop, listed in Sec.~\ref{sec:cloop}.  For the sake of illustration,
we consider the contribution of panel (f) in Fig.~\ref{fig:f3}, given
by (in the notation of Ref.~\cite{Pastore11})
\begin{eqnarray}
\label{eq:cloop1}
 \rho_{\rm f}^{(1)}\!\!\!&=\!\!\!&-e\,\frac{2\,g_A^4}{F_{\pi}^4} 
 \int_{  {\bf q}_1,{\bf q}_2,{\bf q}_3  }\!\!\!
 \overline{\delta}({\bf q}_2+{\bf q}_3-{\bf k}_2)
 \overline{\delta}({\bf q}_1-{\bf q}_2-{\bf k}_1) \nonumber\\
&&\times
 \overline{\delta}({\bf q}_1+{\bf q}_3-{\bf q})\,
 \frac{1}{\omega_1^2\, \omega_2^2\, \omega_3^2} 
\, \Big [2\, \tau_{1,z}\, 
({\bf q}_2\cdot{\bf q}_1\, \, {\bf q}_2\cdot {\bf q}_3 \nonumber\\
&&-{\bm \sigma}_1\cdot {\bf q}_2\times {\bf q}_1
\,\, {\bm \sigma}_2\cdot{\bf q}_3\times{\bf q}_2) 
-\left({\bm \tau}_1\times{\bm \tau}_2\right)_z\, 
{\bf q}_1\cdot{\bf q}_2 \nonumber\\
&&\times {\bm \sigma}_2\cdot{\bf q}_3\times{\bf q}_2 \Big]+ 1 \rightleftharpoons 2 \ ,
\end{eqnarray}
which can conveniently be written as
\begin{equation}
\label{eq:cloop5}
 \rho_{\rm f}^{(1)}=-e\,\frac{2\,g_A^4} {F_{\pi}^4}
 \int_{\bf p} \frac{N({\bf q},{\bf k},{\bf p})}
 { \omega_{{\bf q}/2+{\bf p}}^2
  \, \omega_{{\bf q}/2-{\bf p}}^2
 \, \omega_{{\bf p}-{\bf k}}^2
 } + 1\rightleftharpoons 2\ ,
\end{equation}
with
\begin{widetext}
\begin{eqnarray}
\label{eq:num1}
 N({\bf q},{\bf k},{\bf p})&=&2\,\tau_{1,z}\big[\left({\bf p}-{\bf k}\right)\cdot
 \left({\bf q}/2+{\bf p}\right)\,\,
 \left({\bf p}-{\bf k}\right)\cdot
 \left({\bf q}/2-{\bf p}\right) 
-{\bm \sigma}_1\cdot\left({\bf p}-{\bf k}\right) \times\left({\bf q}/2+{\bf p}\right)\,\,
 {\bm \sigma}_2\cdot\left({\bf q}/2-{\bf p}\right)\times\left({\bf p}-{\bf k}\right)\big]\nonumber\\
&&-\left({\bm \tau}_1\times{\bm \tau}_2\right)_z \, 
\left({\bf q}/2+{\bf p}\right)\cdot\left({\bf p}-{\bf k}\right)
\,\, {\bm \sigma}_2\cdot\left({\bf q}/2-{\bf p}\right)\times\left({\bf p}-{\bf k}\right)\ ,
\end{eqnarray}
\end{widetext}
and the momentum ${\bf k}$ defined as in Eq.~(\ref{eq:ppp}).  We now use standard
techniques~\cite{Gross93} to express the product of energy denominators in the following way
\begin{eqnarray}
\label{eq:cloop6}
&& \frac{1}{ \omega_{{\bf q}/2+{\bf p}}^2
\, \omega_{{\bf q}/2-{\bf p}}^2
\, \omega_{{\bf p}-{\bf k}}^2 } 
=2\int_0^1 {\rm d}z_1 \int_0^{1-z_1} {\rm d}z_2\nonumber\\
&&\times \Big[\left[\left({\bf q}/2+{\bf p}\right)^2+m_{\pi}^2\right]z_1
+\left[\left({\bf q}/2-{\bf p}\right)^2+m_{\pi}^2\right] z_2\nonumber\\
&&+\left[\left({\bf p}-{\bf k}\right)^2+m_{\pi}^2\right]\, (1-z_1-z_2)\Big]^{-3}\ ,
\end{eqnarray}
which, in terms of
\begin{equation}
\label{eq:pppp}
 {\bf p}^\prime={\bf p}+\left(z_1-z_2 \right){\bf q}/2-\left(1-z_1-z_2\right){\bf k}\ ,
\end{equation}
simply reads
\begin{eqnarray}
&& \frac{1}{ \omega_{{\bf q}/2+{\bf p}}^2
\, \omega_{{\bf q}/2-{\bf p}}^2
\, \omega_{{\bf p}-{\bf k}}^2 } 
=2\int_0^1 {\rm d}z_1 \int_0^{1-z_1}  {\rm d}z_2 \nonumber\\
&&\times \left[\, p^{\,\prime\, 2}+\lambda^2(z_1,z_2)\,\right]^{-3}\ ,
\end{eqnarray}
where
\begin{eqnarray}
\label{eq:ldef}
 \lambda^2(z_1,z_2)\!\!&=&\!\!
 \left(z_1+z_2\right){\bf q}^2/4 -\big[ \left(z_1-z_2\right){\bf q}/2-(1-z_1\nonumber\\
&&-z_2)\,{\bf k}\, \big ]^2+\left(1-z_1-z_2\right){\bf k}^2 +m_{\pi}^2 \ .
\end{eqnarray}
After these manipulations, the charge operator can finally be written as
\begin{eqnarray}
 \rho_{\rm f}^{(1)}&=&-e\,\frac{4\,g_A^4} {F_{\pi}^4}
 \int_0^1 {\rm d}x\, x \int_{-1/2}^{1/2} \!\!\! {\rm d}y \int_{\bf p^{\, \prime}} 
N^\prime({\bf q},{\bf k},{\bf p}^{\, \prime})\nonumber\\
&&\times \left[\, p^{\,\prime\, 2}+\lambda^2(x,y)\,\right]^{-3}
 + 1\rightleftharpoons 2\ ,
\end{eqnarray}
where the function $N^\prime$ is obtained from $N$ by expressing ${\bf p}$ in
terms of ${\bf p}^\prime$ via Eq.~(\ref{eq:pppp}).  We have also changed
variables in the parametric integrals by introducing~\cite{Gross93}
\begin{equation}
 x=z_1+z_2\ , \qquad x\,y=(z_1-z_2)/2 \ ,
\end{equation}
such that
\begin{equation}
 \int_0^1 {\rm d}z_1\int_0^{1-z_1}\!\!\!{\rm d}z_2
 \longrightarrow\int_0^1{\rm d}x\, x \int_{-1/2}^{1/2}{\rm d} y \ .
\end{equation}
The function $N^\prime$ is a polynomial in ${\bf p}^{\, \prime}$, and
the ${\bf p}^{\, \prime}$-integrations are carried out in dimensional
regularization (see App.~A of Ref.~\cite{Pastore09}).  They are
finite and lead to the
charge operator given in Eq.~(\ref{eq:c_f}).

\begin{thebibliography}{100}
%
%
\bibitem{Park96}
T.-S.\ Park, D.-P.\ Min, and M.\ Rho,
Nucl.\ Phys.\ A {\bf 596}, 515 (1996).
%
\bibitem{Pastore09}
S.\ Pastore, L.\ Girlanda, R.\ Schiavilla, M.\ Viviani, and R.B.\ Wiringa,
Phys.\ Rev.\ C {\bf 80}, 034004 (2009).
%
\bibitem{Pastore11}
S.\ Pastore, L.\ Girlanda, R.\ Schiavilla, and M.\ Viviani,
Phys.\ Rev.\ C {\bf 84}, 024001 (2011).
%
\bibitem{Koelling09}
S.\ K\"olling, E.\ Epelbaum, H.\ Krebs, and U.-G.\ Meissner,
Phys.\ Rev.\ C {\bf 80}, 045502 (2009).
%
\bibitem{Koelling11}
S.\ K\"olling, E.\ Epelbaum, H.\ Krebs, and U.-G.\ Meissner,
Phys.\ Rev.\ C {\bf 84}, 054008 (2011).
%
\bibitem{Walzl01}
M.\ Walzl and U.-G.\ Meissner,
Phys.\ Lett.\ B {\bf 513}, 37 (2001).
%
\bibitem{Phillips03}
D.R.\ Phillips,
Phys.\ Lett.\ B {\bf 567}, 12 (2003).
%
\bibitem{Phillips07}
D.R.\ Phillips,
J.\ Phys.\  G {\bf 34}, 365 (2007). 
%
\bibitem{Koelling12}
S.\ K\"olling, E.\ Epelbaum, and D.R.\ Phillips,
arXiv:1209.083.
%
%
%
\bibitem{Hyde04}
C.E.\ Hyde-Wright and K.\ de Jager,
Ann.\ Rev.\ Nucl.\ Part.\ Sci.\ {\bf 54}, 217 (2004).
%
\bibitem{Hohler76}
G.\ H\"ohler {\it et al.},
Nucl.\ Phys.\ B {\bf 114}, 505 (1976).
%
\bibitem{Kubis01}
B.\ Kubis and U.-G.\ Meissner,
Nucl.\ Phys.\ A {\bf 679}, 698 (2001)
%
\bibitem{Carlson98}
J.\ Carlson and R.\ Schiavilla,
Rev.\ Mod.\ Phys.\ {\bf 70}, 743 (1998).
%
\bibitem{Riska89}
D.O.\ Riska,
Phys.\ Rep.\ {\bf 181}, 207 (1989).
%
\bibitem{Friar77}
J.L.\ Friar,
Ann.\ Phys.\ (N.Y.) {\bf 104}, 380 (1977);
Phys.\ Rev.\ C {\bf 22}, 796 (1980).
%
\bibitem{Wiringa95}
R.B.\ Wiringa, V.G.J.\ Stoks, and R.\ Schiavilla,
Phys.\ Rev.\ C {\bf 51}, 38 (1995).
%
\bibitem{Entem03} 
D.R.\ Entem and R.\ Machleidt,
Phys.\ Rev.\ C {\bf 68}, 041001 (2003);
R.\ Machleidt and D.R.\ Entem,
Phys.\ Rep.\ {\bf 503}, 1 (2011).
%
\bibitem{Carlson86}
C.E.\ Carlson,
Phys.\ Rev.\ D {\bf 34}, 2704 (1986).
%
\bibitem{Girlanda10a}
L.\ Girlanda, A.\ Kievsky, L.E.\ Marcucci, S.\ Pastore, R.\ Schiavilla, and M.\ Viviani,
Phys.\ Rev.\ Lett.\ {\bf 105}, 232502 (2010).

%
%
\bibitem{Schiavilla02}
R.\ Schiavilla and V.R.\ Pandharipande,
Phys.\ Rev.\ C {\bf 65}, 064009 (2002).
%
\bibitem{Schiavilla89}
R.\ Schiavilla, V.R.\ Pandharipande, and D.O.\ Riska,
Phys.\ Rev.\ C {\bf 40}, 2294 (1989).
%
\bibitem{Kievsky97} A.\ Kievsky, L.E.\ Marcucci, S.\ Rosati, and M.\ Viviani, 
Few-Body Syst.\ {\bf 22}, 1 (1997).
%
\bibitem{Viviani06} M.\ Viviani, L.E.\ Marcucci, S.\ Rosati, A.\ Kievsky, and L.\ Girlanda,
Few-Body Syst.\ {\bf 39}, 159 (2006).
%
\bibitem{Kievsky08} A.\ Kievsky, S.\ Rosati, M.\ Viviani, L.E.\ Marcucci, and L.\ Girlanda,
J.\ Phys.\ G: Nucl.\ Part.\ Phys. {\bf 35}, 063101 (2008).
%
\bibitem{Marcucci09}
L.E.\ Marcucci, A.\ Kievsky, L.\ Girlanda, S.\ Rosati, and M.\ Viviani,
Phys.\ Rev.\ C {\bf 80}, 034003 (2009).
%
\bibitem{deSwart98}
J.J.\ de Swart, M.C.M.\ Rentmeester, and R.G.E.\ Timmermans,
Triumf report TRI-97-1, 96 (1997), arXiv:nucl-th/9802084.
%
\bibitem{Pudliner95}
B.S.\ Pudliner, V.R.\ Pandharipande, J.\ Carlson, and R.B.\ Wiringa,
Phys.\ Rev.\ Lett.\ {\bf 74}, 4396 (1995).
%
\bibitem{Navratil07}
P.\ Navratil,
Few-Body Syst.\ {\bf 41}, 117 (2007).
%
\bibitem{Marcucci12}
L.E.\ Marcucci, A.\ Kievsky, S.\ Rosati, R.\ Schiavilla, and M.\ Viviani,
Phys.\ Rev.\ Lett.\ {\bf 108}, 052502 (2012).
%
\bibitem{Park00}
T.-S.\ Park {\it et al.}, Phys.\ Lett.\ B {\bf 472}, 232 (2000);
Y.-H.\ Song {\it et al.}, Phys.\ Lett.\ B {\bf 656}, 174 (2007);
Y.-H.\ Song, R.\ Lazauskas, and T.-S.\ Park, Phys.\ Rev.\ C {\bf 79}, 064002 (2009);
R.\ Lazauskas, Y.-H.\ Song, and T.-S.\ Park, Phys.\ Rev.\ C {\bf 83}, 034006 (2011).
%
%
%
\bibitem{Mohr05}
P.J.\ Mohr and B.N.\ Taylor,
Rev.\ Mod.\ Phys.\ {\bf 77}, 1 (2005).
%
\bibitem{Bishop79}
D.M.\ Bishop and L.M.\ Cheung,
Phys.\ Rev.\ A {\bf 20}, 381 (1979).
%
\bibitem{STANFORD65}
C.D.\ Buchanan and M.R.\ Yearian,
Phys.\ Rev.\ Lett.\ {\bf 15}, 303 (1965).
%
\bibitem{ORSAY66}
D.\ Benaksas, D.\ Drickey, and D.\ Fr\`{e}rejacque,
Phys.\ Rev.\  {\bf 148}, 1327 (1966).
%
\bibitem{CEA69}
J.E.\ Elias {\it et al.},
Phys.\ Rev.\ {\bf 177}, 2075 (1969).
%
\bibitem{DESY71}
S.\ Galster {\it et al.}, 
Nucl.\ Phys.\ B {\bf 32}, 221 (1971).
%
\bibitem{MONTEREY73}
R.W.\ Berard {\it et al},
Phys.\ Lett.\ B {\bf 47}, 355 (1973).
%
\bibitem{SLAC75}
R.G.\ Arnold {\it et al.},
Phys.\ Rev.\ Lett.\ {\bf 35}, 776 (1975).
%
\bibitem{MAINZ81}
G.G.\ Simon, C.\ Schmitt, and V.H.\ Walther,
Nucl.\ Phys.\ A {\bf 364}, 285 (1981).
%
\bibitem{BONN85}
R.\ Cramer {\it et al.},
Z.\ Phys.\ C {\bf 29}, 513 (1985).
%
\bibitem{SACLAY90}
S.\ Platchkov {\it et al.},
Nucl.\ Phys.\ A {\bf 510}, 740 (1990).
%
\bibitem{JLABHALLA99}
D.\ Abbott {\it et al.},
Phys.\ Rev.\ Lett.\ {\bf 82}, 1379 (1999).
%
\bibitem{JLABHALLC99}
L.C.\ Alexa {\it et al.},
Phys.\ Rev.\ Lett.\ {\bf 82}, 1374 (1999).
%
\bibitem{BATES84}
M.\ E.\ Schulze {\it et al.},
Phys.\ Rev.\ Lett.\ {\bf 52}, 597 (1984).
%
\bibitem{BATES91}
I.\ The {\it et al.},
Phys.\ Rev.\ Lett.\ {\bf 67}, 173 (1991).
%
\bibitem{BATES2011}
C.\ Zhang {\it et al.},
Phys.\ Rev.\ Lett.\ {\bf 107}, 252501 (2011).
%
\bibitem{VEPP85}
V.F.\ Dmitriev {\it et al.},
Phys. Lett. {\bf 157B}, 143 (1985).
%
\bibitem{VEPP86}
B.B.\ Wojtsekhowski {\it et al.},
Pis'ma\ Zh.\ Eksp.\ Teor.\ Fiz.\ {\bf 43}, 567 (1986).
%
\bibitem{VEPP90}
R.\ Gilman {\it et al.},
Phys.\ Rev.\ Lett.\ {\bf 65}, 1733 (1990).
%
\bibitem{VEPP2003}
D.M.\ Nikolenko {\it et al.}, 
Phys.\ Rev.\ Lett.\ {\bf 90}, 072501 (2003).
%
\bibitem{BONN91}
B.\ Boden {\it et al.}, 
Z.\ Phys.\ C {\bf 49}, 175 (1991).
%
\bibitem{NIKHEF96}
M.\ Ferro-Luzzi {\it et al.}, 
Phys.\ Rev.\ Lett.\ {77}, 2630 (1996).
%
\bibitem{NIKHEF99}
M.\ Bouwhuis  {\it et al.}, 
Phys.\ Rev.\ Lett.\ {\bf 82}, 3755 (1999).
%
\bibitem{JLAB2000}
D.\ Abbott  {\it et al.}, 
Phys.\ Rev.\ Lett.\ {\bf 84}, 5053 (2000).
%
\bibitem{SACLAY85}
S.\ Auffret {\it et al.},
Phys.\ Rev.\ Lett.\ {\bf 54}, 649 (1985).
%
\bibitem{SLAC90}
P.\ E.\ Bosted {\it et al.}, 
Phys.\ Rev.\ C 42, 38 (1990).
%
\bibitem{SICK2001}
I.\ Sick, 
Prog.\ Part.\ Nucl.\ Phys.\ {\bf 47}, 245 (2001).
%
\bibitem{AMROUN}
A.\ Amroun {\it et al.}, 
Nucl.\ Phys.\ A {\bf 579}, 596 (1994).
%
\bibitem{Epelbaum05}
E.\ Epelbaum, W.\ Gl\"ockle, and U.-G.\ Meissner,
Nucl.\ Phys.\ {\bf A747}, 362 (2005).
%
\bibitem{Pastore12}
S.\ Pastore, S.C.\ Pieper, R.\ Schiavilla, and R.B.\ Wiringa,
in preparation.
%
\bibitem{Pieper08}
S.C.\ Pieper,
AIP Conf.\ Proc.\ {\bf 1011}, 143 (2008).
%
\bibitem{Marcucci98}
L.E.\ Marcucci, D.O.\ Riska, and R.\ Schiavilla,
Phys.\ Rev.\ C {\bf 58}, 3069 (1998).
%
\bibitem{Marcucci05}
L.E.\ Marcucci, M.\ Viviani, R.\ Schiavilla, A.\ Kievsky, and S.\ Rosati,
Phys.\ Rev.\ {\bf 72}, 014001 (2005).
%
\bibitem{Okubo54}
S.\ Okubo,
Prog.\ Theor.\ Phys.\ {\bf 12}, 603 (1954).
%
\bibitem{Pastore08}
S.\ Pastore, R.\ Schiavilla, J.L.\ Goity,
Phys.\ Rev.\ C {\bf 78}, 064002 (2008).
%
%
\bibitem{Girlanda10}
L.\ Girlanda, S.\ Pastore, R.\ Schiavilla, and M.\ Viviani,
Phys.\ Rev.\ C {\bf 81}, 034005 (2010).
%
\bibitem{Gross93}
F.\ Gross,
{\it Relativistic Quantum Mechanics and Field Theory}
(John Wiley and Sons, Inc., 1993).
%
%
%
\end{thebibliography}
\end{document}